\documentclass[
reprint,
groupedaddress,
amsmath,amssymb,
aps,
prb,
floatfix,
]{revtex4-2}

\usepackage{graphicx}
\usepackage[caption=false]{subfig}
\usepackage{dcolumn}
\usepackage{bm}
\usepackage[hidelinks]{hyperref}
\usepackage[mathlines]{lineno}
\usepackage[capitalise]{cleveref}

\usepackage{physics}
\usepackage{xcolor}
\usepackage[caption=false]{subfig}
\usepackage{verbatim}
\usepackage{tikz}
\usepackage{tikz}
\usetikzlibrary{arrows.meta, positioning, shapes.multipart}

\usepackage{soul}

\newcommand{\blue}[0]{\color{black}}

\begin{document}

\preprint{APS/123-QED}

\title{Quantum Entanglement Phase Transitions and Computational Complexity: Insights from Ising Models}

\author{Hanchen Liu}
\email{hanchen.liu@bc.edu}
\author{Vikram Ravindranath}
\author{Xiao Chen}
\affiliation{Department of Physics, Boston College, Chestnut Hill, Massachusetts 02467, USA}

\date{\today}

\begin{abstract}

In this paper, we construct 2-dimensional bipartite cluster states and perform single-qubit measurements on the bulk qubits. We explore the entanglement scaling of the unmeasured 1-dimensional boundary state and show that under certain conditions, the boundary state can undergo a volume-law to an area-law entanglement transition driven by variations in the measurement angle. We bridge this boundary state entanglement transition and the measurement-induced phase transition in the non-unitary  circuit via the transfer matrix method. We also explore the application of this entanglement transition on the computational complexity problems. Specifically, we establish a relation between the boundary state entanglement transition and the sampling complexity of the bipartite 2D cluster state, which is directly related to the computational complexity of the corresponding Ising partition function with complex parameters. By examining the boundary state entanglement scaling, we numerically identify the parameter regime for which the  quantum state can be efficiently sampled, which indicates that the Ising partition function can be evaluated efficiently in such a region.
\end{abstract}

\maketitle

\tableofcontents

\section{Introduction}

\begin{figure*}[t!]
  \centering
  \includegraphics[width=0.85\textwidth]{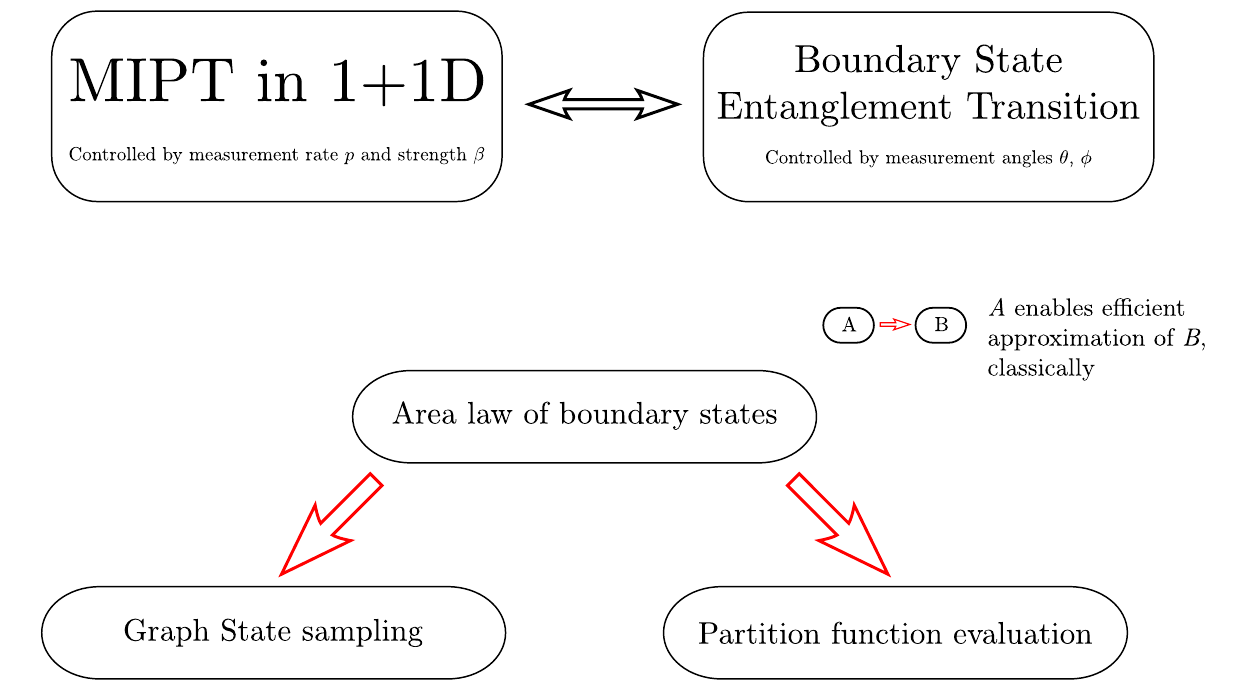}
  \caption{The roadmap of this study. We start by studying the sampling problem of the 2D toric code and the 2D graph state, relating it to the computation of Ising partition functions with complex weights. We then show that this 2D problem can be understood in a $(1 + 1)\mathrm{D}$ setup via the Ising transfer matrix method and the tensor-network-based method, thus relating it to the $(1 + 1)\mathrm{D}$ hybrid-circuit problem. The entanglement phase transition in $(1 + 1)\mathrm{D}$ hybrid circuits reflects the hardness of classical simulation of the sampling problem.} \label{fig: road_map}.
\end{figure*}

Measurement-induced phase transitions (MIPT) occurring within monitored quantum systems have garnered significant attention, inspiring extensive research to unravel their profound implications. Numerous studies have explored these transitions in various quantum systems, investigating their fundamental properties and experimental realization \cite{PhysRevB.98.205136, PhysRevB.99.224307, PhysRevB.100.134306,PhysRevLett.125.030505, PhysRevB.101.104301, PhysRevX.10.041020, PhysRevB.101.104302, skinner2019measurement, PhysRevB.103.104306,google2023measurement, PhysRevLett.130.220404, lavasani2021measurement, lavasani2022monitored, PhysRevLett.127.235701, PhysRevResearch.3.023200, PhysRevX.11.011030, PRXQuantum.2.030313,PhysRevB.108.094304}. One type of phase transition exists in the hybrid quantum circuit composed of entangling unitary gates and disentangling measurements. An entanglement transition manifests by tuning the measurement rate, transitioning from a volume-law phase to an area-law phase \cite{PhysRevB.98.205136, PhysRevB.99.224307, PhysRevB.100.134306,PhysRevLett.125.030505, PhysRevB.101.104302, skinner2019measurement, PhysRevX.10.041020, PhysRevB.101.104301, PhysRevB.103.104306}. Another example of entanglement phase transitions occurs in measurement-only circuits, where the transition is between different area-law phases, with a deviation from area-law scaling at the critical point \cite{lavasani2022monitored, lavasani2021measurement, PhysRevLett.127.235701, PhysRevResearch.3.023200, PRXQuantum.2.030313, PhysRevX.11.011030, PhysRevB.108.094304}. An illustrative example of such a system is the $ZZ/X$ competing measurement model in \cite{PRXQuantum.2.030313}, whose transition can be understood in classical percolation.

MIPT in the {\blue \((1+1)\mathrm{D}\) } hybrid quantum circuit naturally implies a complexity phase transition regarding simulability. The entanglement structure of a quantum system plays a crucial role in determining the computational difficulty of simulating the system on a classical computer. Simulating general volume-law entangled systems are widely recognized as challenging. Generic one-dimensional quantum systems with area-law entanglement scaling are often amenable to efficient classical simulation methods, such as matrix product states (MPS). However, higher-dimensional systems with area-law entanglement scaling (e.g., projected entangled pair states (PEPS)) or systems with volume-law entanglement scaling (e.g., high-depth random quantum circuits) tend to present significant challenges for efficient classical simulation due to the rapid growth of entanglement or computational complexity~\cite{ORUS2014117, PhysRevLett.100.030504}.

Moreover, MIPT is closely related to other computational challenges. In the context of the Random Circuit Sampling (RCS) problem, which is commonly employed to assess quantum advantage~\cite{arute2019quantum}, classical simulation is believed to be computationally arduous~\cite{terhal2004adaptive, bouland2019complexity}. Recent findings~\cite{PhysRevX.12.021021, cheng2023efficient} indicate that for RCS with a finite depth $d$, approximate sampling becomes viable for circuits with depths up to $d_c$, a finite critical depth. Conversely, approximating the outcomes becomes computationally demanding when the circuit depth exceeds $d_c$. The complexity of this sampling problem can be understood by treating one spatial dimension as the time axis and interpreting circuit depth as the inverse of the measurement rate $p = 1/d$ in a hybrid circuit. As a result, a connection emerges between the challenge of approximating outputs in shallow 2D circuits and the transition from a {\blue \((1+1)\mathrm{D}\) }  to an area-law entanglement phase. Recent research on MIPT in shallow circuits also suggests the existence of entanglement phase transitions induced by measurements on 2D quantum states prepared by shallow circuits~\cite{PhysRevB.106.144311, bao2022finite}.

Motivated by these advancements, in our recent work~\cite{PhysRevB.106.144311}, we explicitly consider 2D quantum states generated by various Clifford shallow circuits. We then investigate the entanglement structure of the boundary 1D quantum state after measuring all the bulk qubits. Remarkably, we demonstrate that this boundary state undergoes a volume-law to area-law entanglement phase transition in various shallow circuits, and similar ideas were also discussed in \cite{bao2022finite}. In particular, we consider a cluster state generated on a square lattice and employ random single qubit X or Z measurements for the bulk qubits. Our findings reveal that by manipulating the ratio between X and Z measurements, the boundary 1D state undergoes the volume-law to area-law entanglement transition.

In this paper, we aim to gain a deeper understanding of the boundary entanglement phase transition in various 2D cluster states and its connection with transitions in computational complexity. To achieve this objective, we take the three following approaches: (1) we establish a connection between the sampling problem of the 2D cluster state defined on the bipartite lattice and the evaluation of the Ising partition function with complex parameters. (2) We reveal that the 1D boundary state carries essential information about the 2D bulk Ising partition function. We also illustrate that the boundary state can be generated through {\blue \((1+1)\mathrm{D}\) } non-unitary dynamics. (3) We undertake numerical evaluations of the entanglement entropy of the 1D boundary state by performing single qubit measurements on the bulk qubits. We observe and analyze various entanglement phase transitions by varying the measurement angle on the Bloch sphere or the ratio between X/Y/Z Pauli measurements. 

Through these three approaches, we effectively demonstrate that the boundary entanglement transition in the cluster state is closely linked to the complexity transition of the 2D Ising models. Specifically, in the area-law phase, the corresponding 2D Ising partition function is easy to evaluate. Our findings shed light on the intricate interplay between boundary entanglement scaling, computational complexity, and Ising models with complex parameters. Moreover, we are able to determine certain regimes where these computational tasks are guaranteed to be easy, and relate them to previously studied problems in quantum information.

The paper is organized as follows. In ~\cref{sec: prelim,sec: boundary_states}, we show how certain resource states and their boundary states can be written as superpositions of states, with amplitudes given by Ising partition functions. In \cref{sec: dynamics}, we describe how the boundary states can be thought of as being generated by a 1D nonunitary circuit. The connections between the aforementioned computational tasks are elucidated in \cref{sec:connections}, with numerical evidence for our results presented in \cref{sec: numerics}. A road map of this work is shown in \cref{fig: road_map}. 

\section{Resource States and Ising Partition Functions}
\label{sec: prelim}

In this section, we examine the relationship between the toric code, cluster state, and the classical Ising partition function, following similar steps as outlined in previous works \cite{PhysRevA.76.022304, PhysRevLett.100.110501, de2009completeness}. We first introduce the measurement basis 
\begin{equation}\label{eq: mea_basis}
    \mathcal B(w) = \{ \ket{\mu, w} \equiv \bigotimes_{i\in \Lambda} \ket{\mu_i, w_{i}}_i \mid  \mu_i \in \mathbb{Z}_2\}
\end{equation} where $\Lambda$ is the set of all qubits, $w \equiv \{w_{i} \} \in \mathbb{C}^{|\Lambda|}$ labels the measurement direction configuration, $\mu \equiv \{\mu_i \} \in \mathbb{Z}_2^{|\Lambda|}$ is the measurement outcome configuration, and $\ket{\ldots}_i$ is the quantum state defined on local Hilbert space of qubit $i$. 

The local projective measurement on qubit $i$ along $\hat O_i = \cos \theta_i Z_i + \sin \theta_i \cos \varphi_i X_i + \sin \theta_i \sin \varphi_i Y_i$, where $(\theta_i, \varphi_i)$ is given by the standard definition of spherical coordinates, collapses the single-site wave function on qubit $i$ to a measurement state
\begin{equation}
\begin{aligned}
    \ket{+, (\theta_i, \varphi_i)}_{i} &= \cos\frac{\theta_i}{2}\ket{0}_i + \exp(i \varphi_i) \sin \frac{\theta}{2}\ket{1}_i \\ 
    \ket{-, (\theta_i, \varphi_i)}_{i} &= \sin\frac{\theta_i}{2}\ket{0}_i - \exp(i \varphi_i) \cos \frac{\theta}{2}\ket{1}_i
\end{aligned}
\end{equation}
where $\ket{0}_i$ and $\ket{1}_i$ are the eigenstates of Pauli $Z$ on qubit $i$, that satisfy $Z_i \ket{0}_i = \ket{0}_i$, $Z_i\ket{1}_i = - \ket{1}_i$. We define 
\begin{equation}\label{eq: measurement parameter}
     w_{i} \equiv \tan (\theta_i/2) \exp(i \varphi_i ),
\end{equation}
and then introduce the weight parameter $W_i$
\begin{equation}
\label{eq: weight_parameter}
    W_{i} \equiv\left\{
    \begin{matrix}
        w_{i} & \text{if} & \mu = +1\\
         -\left[w_{i}^*\right]^{-1} & \text{if}& \mu = -1
    \end{matrix}
    \right.,
\end{equation}
where $(\cdots)^*$ is the complex conjugate. The measurement state is further written more compactly as
\begin{equation}
\ket{\mu_i, w_{i}}_i = N_{i} \left(\ket{0}_i + W_{i}\ket{1}_i\right)
\label{eq: local_basis}
\end{equation}
where $N_i = 1/\sqrt{1 + |W_{i}|^{2}}$ is the normalization factor, $w_i$ is the measurement direction parameter defined in \cref{eq: measurement parameter}, and $\mu$ is the measurement outcome. It is easy to show that for a given measurement direction configuration $\{\theta_i, \varphi_i \}$, the set of measurement states $\{\bigotimes_{i\in\Lambda} \ket{\mu_i, w_i}_i, \mu_i = \pm 1\}$ form  an orthogonal basis $\mathcal{B}(w)$ of the Hilbert space of qubit set $\Lambda$, and we define such basis to be the measurement basis. One thing worth noticing is that the weight parameter $W_i$ stands for the ratio between the coefficients of $\ket{0}$ and $\ket{1}$, meaning that $W_i$ is independent of the normalization factor. 

We now try to expand a multi-qubit wave function $\ket{\Psi}$ in the measurement basis. Formally, it can be written as
\begin{equation}
    \ket{\Psi} = \sum_{\mu} \Omega_{w,\mu} \ket{\mu,w},
    \label{eq:wf_2d}
\end{equation}
where $\Omega_{w,\mu} = \braket{\mu,w}{\Psi}$ is the overlap between measurement  basis state $\ket{\mu,w} \in \mathcal{B}(\omega)$ and the multi-qubit wave function $|\Psi\rangle$. The summation is over all the allowed measurement outcome configurations $\mu$. We will show later in this section that the overlap function can be written in the form of the partition function of classical spin systems
\begin{equation}
    \Omega_{w,\mu} \propto \mathcal{Z}(K, h).
\end{equation}
with the effective coupling $K$ and effective magnetic field $h$ determined by the measurement outcome $\mu$, and measurement direction $\omega$.

\subsection{The Toric Code State}
\begin{figure}[ht]
    \centering
    \includegraphics[width=0.8\linewidth]{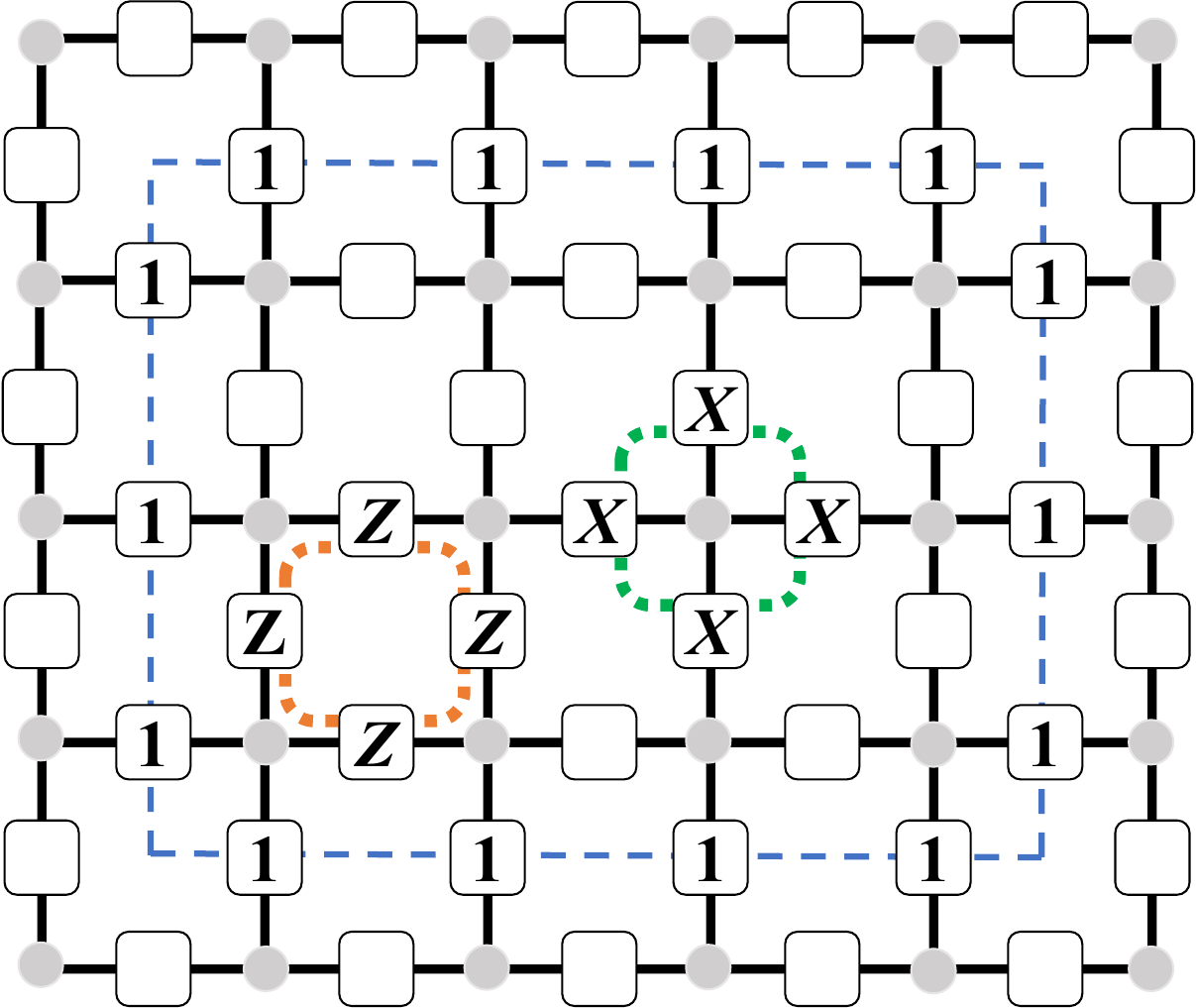}
    \caption{Toric code conventions: green dashed line represents the vertex operator $A_v$, the orange dashed line represents the plaquette operator $B_p$, and the blue dashed line gives an example of the quantum state related to closed-loop $l$.}
    \label{fig:toric_code_conven}
\end{figure}

 The toric code state gives a simple example. The toric code state, as first introduced in \cite{KITAEV20032}, is the ground state of the Hamiltonian given by
\begin{equation}
    H = -\sum_v A_v - \sum_p B_p,
\end{equation}
where the qubits are defined on the edges of a square lattice. Here, $A_v = \prod_{i\in v} X_i$ represents the vertex operator, which is the product of Pauli X operators enclosing the vertex $v$, and $B_p = \prod_{j \in p} Z_j$ represents the plaquette operator, which is the product of Pauli Z operators associated with the  plaquette $p$. The ground state of such a Hamiltonian takes the form
\begin{equation}\label{eq: tc_ground_state}
    \ket{\Psi}^{TC} = \frac{1}{\sqrt{|\mathcal{C}|}} \sum_{l \in \mathcal{C}} \ket{l},
\end{equation}
 where $\mathcal{C}$ is the set of configurations of  loops closed with respect to the boundary and formed by connecting edges on the dual lattice, $|\mathcal{C}|$ is the number of the loops, and $\ket{l}$ is the quantum state related to loop configuration $l \in \mathcal C$. The explicit form of $|l\rangle$ is 
\begin{equation}
    \ket{l} = \bigotimes_{i \in l, j \not \in l} \ket{1}_i \otimes \ket{0}_j.
\end{equation}
An example of such a loop is shown in Figure.~\ref{fig:toric_code_conven}.

We now calculate the overlap function $\Omega_{w,\mu}$ for the toric code state. Utilizing \cref{eq: local_basis}, we can demonstrate that the overlap function is given by:
\begin{equation}
    \Omega_{w,\mu} = \frac{ \prod_{i} N_i }{\sqrt{|\mathcal C|}} \sum_{l \in \mathcal{C}} W_l.
\end{equation}
Each loop configuration takes a weight of $W_l = \prod_{i \in l} W_{i}$ that can be effectively treated as the Boltzmann weight for a domain wall configuration in the Ising model. The overlap function is thus
\begin{equation}\label{eq: toric_ising}
    \Omega_{w,\mu} \propto \sum_{\{s\}} \exp\left(\sum_{i = \langle m, n\rangle } K_i s_m s_n\right),
\end{equation}
where 
\begin{equation}\label{eq: toric_coupling}
    \exp(-2 K_i) = W_{i},
\end{equation}
with $i = \langle m, n \rangle$ being the edge connecting site neighboring site  $m$ and $n$ and the classical spins $\{s\}$ are defined on the vertices of the square lattice given in Fig.~\ref{fig:toric_code_conven}.  Here the classical spin variables $\{s\}$  take $\mathbb{Z}_2$ value $\pm 1$.

\subsection{The Bipartite Cluster State}\label{sec: bipartite_graph}
\begin{figure}[ht]
\centering
\subfloat[Lieb lattice]{\label{fig: lieb_bip}\includegraphics[width=0.2\textwidth]{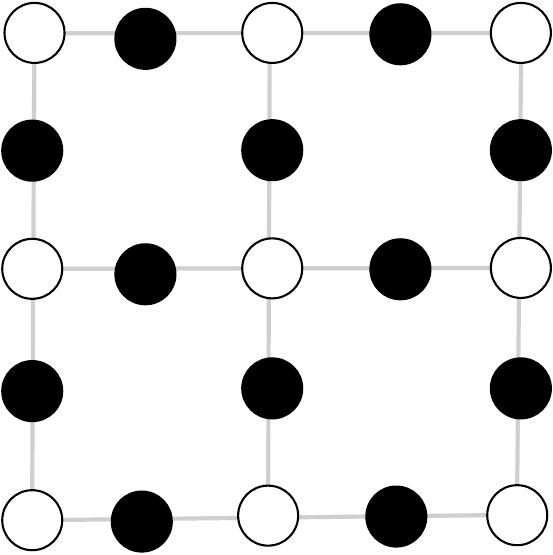}}\qquad
\subfloat[Square lattice]{\label{fig: sq_lattice}\includegraphics[width=0.2\textwidth]{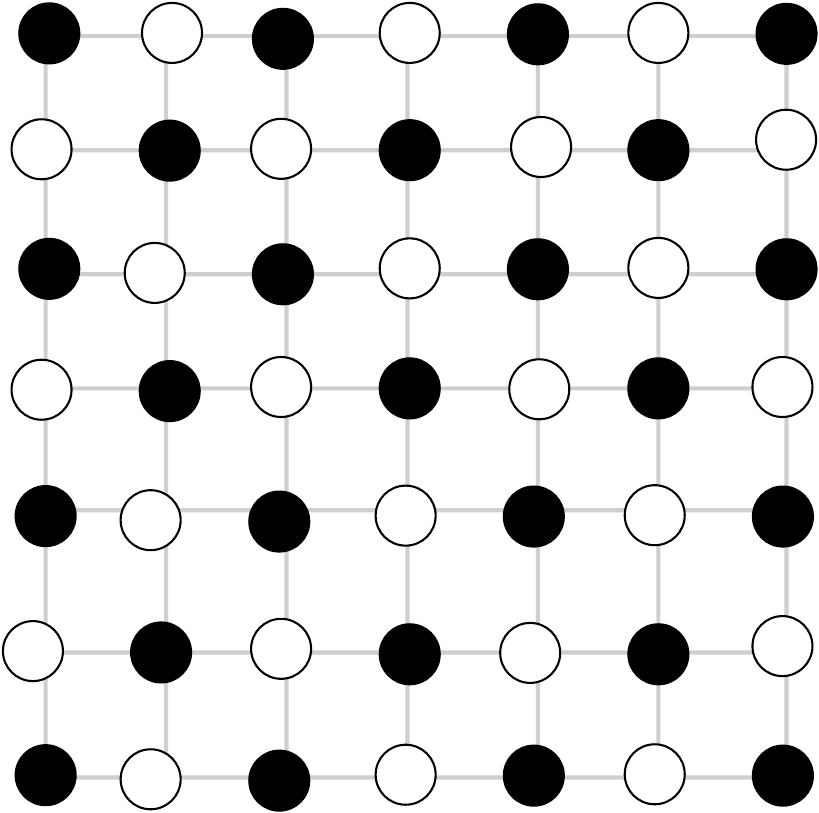}}
\caption{\label{fig: latt_bipartition} Lattice bipartitions}
\end{figure}

We now consider cluster states defined on bipartite graphs. The cluster state is a stabilizer state defined on a graph $G = (V, E)$, where the stabilizer generator takes the form
\begin{equation}
    g_{i} = X_i \prod_{j \in \mathcal N_i} Z_j \quad i \in V.
    \label{eq:clusDefn}
\end{equation}
where $j \in \mathcal{N}_i$ is the neighbor of vertex $i$, and the cluster state $\ket{\Psi}$ satisfies the stabilizer condition $g_i \ket{\Psi} = \ket{\Psi}$, for all sites $i$. For a bipartite graph $G$, the stabilizer generators naturally split into two sets:
\begin{equation}\label{eq: cluster_stab}
\begin{aligned}
     g_i &= X_i \prod_{j \in \mathcal{N}_i} Z_j \quad \text{for } i \in \circ\\
     g_m &= X_m \prod_{n \in \mathcal{N}_m} Z_n \quad \text{for } m \in \bullet,
\end{aligned}
\end{equation}
where $\mathcal{N}_{i/m}$ denotes the neighbors of site $i/m$, and $\circ, \bullet \subset V$ labels the bipartition of graph vertices. Two examples of graph bipartition are shown in Fig.~\ref{fig: latt_bipartition}.

Applying Hadamard rotation $\operatorname{H}$ on the $\bullet \in V$ sites of the bipartite cluster state, we have the Hadamard rotated state $\ket{\Psi}_H = \prod_{i\in \bullet} \operatorname{H}_i \ket{\Psi}$, which is stabilized under
\begin{equation}
\begin{aligned}
    g_i^X &= X_i \prod_{j \in \mathcal{N}_i} X_j \quad \text{for } i \in \circ\\
    g_m^Z &= Z_m \prod_{n \in \mathcal{N}_m} Z_n \quad \text{for } m \in \bullet,
\end{aligned}
\end{equation}
as presented in \cite{PhysRevLett.100.110501}. The rotated quantum state $\ket{\Psi}_H$ can then be expressed as
\begin{equation}
\begin{aligned}
    \ket{\Psi}_H &= \frac{1}{2^N}\prod_{m \in \bullet } (1 + g_m^Z)\prod_{i\in \circ} (1 + g_i^X)\bigotimes_{k \in \Lambda} \ket{0}_k\\
    &= \frac{1}{2^{|\circ|}}\prod_{i\in \circ} (1 + g_i^X)\bigotimes_{k \in \Lambda} \ket{0}_k\\
    &= \frac{1}{2^{|\circ|}} \sum_{A} g^X_A \bigotimes_{k \in \Lambda} \ket{0}_k
\end{aligned}
\end{equation}
  all possible regions $A$ consisting of the $\circ$ sites and $g_A^X = \prod_{i\in A} g_i^X$. Here the rotated cluster state $\ket{\Psi}_H$ is given by applying projection operators $ 1/2 (1 + g_m^Z)$ and $1/2(1 + g_i^X)$ to an initial state $\bigotimes_{k \in \Lambda} \ket{0}_k$. Taking advantage of the fact that $X \ket{ 0 } = \ket{1}$ and $Z\ket{0} = \ket{0}$, we write the rotated cluster state in the following form
\begin{equation}
    \ket{\Psi}_H = \frac{1}{2^{|\circ|}}\sum_{A} \ket{A}\otimes \ket{\partial A}\otimes\ket{\overline{A}}.
\end{equation}
We use $A$ to denote $\circ$ sites that support $g_A^X$ and $\partial A$ to denote neighboring sites of $A$, which, because the lattice is bipartite, are $\bullet$, and $\overline{A}$. The quantum states $\ket{A} = \bigotimes_{A} \ket{1}$  and $\ket{\overline{A}} = \bigotimes_{\overline{A}} \ket{0}$. As for $\ket{\partial A}$, we take
\begin{equation}
    \ket{\partial A} = \bigotimes_{i\in \mathcal N_E, j \in \mathcal N_O} \ket{0}_i\ket{1}_j
\end{equation}
where $\mathcal N_{E/O}$ denotes $\bullet$ sites that have even/odd number of $g_i^X$'s operating on the its $\circ$ neighbours. The overlap function is
\begin{equation}
    \Omega_{w,\mu} \propto \sum_{A} W_A \times W_{\partial A}
\end{equation}
where 
\begin{equation*}
\begin{aligned}
     W_A &= \braket{\mu_{A}, w_{A} }{A} = \prod_{i \in A} W_{i} \\
     W_{\partial A} &= \braket{\mu_{\partial A}, w_{\partial A}}{\partial A} = \prod_{i \in \mathcal{N}_O} W_{i}
\end{aligned}
\end{equation*}
with $\ket{\mu_A, w_A}$ and $\ket{\mu_{\partial A}, w_{\partial A}}$ being the measurement basis on $A$ and $\partial A$
\begin{equation}
    \begin{aligned}
        \ket{\mu_{A}, w_{A}} &= \bigotimes_{j\in A} \ket{\mu_j ,w_j}_j\\
        \ket{\mu_{\partial A}, w_{\partial A}} &= \bigotimes_{i\in \partial A} \ket{\mu_i ,w_i}_i.
    \end{aligned}
\end{equation}
When introducing the Ising degrees of freedom $\{s\}$ on $\circ$ sites, $W_A$  can be treated as the onsite magnetic field term whereas the boundary $W_{\partial A}$ contributes to the spin interaction. Thus, the overlap function can be expressed in the form of an Ising partition function,
\begin{equation}\label{eq: cluster_ising}
\begin{aligned}
    \braket{\mu, w}{\Psi}_H \propto &\\
     \sum_{s} &\exp\left(\sum_{i \in \bullet} K_i \prod_{j \in \mathcal{N}_i | j \in \circ} s_j + \sum_{k \in \circ} h_k s_k\right).
\end{aligned}
\end{equation}
where the classical spins $\{s_i\}$ are defined on the $\circ$ sublattice. The effective multi-spin coupling constant $K_i$ is given by $\exp(-2 K_i ) = W_{i}$ with $i\in \bullet$, and the onsite effective magnetic field is obtained from $\exp(-2 h_k) = W_k $ with $k\in \circ$. The overlap function of $\ket{\Psi}$ without Hadamard rotation can be obtained using the identity
\begin{equation}
    \braket{\mu,w}{\Psi} = \bra{\mu, w} \prod_{i\in \bullet} H_i^\dagger H_i \ket{\Psi} = \bra{\mu, w'}\ket{\Psi}_H
\end{equation}
where $W'_i = W_{i} $ for $i\in \circ$, and $W'_i = (1 - W_{i})/(1 + W_{i})$ for $i \in \bullet$ due to the Hadamard rotation on site $i \in \bullet$. The correspondence between the measurement parameter $\{W_{i}, W_k\}$ and the Ising parameters $\{K_i, h_k\}$ are now
\begin{equation}\label{eq: ising_para}
    \begin{matrix}
        &\exp(-2 h_k) = W_{k} &\quad k \in \circ\\
        &\exp(-2 K_i) =\frac{ 1 - W_{i}}{1 + W_{i}}&\quad i \in \bullet
    \end{matrix}.
\end{equation}
It should be noted that the variable $W_{i}$ is generally complex, implying that both $K_j$ and $h_j$ can take complex values. In Appendix~\ref{ap: graph_ground}, an alternative approach for establishing a connection between the cluster state overlap function $\Omega_{w,\mu}$ and the Ising partition function $\mathcal{Z}$ is presented.

\begin{figure}[ht]
    \includegraphics[width=0.8\linewidth]{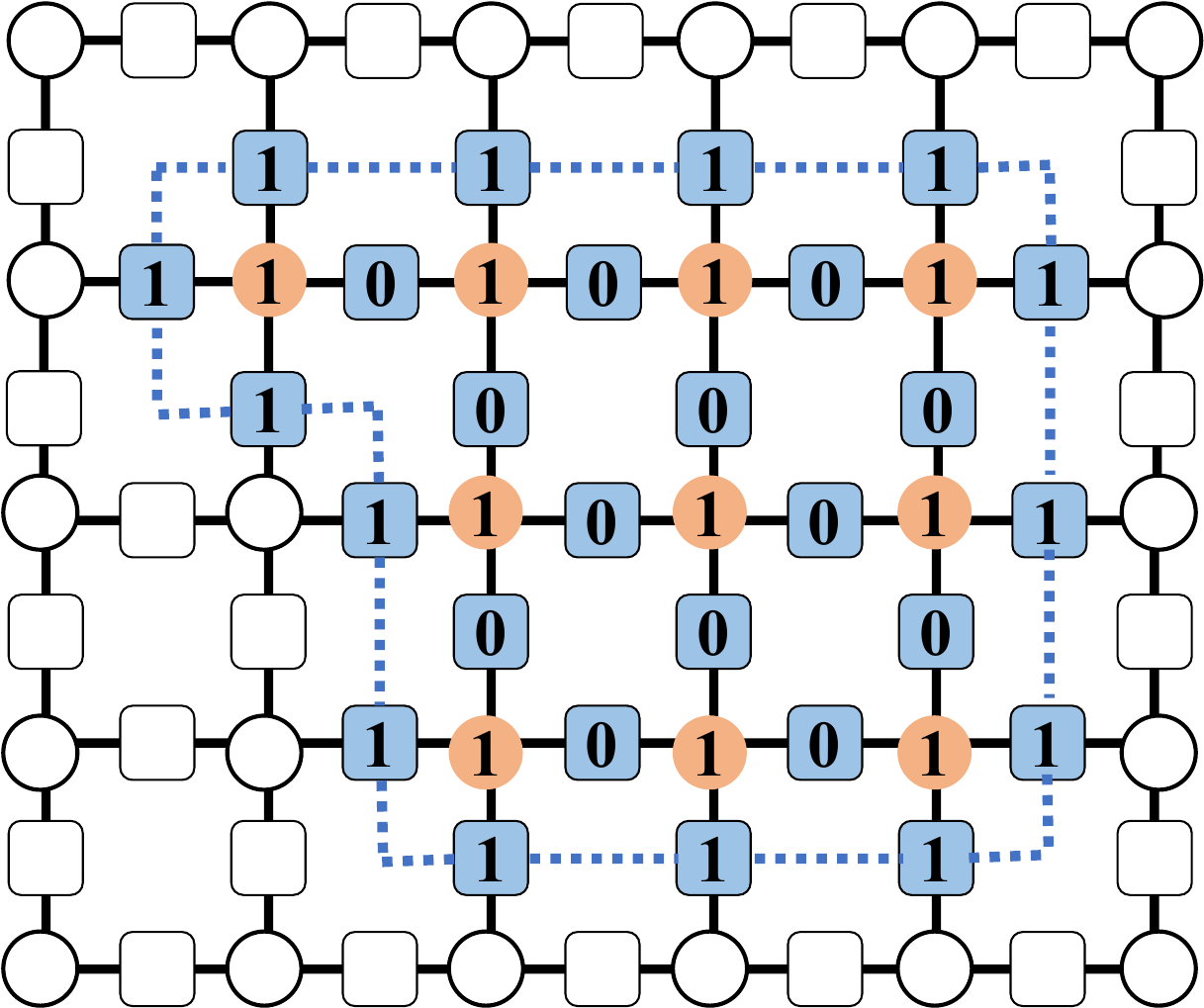}
    \caption{Example structure of wave function $\ket{A} \otimes \ket{\partial A} \otimes \ket{\overline{A}}$. The orange dots are $\ket{A}$, $\ket{\partial A}$ is colored in Blue, and the un-colored part labels $\ket{\overline{A}}$. The blue dashed line labels the $\prod_{j\in\mathcal N_O}\ket{1}_j$ part of $\ket{\partial A}$}
    \label{fig: lieb_conven}
\end{figure}

One simple example is the cluster state defined on the Lieb lattice, the bipartition of such lattice is shown in Fig.~\ref{fig: lieb_bip}, where the vertices are labeled ``$\circ$" while the edges are ``$\bullet$", and an example of $\ket{A} \otimes \ket{\partial A} \otimes \ket{\overline{A}}$ is shown in Fig.~\ref{fig: lieb_conven}. Following the same logic, we write down the overlap function as
\begin{equation}
    \Omega_{w,\mu} \propto \sum_{s} \exp\left(\sum_{m = \langle i, j \rangle} K_m s_i s_j + \sum_k h_k s_k\right).
    \label{eq: lieb_ising}
\end{equation}
This is the partition function of the Ising model with nearest neighbor coupling in the presence of the magnetic field. The classical Ising spin is defined on the $\circ$ sublattice, which forms a square lattice. The parameters $K_m$ and $h_k$ are given by Eq. \ref{eq: ising_para}.

An interesting scenario is when taking the measurement on the vertex to be  in direction $X$ and forcing the measurement outcome to be $\mu_i = +$, which means that for vertex sites $w_i = 1$ and thus $h_i = 0$. The wave function with forced $X$ measurements on the vertices takes the form
\begin{equation}
    \ket{\Psi}^+ = \left(\sum_{\mu_\bullet} \Omega_{w,\mu,h = 0} \ket{\mu, w}_\bullet \right ) \otimes\ket{+}_\circ
    \label{eq: tc1}
\end{equation}
where $\{\ket{\mu, w}_\bullet\}$ is the edge measurement basis, and $\ket{+}_\circ \equiv \bigotimes_{i \in \circ} \ket{+}_i$ is the tensor product of $+$ eigenstates of Pauli $X$ on the vertices. As $\Omega_{w,\mu,h = 0}$ is the same as the overlap function of the toric code shown in \cref{eq: toric_ising} up to Hadamard rotations, we thus have
\begin{equation}\label{eq: tc2}
    \ket{\Psi}^+ = \ket{\Psi}^{TC}_H\otimes\ket{+}_\circ
\end{equation}
with
\begin{equation}\label{eq: tc3}
    \ket{\Psi}^{TC}_H = \frac{1}{\sqrt{|\mathcal{C}|}} \sum_{l \in \mathcal{C}} \ket{l}_H ~\text{and}~  \ket{l} = \bigotimes_{i \in l, j \not \in l} \ket{+}_i \otimes \ket{-}_j
\end{equation}
where $X_i\ket{\pm}_i = \pm \ket{\pm}_i$. One subtlety here is that the quantum state obtained via vertex $X$ measurement differs from the toric code state shown in \cref{eq: tc_ground_state} by Hadamard rotations on each unmeasured site, and the loops are represented in $\ket{\pm}$ instead of $\ket{0},\ket{1}$. 

The $X$ measurements on the vertices can also be interpreted as enforcing flux constraints in $\mathbb{Z}_2$ gauge theory:

\begin{equation}
\begin{aligned}
    \mathcal{Z} = \sum_{\tau} \exp\left(\sum_{p = \langle a, b\rangle'} \tilde{h_p} \tau_p \right)\prod_{i}(1 + m_i \prod_{q\in \square_i} \tau_q).
\end{aligned}
\end{equation}
Here, $\tau_p$ represents the $\mathbb{Z}_2$ gauge field defined on edge $p$, $a,b$ are neighboring plaquettes that share the same edge $p$, and $\prod_{i}(1 + m_i \prod_{q\in \square_i} \tau_q)$ denotes the flux constraint associated with the $X$ measurement outcome $m_i = \pm 1$ on vertex $i$, with $\prod_{q\in \square_i}$ being the product over edges sharing the same vertex $i$. By enforcing $m_i = 1$ for all vertices $i$, we recover the Ising partition function by directly setting $\tau_q = \sigma_a \sigma_b$, where $\sigma_a \sigma_b$ are $\mathbb{Z}_2$ variables assigned to plaquettes $a$ and $b$. {\blue A similar construction is previously studied in the context of the symmetry protected topological states in~\cite{lee2022measurement, zhu2022nishimoris}.} A detailed discussion can be found in Appendix~\ref{ap: graph_dual}.

Another example is the cluster
state defined on the square lattice, found in our previous work~\cite{PhysRevB.106.144311}. With the bipartition shown in Fig.~\ref{fig: sq_lattice}, the overlap function for this case is
\begin{equation}\label{eq: square_ising}
    \Omega_{w, \mu} \propto\sum_{s} \exp (\sum_{i\in \bullet}  K_{i}s_a s_b s_c s_d + \sum_{r \in \circ} h_r s_r)
\end{equation}
where the Ising spins $\{s\}$ locate on $\circ$ sites. The four-body Ising interaction $K_{i}s_a s_b s_c s_d$ is defined on four neighboring sites $a, b, c, d \in \circ$ enclosing site $i\in \bullet$. The effective Ising coupling $\{K_i\}$ and magnetic field $\{h_r\}$ are determined by \cref{eq: ising_para}. 

In summary, we show that the overlap function for the ground state of the toric code model and cluster state defined on the bipartite graph can be expressed in terms of classical Ising partition function $\Omega_{w, \mu} = \mathcal{Z}(K, h)$, as shown in \cref{eq: toric_ising,eq: lieb_ising,eq: square_ising}. In the subsequent sections of this paper, we will establish the connection between the ease of approximating this partition function and the entanglement scaling of the 1D boundary state. We will show that by varying the measurement direction $\{w\}$, the boundary state can undergo a volume-law to an area-law entanglement phase transition. This transition, in turn, uncovers a regime of measurement directions where the resource state sampling problem .

\section{Boundary States and Ising Partition Functions}\label{sec: boundary_states}

\begin{figure}[ht]
    \includegraphics[width=0.8\linewidth]{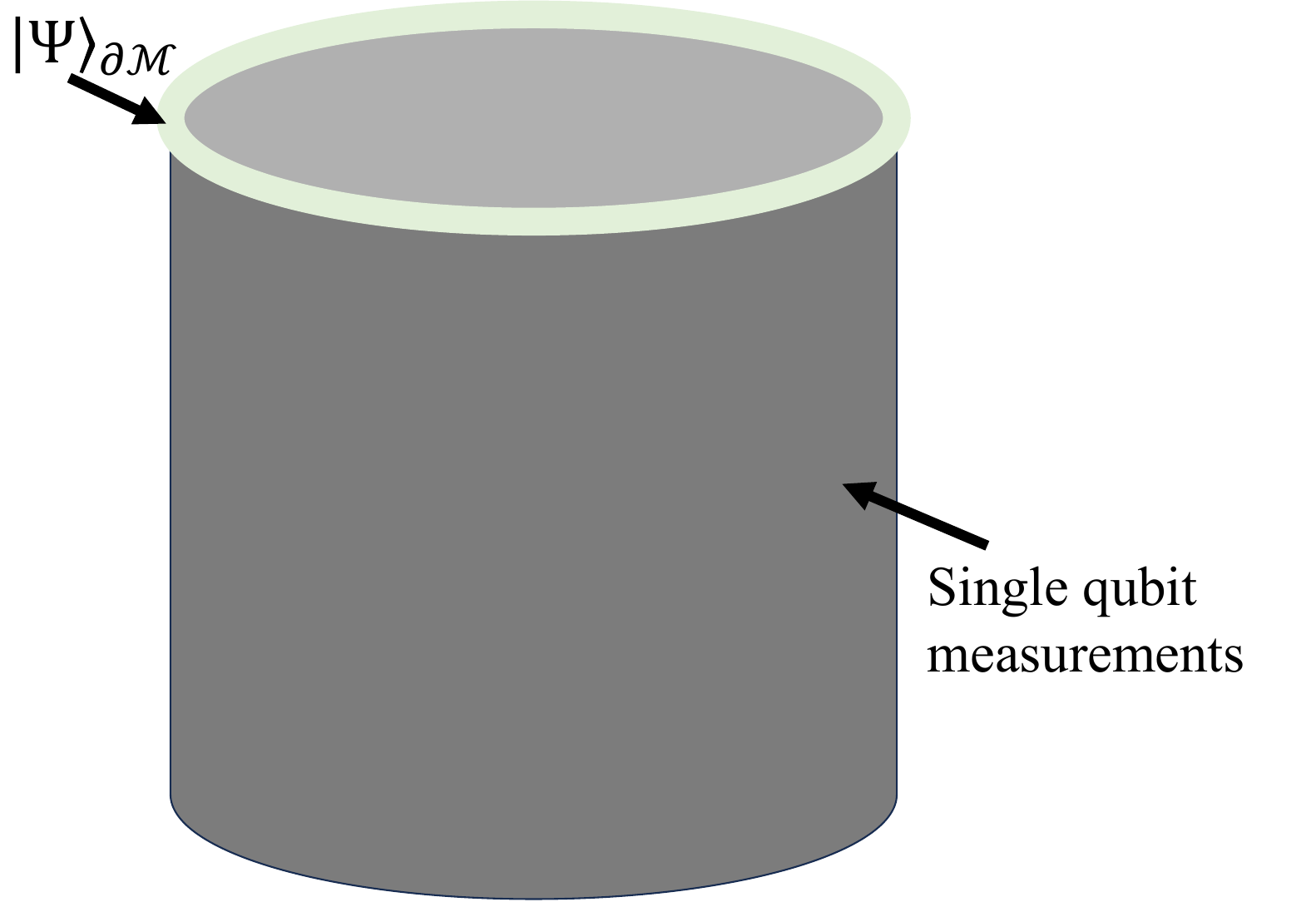}
    \caption{The 2D quantum state is defined on the half-cylinder $\mathcal M$. We perform single-qubit  measurements in the bulk and leave the boundary $\partial \mathcal M$ (top circle) qubits unmeasured. $\ket{\Psi}_{\partial \mathcal M}$ denotes the 1D wavefunction on the unmeasured boundary.}
    \label{fig: cylinder} 
\end{figure}

In this section, we explore the boundary state that arises from measuring the bulk qubits in the previously discussed 2D quantum states. Specifically, we examine the 2D quantum state supported on a half-cylinder denoted as $\mathcal M$, as illustrated in Fig.~\ref{fig: cylinder}. By using Eq.\eqref{eq:wf_2d}, we have 
\begin{equation}
\ket{\Psi}_{\partial \mathcal M} \propto\sum_{\mu_{\partial \mathcal M}}\sum_s \exp{(H[s])} \ket{\mu_{\partial\mathcal M}, w_{\partial \mathcal M}},
\end{equation}
where the summation $ \sum_{\mu_{\partial \mathcal M}} \sum_{s}$ is over all the spin configurations and boundary basis, and 
\begin{equation}
    \sum_s \exp{(H[s])}  \propto \braket{\mu_{\mathcal M},w_{\mathcal M}, \mu_{\partial \mathcal M}, w_{\partial \mathcal M}}{\Psi}
\end{equation}
where $\ket{\mu_{\mathcal M},w_{\mathcal M}, \mu_{\partial \mathcal M}, w_{\partial \mathcal M}} \equiv \ket{\mu_{\mathcal M},w_{\mathcal M}}\otimes \ket{\mu_{\partial \mathcal M}, w_{\partial M }}$ is the tensor product of the bulk measurement basis $\{\ket{\mu_{\mathcal M}, w_{\mathcal M}}\}$ and the boundary basis $\{\ket{\mu_{\partial \mathcal M}, w_{\partial \mathcal M}}\}$. We will further show that by properly choosing the boundary basis $\{\ket{\mu_{\partial \mathcal M}, w_{\partial \mathcal M}}\}$, the boundary state can be written in a more compact form as
\begin{equation}
\label{eq: general_boundary_state}
\ket{\Psi}_{\partial \mathcal M} \propto \sum_{s} \exp{(H[s])} \ket{s_{\partial \mathcal M}},
\end{equation}
where $\{\ket{s_{\partial \mathcal M}}\}$ is a set of complete basis  states of the boundary labeled by classical boundary spin configuration $\{s_{\partial \mathcal M}\}$.

\begin{figure}[ht]
\centering\includegraphics[width=\linewidth]{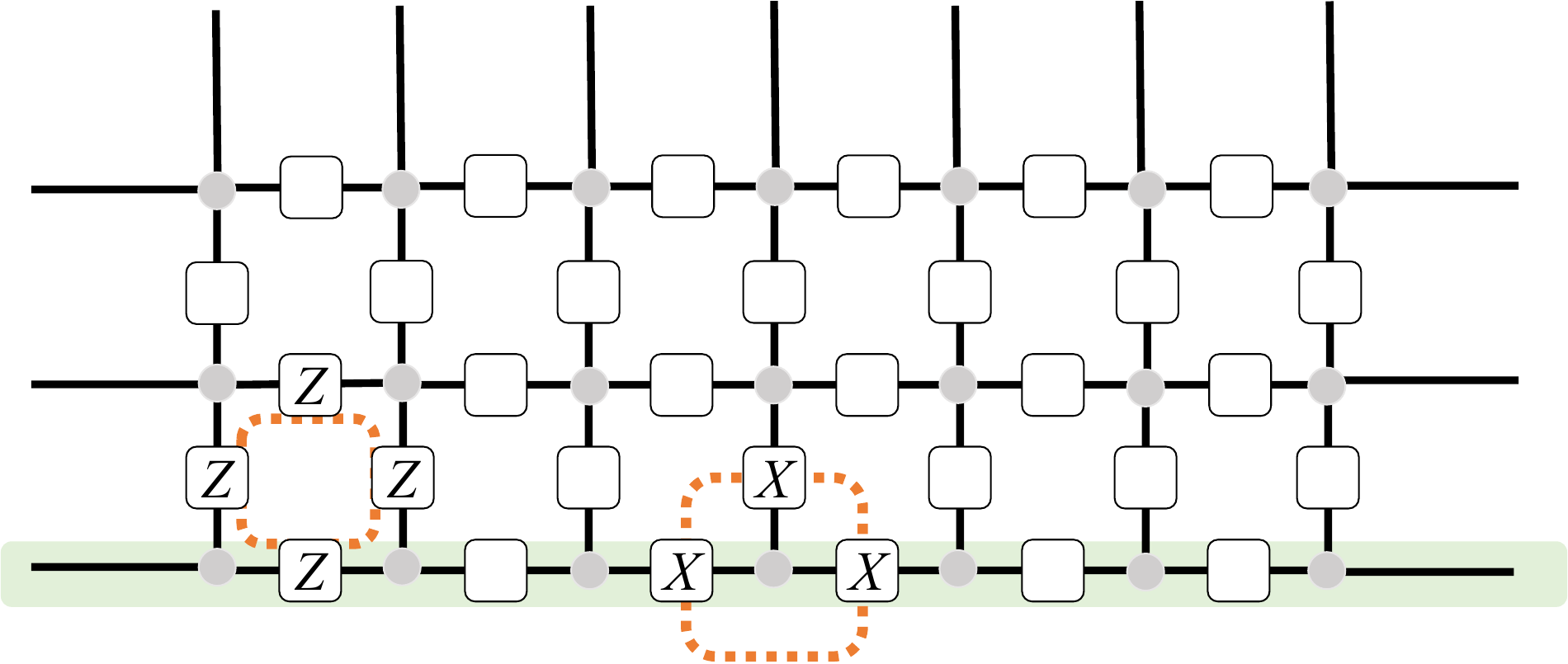}
\caption{Toric code with smooth boundary conditions. Green shaded area denotes the unmeasured boundary qubits. The vertex stabilizer $A_v = \prod_i X_i$ and plaquette stabilizer $B_p = \prod_i Z_i $ living on the boundary are labeled by orange dashed lines.} 
\label{fig: toric_code_bounadry_conven} 
\end{figure}

We first consider the toric code model with a smooth boundary as shown in Fig.~\ref{fig: toric_code_bounadry_conven}. We now express the boundary state in Pauli-$Z$ basis $\ket{\mu_{\partial \mathcal M}}\equiv \bigotimes_{i \in \partial \mathcal M } \ket{\mu_i}_i$, with $\mu_{\partial \mathcal M} \equiv (\mu_1, \mu_2, \ldots, \mu_L)$, and $Z_i \ket{\mu_i}_i = \mu_i \ket{\mu_i}_i$. The boundary basis is effectively taking $\theta_i \to 0$ in \cref{eq: measurement parameter} on the boundary, and the local weight parameter $W_i$ defined in Eq~\ref{eq: weight_parameter} is then 

\begin{equation}
    W_i = \lim_{\theta_i \to 0}\left\{
    \begin{matrix}
        \tan (\theta_i/2) \exp(i \varphi_i ) & \text{if} & \mu = +1  \\
        \cot (\theta_i/2) \exp(-i \varphi_i)  & \text{if} & \mu = -1
     \end{matrix}\right..
\end{equation}
For simplicity, we choose the limit direction to be $0^+$ to fix the sign of the weight parameter when the measurement outcome is negative $\mu = -1$ 

\begin{equation}
    W_i = \lim_{\theta_i \to 0^+}\left\{
    \begin{matrix}
       0 & \text{if} & \mu = + 1 \\
       + \infty & \text{if} & \mu = -1
     \end{matrix}\right.
\end{equation}
and we thus have the boundary coupling
\begin{equation}
     K_i = \left\{
    \begin{matrix}
       +\infty & \text{if} & \mu = +1 \\
       -\infty & \text{if} & \mu = -1
     \end{matrix}\right. .
\end{equation}
This can be further written as
\begin{equation}\label{eq: boundary_coupling}
    \tanh K_i = \mu_i \quad i \in \partial \mathcal M
\end{equation}

The boundary state of the toric code can be expressed as follows: 
\begin{equation}\label{eq: toric_boundary_overlap}
\begin{aligned}
    \ket{\Psi}_{\partial \mathcal M} \propto &\sum_{s, \mu_{\partial \mathcal M }} \exp\left(\sum_{j = \langle c, d \rangle} K_j s_c s_d\right)\times \\
    &\exp\left(\sum_{ i = \langle a, b \rangle} K_i s_a s_b\right) \ket{\mu_{\partial \mathcal M}}.
\end{aligned}
\end{equation}
Here, the first sum, denoted by $\sum_{j = \langle c, d \rangle}$, represents the summation over the links of the bulk of the square lattice, while the second summation $\sum_{i = \langle a, b \rangle}$ is the sum over the links that lie on the boundary. Taking advantage of \cref{eq: boundary_coupling}, we further write the boundary summation as
\begin{equation}\label{eq: toric_boundary_base}
\begin{aligned}
     &\exp\left(\sum_{i = \langle a, b \rangle} K_i s_a s_b\right) \propto \prod_{i = \langle a, b\rangle}(1 + \tanh K_i s_a s_b)\\
     &=\prod_{i = \langle a, b \rangle}(1 + \mu_i s_a s_b) \propto \prod_{i \in \partial \mathcal M}\delta_{\mu_i, s_a s_b}
\end{aligned}
\end{equation}
where we used the fact that $\prod_{i = \langle a, b\rangle}(1 + \mu_i s_a s_b) \neq 0$ if and only if $\mu_i = s_a s_b$ for all $i$. The boundary state for toric code can now be written as
\begin{equation}
\begin{aligned}\label{eq: toric_code_boundary_state}
     \ket{\Psi}_{\partial \mathcal M} &\propto \sum_{\mu_{\partial \mathcal M}} \sum_{s} \exp\left(\sum_{ j = \langle c, d \rangle} K_{j} s_c s_d\right)\prod_{i\in\partial \mathcal M}\delta_{\mu_i, s_a s_b}\ket{\mu_{\partial \mathcal M}} \\
     &\propto \sum_{s} \exp\left(\sum_{j = \langle c, d \rangle} K_j s_c s_d\right) \bigotimes_{i = \langle a, b\rangle} \ket{s_a s_b}_i.
\end{aligned}
\end{equation}
where $\ket{s_a s_b}_i$ represents the quantum state on the $i$-th qubit of the boundary, satisfying $Z_i \ket{s_a s_b}_i = s_a s_b \ket{s_a s_b}_i$, with $i$ being the edge connecting neighboring sites $a$ and $b$. This quantum state is determined by the corresponding classical Ising domain wall configuration on the boundary, denoted as $\{s_a s_b\}$ with $a,b$ being neighbors. With this, we complete the construction of the boundary state for the toric code model.

For the boundary state of the Lieb cluster, similar to the toric code case, we take the basis 
\begin{equation}
    \ket{\mu_{\partial \mathcal M }} \equiv \big(\bigotimes_{i \in \partial \mathcal M_\bullet }\ket{\mu_i}_i \big)\otimes \big(\bigotimes_{a \in \partial \mathcal M_\circ }\ket{\mu_a} _a\big).
\end{equation}
For the boundary edge qubit $i \in \partial \mathcal M_\bullet $, we define the basis as follows
\begin{equation}
    X_i\ket{\mu_i}_i = \mu_i\ket{\mu_i }_i.
\end{equation}
For the boundary vertex qubit $a \in \partial \mathcal M_\circ$, the basis is
\begin{equation}
    Z_a\ket{\mu_a}_a = \mu_a \ket{\mu_a}_a.
\end{equation}
To obtain  this basis via the measurement basis defined in \cref{eq: mea_basis}, we take $\theta_i\to \pi/2$ and $\phi_i\to 0$ in the \cref{eq: measurement parameter}.  On the other hand, the basis for the vertex qubit $a \in \partial \mathcal M_\circ$ is acquired by performing a Pauli $Z$ measurement with $\theta_i\to 0$.

For the edge qubit, the corresponding weight parameter, defined in \cref{eq: weight_parameter}, of such basis is then
\begin{equation}
W_i = \left\{
     \begin{matrix}
        1 & \text{if} & \mu_i = +1  \\
        -1 & \text{if} & \mu_i = -1
     \end{matrix}\right..
\end{equation}
As $\exp(-2K_i ) = (1 - W_i )/ (1 + W_i)$, the corresponding boundary coupling is
\begin{equation}
    K_i = \left\{
     \begin{matrix}
        +\infty & \text{if} & \mu_i = +1  \\
        -\infty & \text{if} & \mu_i = -1
     \end{matrix}\right..
\end{equation}
Again we write
\begin{equation}
    \tanh K_i = \mu_i  \quad i \in \partial \mathcal M_\bullet
\end{equation}
As for the boundary vertex qubit, we have
\begin{equation}
W_a = \left\{
     \begin{matrix}
        0 & \text{if} & \mu_a = +1  \\
        +\infty & \text{if} & \mu_a = -1
     \end{matrix}\right..
\end{equation}
The corresponding magnetic field is then
\begin{equation}
       h_a = \left\{
    \begin{matrix}
       +\infty & \text{if} & \mu_a = +1 \\
       -\infty & \text{if} & \mu_a = -1
     \end{matrix}\right. ,
\end{equation}
which can be further written as

\begin{equation}
    \tanh h_a = \mu_a \quad a\in \partial \mathcal M_\circ.
\end{equation}

We now write the boundary state of the Lieb cluster state as
\begin{equation}
\begin{aligned}
    \ket{\Psi}_{\partial \mathcal M} &\propto \sum_{\mu_{\partial \mathcal M }} \sum_{s} \exp{\sum_{j = \langle c, d \rangle} K_j s_c s_d + \sum_{e}  h_e s_e} \\
    \times &\exp{\sum_{i  = \langle a, b \rangle} K_i s_a s_b + \sum_{c}  h_c s_c}
    \ket{\mu_{\partial \mathcal M}},
\end{aligned}
\end{equation}
where the first line is the bulk term and the second term is the boundary term. The boundary terms can be further written as
\begin{equation}
\begin{aligned}
   & \exp{\sum_{i  = \langle a, b \rangle} K_i s_a s_b + \sum_{c}  h_c s_c} \\
    &\propto \prod_{i = \langle a, b \rangle }(1 + \tanh K_i s_a s_b)\prod_{c}(1 + \tanh h_c s_c) \\
    &\propto \prod_{i = \langle a, b \rangle }\delta_{\mu_i s_a s_b}\prod_{c}\delta_{\mu_c s_c}.
\end{aligned}
\end{equation}
where we used the same derivation in \cref{eq: toric_boundary_base}. We may thus relabel the boundary basis using the boundary Ising spin variables
\begin{equation}
    \begin{matrix}
        \mu_i = s_a s_b & &i  \in \partial \mathcal M_\bullet\\
        \mu_c = s_c & &c  \in \partial \mathcal M_\circ 
    \end{matrix}
\end{equation}
where $i = \langle a, b\rangle$ is the edge connecting neighboring sites $a, b$. The boundary state of the Lieb lattice model is thus
\begin{equation}\label{eq: lieb_boundary_state}
\begin{aligned}
     \ket{\Psi}_{\partial \mathcal M} &\propto  \sum_{s} \exp{\sum_{j = \langle c, d \rangle} K_j s_c s_d + \sum_{e}  h_e s_e}\times \\
    & \bigotimes_{i = \langle a, b \rangle , c}\ket{s_{a}s_{b}}_{i}\ket{s_c}_d.
\end{aligned}
\end{equation}
where the boundary basis satisfies
\begin{equation}\label{eq: lieb_boundary_basis}
    \begin{matrix}
         X_i\ket{s_a s_b}_i = s_a s_b \ket{s_a s_b }_i &i  \in \partial \mathcal M_\bullet\\
        Z_c \ket{s_c}_c = s_c\ket{s_c}_c & c \in \partial \mathcal M_\circ 
    \end{matrix}.
\end{equation}
It is thus obvious that the boundary state is the eigenstate of $g_i = X_i Z_a Z_b$ with $a, b$ being neighboring boundary sites connected by edge $i$
\begin{equation}\label{eq: gauge}
    g_i \ket{\Psi}_{\partial \mathcal M} = \ket{\Psi}_{\partial \mathcal M} \quad \forall i \in \partial \mathcal M_\bullet.
\end{equation}

\begin{figure}[ht]
\centering
\subfloat[Lieb lattice with smooth boundary. Green shaded area: boundary qubits. $a, b \in \circ$ label the neighboring sites of $i \in \bullet$. $c$ labels the boundary $\circ$ sites. ]{\label{fig: lieb_bounadry_conven}\includegraphics[width=0.4\textwidth]{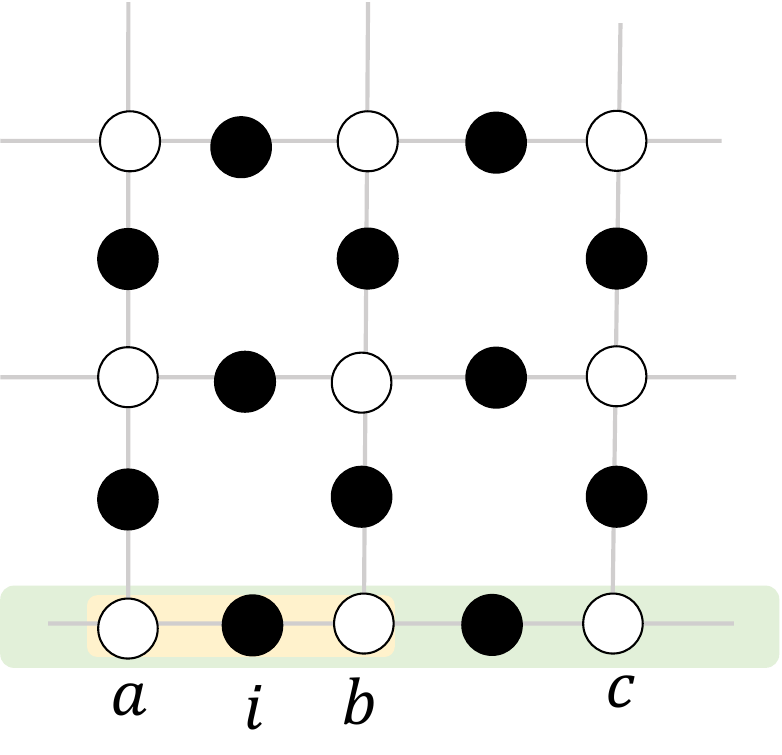}}\qquad
\subfloat[The boundary of the square lattice cluster state. $a,b,c$ label the neighboring sites of $i \in \bullet$. $d$ labels the boundary $\circ$ sites.]{\label{fig: sq_bounadry_convenn}\includegraphics[width=0.4\textwidth]{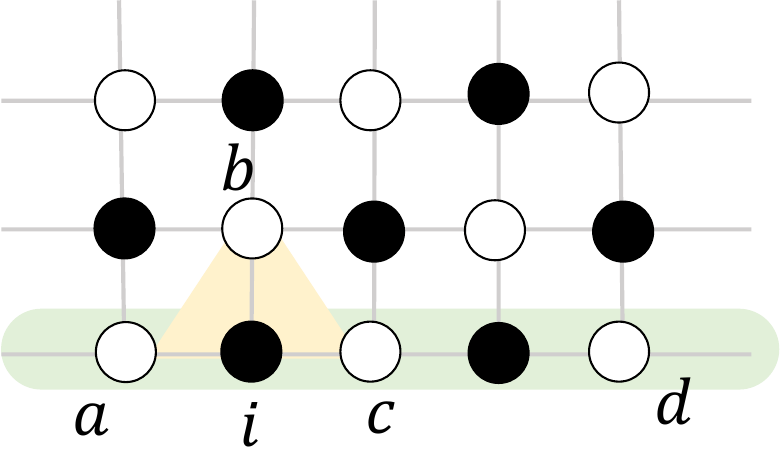}}
\caption{\label{fig: bound_conven} Boundary conventions}
\end{figure}

The boundary state of the cluster state on the square lattice is obtained using a similar approach as the one defined on the Lieb lattice. The boundary state is 
\begin{equation}
\begin{aligned}
    &\ket{\Psi}_{\partial \mathcal M} \propto \\
    \sum_{s} &\exp{\sum_{j \in \bullet }  K_{j}s_e s_f s_g s_h +  \sum_{r \in \circ} h_r s_r } \\
    &\bigotimes_{i, d} \ket{s_{a} s_{b} s_{c}}_i \ket{s_d}_d
\end{aligned}
\label{eq:sq_lattice_boundary_state}
\end{equation}
where $j$ is the bulk $\bullet$ site enclosed by neighboring $\circ$ sites $e,f,g,h$, and $i$ labels the boundary $\bullet$ sites enclosed by $a,b,c\in \circ$, as shown in Fig.~\ref{fig: sq_bounadry_convenn}. The boundary basis satisfies
\begin{equation}
    \begin{aligned}
    X_{i} \ket{s_{a} s_{b} s_{c}}_i &= s_{a} s_{b} s_{c}\ket{s_{a} s_{b} s_{c}}_i\\
    Z_{d}\ket{s_d}_d &= s_d \ket{s_d}_d
\end{aligned}
\end{equation}
We thus finish the derivation of the boundary state in terms classical Ising model.

In this section, we have demonstrated that the boundary state, which is obtained through single qubit projective measurements on the bulk qubits, carries crucial information about the partition function of the 2D classical Ising model. Consequently, if we can successfully compute the 1D boundary state for a large system, we can effectively solve the 2D state sampling problem. One promising method for approximately simulating 1D quantum states is the MPS approach. The complexity of this representation relies on the entanglement scaling of the wave function. If this 1D state follows an area-law scaling of entanglement, it can be efficiently approximated on a classical computer, making approximately computing the corresponding partition function easy. In the next section, we will elaborate on the formalism to compute the boundary state by mapping it to a {\blue \((1+1)\mathrm{D}\) } dynamical problem. This approach will provide further insights into understanding and addressing the complexity of the quantum sampling problem.

\section{Bridging 2D Sampling Problem and \((1+1)\mathrm{D}\) Circuit Dynamics}\label{sec: dynamics}

In the previous section, we demonstrated that for a cluster state defined on a bipartite graph, the boundary 1D quantum state can be expressed as follows after performing single qubit measurements on the bulk qubits:
\begin{equation}
\ket{\Psi}_{\partial \mathcal M} \propto \sum_{s} \exp{(H[s])} \ket{s_{\partial \mathcal M}}
\label{eq:1d_bs}
\end{equation}
Here $H[s]$ represents a 2D classical spin Hamiltonian with parameters that can take complex values, and $\ket{s_{\partial \mathcal M}}$ is the basis for the boundary state related to the classical-spin configurations on the boundary. This boundary state exhibits entanglement transitions and establishes an interesting connection with the {\blue \((1+1)\mathrm{D}\) } hybrid quantum dynamics. In this section, we provide two approaches to dynamically generate this boundary state.

\begin{figure}[ht]
\centering\includegraphics[width=0.8\linewidth]{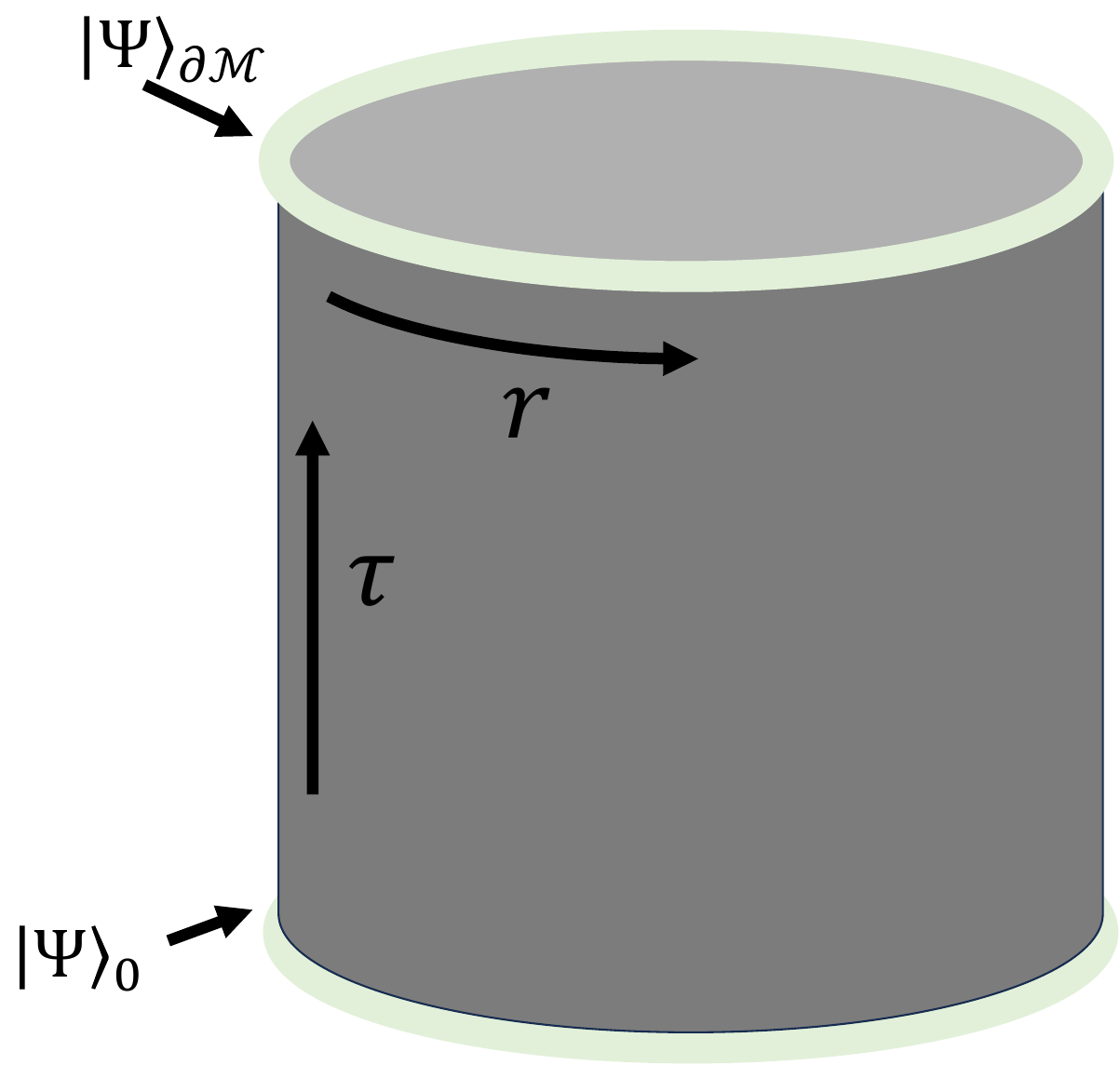}
\caption{The two ``dimensions" are now labeled as the spatial dimension $r$ and the temporal dimension $\tau$. $\ket{\Psi}_{\partial \mathcal M}$ is obtained by ``time" evolution of initial state $\ket{\Psi}_0$.} 
\label{fig: space_time} 
\end{figure}

\subsection{Transfer Matrix Method}\label{sec: transfer_matrix}
It is well-established that the partition function of the 2D classical spin model is connected to the {\blue \((1+1)\mathrm{D}\) } quantum dynamics using the transfer matrix method as in \cref{fig: space_time}. Consequently, the boundary wave function in \cref{eq:1d_bs} can be expressed as follows:
\begin{equation}
    \ket{\Psi}_{\partial \mathcal M} = \mathcal T \prod_{\tau} \mathbf{T} (\tau) \ket{\Psi_0}
\end{equation}
where $\mathcal T$ is the time ordering operator, $\mathbf{T}(\tau)$ is the transfer matrix acting on the initial state $\ket{\Psi_0}$. The initial state $\ket{\Psi_0}$ is determined by the shape of the bottom boundary as shown in Fig.~\ref{fig: space_time}.

\begin{figure}[ht]
\centering
\subfloat[The "smooth" boundary]{\label{fig: lieb_bounadry}\includegraphics[width=0.22\textwidth]{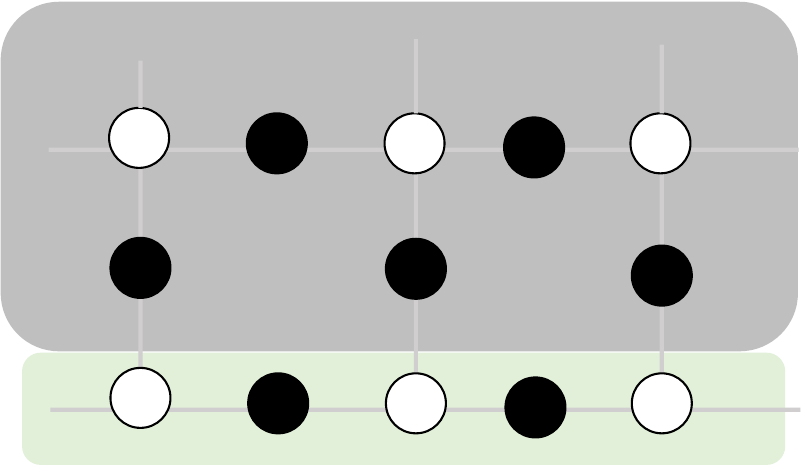}}\qquad
\subfloat[The "rough" boundary]{\label{fig: sq_bounadry_conven}\includegraphics[width=0.22\textwidth]{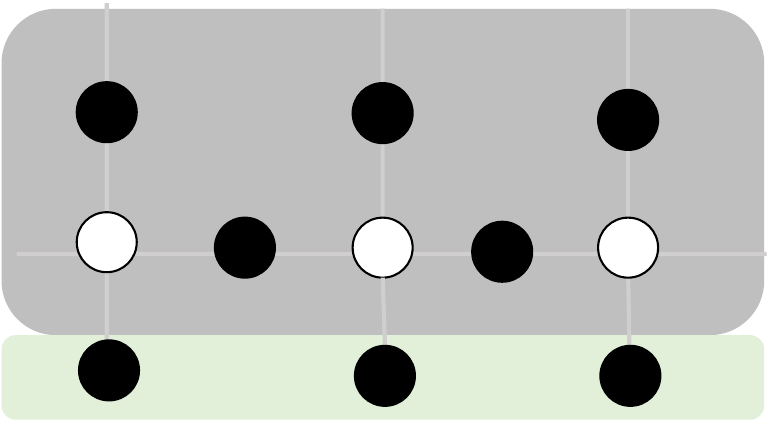}}
\caption{\label{fig: lieb_boundary_condition} Typical boundary shapes of Lieb lattice}
\end{figure}

For the cluster state defined on the Lieb lattice, two typical boundaries are the ``smooth" and ``rough" boundaries, as shown in Fig.~\ref{fig: lieb_boundary_condition}. For "smooth boundary," $\ket{\Psi_0}$ is the equal weight superposition of quantum states labeled by all possible boundary classical spin configurations $\{s_{\partial\mathcal M }\}$
\begin{equation}
    \ket{\Psi_0} = \frac{1}{2^{|\{s_{\partial \mathcal M}\}|/2}} \sum_{s_{\partial \mathcal M}}\ket{s_{\partial\mathcal M}},
\end{equation}
while for ``rough" boundary conditions, we have
\begin{equation}
    \ket{\Psi_0} = \ket{s_{\partial_M} = + 1}.
\end{equation}
A detailed discussion of the boundary condition is shown in App.~\ref{ap: graph_boundary}. In the following, we will use the term ``top boundary" to refer to qubits  that support the wave function induced by bulk measurement $\ket{\Psi}_{\partial \mathcal M}$, and ``bottom boundary" to refer qubits on the other end of the cylinder $\mathcal M$ which determines the initial state $\ket{\Psi_0}$.

The 1D boundary state $\ket{\Psi}_{\partial \mathcal M}$ is defined in \cref{eq:  lieb_boundary_state}. We take the perpendicular axis as the temporal direction $\hat{\tau}$ and the horizontal axis as the spatial direction $\hat{r}$ shown in Fig.~\ref{fig: space_time}. We will use ``time slice $\tau$" to refer to the row position of a vertex site. The indices $a, b, c, \ldots$ denote the column position, and $i, j, k, \ldots$ are used to label the edges connecting neighboring vertex sites. We will use $i = \langle a, b \rangle$ to label the edge $i$ connecting neighboring sites $a$ and $b$, and $e = (k, l)$ to label vertex $e$ shared by neighboring edges $k$ and $l$.

For the Ising model defined on the square lattice, it is well-known that by associating each vertex with a qubit, we obtain the spin basis for each time slice $\tau$
\begin{equation}\label{eq:vertex_basis}
    \ket{s}_\tau = \bigotimes_{a\in\tau}\ket{s_a}_a \quad Z_a\ket{s_a}_a = s_a \ket{s_a}_a,
\end{equation}
which spans the Hilbert space $\mathcal{H}_\tau =\operatorname{span} \{\ket{s}_\tau\}$. The transfer matrix that maps $\mathcal{H}_{\tau}$ to $\mathcal{H}_{\tau + 1}$ can now be written in the form
\begin{align}
T(\tau)=T_2(\tau)T_1(\tau)
\end{align}
with 
\begin{equation}
    \begin{aligned}\label{eq:Ising_transf}
        &T_1(\tau)\propto\exp\left(\sum_{\substack{i = \langle a, b \rangle}} K^{\tau}_i Z_a^\tau Z_b^\tau + \sum_{c} h^\tau_c Z_c^\tau \right)\\
&T_2(\tau)\propto\exp( \sum_{e} \tilde{K}_{e}^\tau X_{e}^\tau)
    \end{aligned}
\end{equation}
with $K_i^\tau$ being the quantum Ising coupling, $\tilde{K}_e^\tau$ the effective transverse field, and $h_c^\tau$ the longitudinal field. The Paulis are defined in the following way
\begin{equation}
\begin{aligned}
      Z_a^\tau &\equiv  \sum_{s_{a_\tau}} s_{a_\tau}\ket{s_{a_\tau}}_{a_\tau}\bra{s_{a_\tau}}_{a_\tau}\\
      X_e^\tau &\equiv \sum_{s_{e_\tau}}  \ket{\overline{s_{e_\tau}}}_{e_{\tau+1}}\bra{s_{e_\tau}}_{s_{e_\tau}}
\end{aligned}
\end{equation}
where $\overline{s_{e_\tau}} \equiv - s_{e_\tau}$. As $\mathcal{H}_\tau$'s are identical for all time slice $\tau$, we will drop the index $\tau$ on the Pauli's as well as on the classical spin indices labeling the quantum states in the remaining part of this section. The relations between the quantum parameters $\{K^\tau ,\tilde K^\tau, h\}$ and the classical parameters $\{K, h\}$ are shown below
\begin{equation}\label{eq: quant_classical_para}
    \begin{aligned}
         K_i^\tau &= K_{i_{\tau}}\\
          h_c^\tau &= h_{c_{\tau}}\\
    \tilde{K}_{e}^\tau &=  - \frac{1}{2} \ln \tanh{K_{e_\tau}}
    \end{aligned}
\end{equation}
where $K_{i_{\tau}}$ is the classical coupling along the spatial direction $\hat r$ on the $i$-th edge of time slice $\tau$, $h_{c_{\tau}}$ is the classical magnetic field on site $c$ at time slice $\tau$, and $K_{e_\tau}$ is the classical coupling along the temporal direction $\hat \tau$ of neighboring sites on time slice $\tau$ and $\tau + 1$ of the $e$-th row of vertex sites.

In the Lieb lattice model, the qubits are on both the vertices and edges of the square lattice. Notice that the $1$d quantum state on the boundary $\ket{\Psi}_{\partial \mathcal M}$ is defined on the Pauli-$X$ basis for $\circ$ vertices and the Pauli-$Z$ basis for $\bullet$ edges as shown in the previous section. The basis for the transfer matrix of the Lieb lattice model is naturally
\begin{align}\label{eq: spin_edge_basis}
\ket{s}_{\text{Lieb}} = \bigotimes_{i , c}\ket{s_a s_b}_i\otimes \ket{s_c}_c    
\end{align}
 as defined in Eq.\ref{eq: lieb_boundary_basis}, with $i \in \bullet$ labeling the boundary edge connecting vertex $a, b$, and $c\in \circ$ the boundary vertex.
 On such basis, the transverse field part is modified to
\begin{align}\label{eq: transverse_field}
    \mathbf{T}_2(\tau)\propto\exp( 
    \sum_{e = (k, l)} \tilde{K}_{e}^\tau X_{e} Z_k Z_l),
\end{align}
where again $\tilde K_e^\tau = -1/2 \ln \tanh K_{e_\tau}$ with $K_{e_\tau}$ being the classical coupling along $\hat\tau$ that connects neighboring sites of $e$-th row of vertices on time slice $\tau$ and $\tau + 1$. Here, $X_e$ is a Pauli $X$ operator on the vertex qubit $e$, and $Z_k$, $Z_l$ are $Z$ Paulis operators on neighboring edges in the spacial direction $k,l \in \hat r$ that share the same vertex $e$. Since the edge qubits
\begin{equation}
    Z_k \ket{s_e s_g}_k = \ket{\overline{s_e s_{g}}}_k,
\end{equation}
where $\ket{s_e s_g}_k$ and $\ket{\overline{s_e s_g}}_k$ are the eigenstates of $X_k$, operator $X_e Z_k Z_l$ flips the classical spin defined on vertex $e$ in the following manner
\begin{equation}
    \begin{aligned}
     X_{e} Z_k Z_l &\ket{s_e}_e \ket{s_e s_g}_k\ket{s_{e}s_{h}}_{l} \\
    &=\ket{\overline{s_e}}_e\ket{\overline{s_e s_{g}}}_k\ket{\overline{s_{e}s_{f}}}_{l}\\
    &=\ket{ \overline{s_e}}_e \ket{\overline{s_e} s_{g}}_k  \ket{\overline{s_{e}}s_{f}}_{l}.
\end{aligned}
\end{equation}
Here $e = (k, l)$ is the site shared by two edges $k$ and $l$.

The spatial coupling of the Lieb lattice model $\mathbf{T}_1(\tau)$ is the same as $T_1(\tau)$ for the Ising model defined on a square lattice, since Pauli $Z$ transforms spin-edge basis \cref{eq: spin_edge_basis} the same way as the spin basis \cref{eq:vertex_basis}. The transfer matrix $\mathbf{T}(\tau) = \mathbf{T}_1 (\tau)\mathbf{T}_2 (\tau)$ at time $\tau$ is thus obtained.

The initial state $\ket{\Psi}_0$ for both ``smooth" and ``rough" boundaries can be expanded using the basis defined in \cref{eq: spin_edge_basis}. Since
\begin{equation}\label{eq: g_T commute}
    [g_i, \mathbf{T}(\tau)] = 0\quad \forall i, \tau
\end{equation}
the boundary state state $\ket{\Psi}_{\partial \mathcal M}$ obtained by evolving $\ket{\Psi}_0$ is still an eigenstate of $g_i$. This is consistent with \cref{eq: gauge}. This concludes the derivation of the transfer matrix of the cluster state defined on the Lieb lattice.

\subsection{Connection with {\blue \((1+1)\mathrm{D}\) } Circuit Dynamics}

The transfer matrix of the Lieb lattice model obtained in Sec.~\ref{sec: transfer_matrix} can be decomposed into three types of local gates in the following way
\begin{equation}
\begin{aligned}
    \mathbf{T}(\tau) \propto& \exp\left(\sum_{\substack{i = \langle a, b \rangle}} K^{\tau}_i Z_a Z_b + \sum_{c} h^\tau_c Z_c\right) \\
    &\times \exp( 
    \sum_{e = (k, l)} \tilde{K}_{e}^\tau X_{e} Z_k Z_l) \\
    =&\prod_i  \mathbf{T}_{ZZ}^{i = \langle a, b\rangle}(\tau) \prod_c  \mathbf{T}_{Z}^c(\tau) \prod_e  \mathbf{T}_X^{e = (k, l)}(\tau),
\end{aligned}
\end{equation}
where
\begin{equation}
    \mathbf{T}_{ZZ}^{i = \langle a, b \rangle}(\tau) = \exp(K_{i}^\tau Z_a Z_b)
\end{equation}
is a two-qubit gate and characterizes the Ising interaction between neighboring qubits $a$ and $b$,
\begin{equation}
    \mathbf{T}_{Z}^c(\tau) =  \exp(h_c Z_c),
\end{equation}
is a single qubit gate and represents the longitudinal field acting on vertex qubit $c$, and
\begin{equation}
    \mathbf{T}_X^{e = (k, l)}(\tau)= \exp(\tilde{K}_{e}^\tau X_{e}Z_{k} Z_{l})
\end{equation}
is a three-qubit gate, 
where $e = (k, l)$ is the vertex connecting two neighboring edges $k$ and $l$.

Taking advantage of Eq.\ref{eq: ising_para} and Eq.\ref{eq: quant_classical_para}, the correspondence between parameters in the quantum gates $\{K^\tau, \tilde K^\tau, h^\tau\}$ and the weight parameters $\{W\}$ defined in \cref{eq: weight_parameter} is then
\begin{equation}
    \begin{aligned}
        \exp(-2 K_{i}^\tau) &= \frac{1 - W_{i_\tau}}{1 + W_{i_\tau}} \\
        \exp(-2 \tilde K_{e}^\tau) &= W_{e_\tau}\\
        \exp(-2 h_c^\tau) &= W_{c_\tau}\\
    \end{aligned},
\end{equation}
where $W_{i_\tau}$ and $W_{e_\tau}$ are weight parameters on edges in spatial $\hat r$ and temporal $\hat\tau$ direction respectively. Here $i_\tau$ is the $i$-th edge on time slice $\tau$ and $e_\tau$ labels the edge connecting neighboring sites on the $e$-th row of vertex sites on time slice $\tau$ and $\tau + 1$. $W_{c_\tau}$ is the measurement weight parameter on vertex site $c$ of time slice $\tau$. For measurements in the $XZ$-plane, the weight parameter $W \in \mathbb{R}$. For measurements containing $Y$ components, the weight parameter $W$ becomes complex, and the parameters in both the classical 2D Hamiltonian and {\blue \((1+1)\mathrm{D}\) } quantum dynamics are complex.

\subsubsection{Pauli Measurements}

To illustrate the correspondence between the 2D measurement process and the {\blue \((1+1)\mathrm{D}\) } circuit dynamics and to understand the physical meaning of the complex Ising parameters, we examine three straightforward cases: measuring all the bulk qubits along the $X$, $Z$, and $Y$ directions. We show that bulk measurements along $X$ and $Z$ directions can be effectively treated as projective measurements in the {\blue \((1+1)\mathrm{D}\) } circuit dynamics, while bulk $Y$ measurements are effective unitary gates.

When measuring the bulk along $X$ direction, we take $W = \pm 1$, where the $\pm$ sign is determined by measurement outcome. For every site in bulk, we have
\begin{equation}
    \begin{aligned}
        K_i^\tau       &=  \pm \infty \\
        \tilde K_e^\tau&= 0 \text{ or } \pm\frac{i\pi}{2}\\
        h_c^\tau       &= 0 \text{ or } \pm\frac{i\pi}{2}\\
    \end{aligned}.
\end{equation}
The local terms are
\begin{equation}
    \begin{aligned}
        \mathbf{T}_{ZZ}^{i = \langle a, b \rangle }(\tau) &\propto (1 \pm Z_a Z_b) \\
        \mathbf{T}_{Z}^c(\tau) & \propto  I \text{ or } Z_c\\
        \mathbf{T}_{X}^{e = (k, l)}(\tau) &\propto I \text{ or } X_e Z_{k} Z_{l} \\
    \end{aligned}.
\end{equation}
In this case, $\mathbf{T}_{ZZ}^{i=\langle a, b \rangle }(\tau)$ is effectively a $Z_a Z_b$ projective measurement on neighboring sites $a$ and $b$ connected by edge $i$, and $\mathbf{T}_{Z}^c(\tau)$ and $\mathbf{T}_{X}^{e = (k, l)}(\tau)$ are local Pauli operations that do not affect the entanglement structure. 

The boundary state $\ket{\Psi}_{\partial \mathcal{M}}$ is a trivial product state stabilized under $\langle X_i, Z_a\rangle$ with $i$ being the boundary edges and $a$ the vertices, if taking the ``rough" boundary on the bottom and is a GHZ state on vertices and product state on edges stabilized under $\langle X_i, Z_a Z_b, \prod_j X_j \rangle $ with $i = \langle a, b\rangle$ being the edges, and $\prod_j X_j$ the $X$-string supported on all the boundary edge qubits if taking the ``smooth" boundary condition. This difference is because ``smooth" bottom boundary condition preserves the global $\mathbb Z_2$ symmetry of the Ising model when measuring every site along $X$, which results in the vertex GHZ state. In contrast ``rough" bottom boundary breaks it, and thus results in a trivial product state on the boundary.

For Pauli $Z$ measurements, we have $W = 0, +\infty$ depending on the measurement outcome $\mu$, and the transfer matrix parameters are
\begin{equation}
    \begin{aligned}
        K_{i}^\tau       &= 0 \text{ or } \pm\frac{i\pi}{2} \\
        \tilde K_{e}^\tau&= \pm \infty \\
        h_c^\tau         &= \pm \infty.
    \end{aligned}
\end{equation}
The local terms of the transfer matrices thus become
\begin{equation}
    \begin{aligned}
        \mathbf{T}_{ZZ}^{i = \langle a, b \rangle }(\tau) &\propto I \text{ or } Z_a Z_b \\
        \mathbf{T}_{Z}^c(\tau)       &\propto (1 \pm Z_c)\\
        \mathbf{T}_{X}^{e = (k, l)}(\tau)   &\propto (1 \pm X_e Z_{k} Z_{l})
    \end{aligned}
\end{equation}
where $a$ and $b$ denote neighboring sites connected by edge $i$, $c$ labels the vertex sites, and $k$, $l$ are edges sharing the same vertex $e$. At the final time, the quantum state is projected onto the $1$-dimensional cluster state up to some Paulis. The reason is that the edges connecting the boundary and bulk are in the temporal direction due to the boundary condition we take, meaning that at the final time the transfer matrix is the temporal part 
\begin{equation}
    \prod_e \mathbf{T}_X^{e = (k, l)} \propto \prod_{e = (k, l)} (1 \pm X_e Z_{k} Z_{l}),
\end{equation}
which projects the quantum state onto the eigenstate of $\{X_e Z_{k} Z_{l}\}$
\begin{equation}
   X_e Z_{k} Z_{l} \ket{\Psi}_{\partial \mathcal M} = \mp \ket{\Psi}_{\partial \mathcal M}\quad \forall e=(k, l).
\end{equation}
The initial state $\ket{\Psi}_0$ is the eigenstate of $\{g_i = X_i Z_a Z_b \}$ with $i = \langle a, b \rangle$ being the edge connecting neighboring sites $a$ and $b$ and $g_i$ commute with all the transfer matrices as shown in \cref{eq: gauge} and \cref{eq: g_T commute}. The boundary state $\ket{\Psi}_{\partial \mathcal M}$ obtained by measuring the bulk in the $Z$ direction is therefore a stabilizer state under group
\begin{equation}
    \mathcal G = \langle X_i Z_a Z_b, \mp X_e Z_{k} Z_{l}\rangle\quad i = \langle a, b),~ e = (k, l)
\end{equation}
where the sign $\mp$ is determined by the measurement outcomes. Such a quantum state can be recognized as a $1$d cluster state up to some Paulis. This result is consistent with direct calculation using the stabilizer measurement formula shown in \cite{PhysRevB.100.134306}.

When taking Pauli $Y$ measurements in bulk, the transfer matrix parameters are
\begin{equation}
    \begin{aligned}
          K_{i}^\tau     &= \pm\frac{i\pi}{4} \\
        \tilde K_{e}^\tau&= \pm\frac{i\pi}{4} \\
        h_c^\tau            &= \pm\frac{i\pi}{4} \\
    \end{aligned},
\end{equation}
and the local terms all become $\frac{\pi}{4}$ Pauli rotations
\begin{equation}
    \begin{aligned}
    \mathbf{T}_{ZZ}^{i = \langle a, b \rangle }(\tau) &= \exp(\pm i \frac{\pi}{4} Z_a Z_b)\\
    \mathbf{T}_{Z}^c(\tau) &=  \exp( \pm i \frac{\pi}{4}Z_c)\\
    \mathbf{T}_{X}^{e = (k, l)}(\tau) &= \exp(\pm i \frac{\pi}{4} X_e Z_{k} Z_{l})
    \end{aligned}.
\end{equation}
By randomly measuring the bulk along the $X$ $Y$ and $Z$, we obtain an effective {\blue \((1+1)\mathrm{D}\) } hybrid random circuit. In the next section, we will further show that such a system exhibits a volume-law-area-law entanglement transition on the unmeasured boundary.

\subsubsection{Free Fermion Dynamics}\label{sec: exact_solvable}

As the $2$-dimensional Ising model is exactly solvable when $h_c = 0$ for all site $c$, we discuss the corresponding  exactly solvable limit of the boundary state $\ket{\Psi}_{\partial \mathcal M}$ of the Lieb lattice model. As is shown in previous sections, measuring all the vertex qubits in $X$ direction effectively turns off the magnetic field. We now consider the case where all the vertex qubits including those on the boundary are measured in $X$ direction. For the bulk measurements, we have $W_c = \pm 1$ in the bulk, which means that
\begin{equation}
     h_c = 0 \text{ or } \frac{i\pi}{2}
\end{equation}
depending on the measurement result. The negative measurement result induces an $i \pi / 2$ magnetic field, which effectively is an single site $Z$ rotation, and that does not change free Fermion nature of the effective quantum dynamics. The transfer matrices are 
\begin{equation}
    \begin{aligned}
\mathbf{T}_{ZZ}^{i = \langle a, b \rangle}(\tau) &= \exp(K_{i} Z_a Z_b),\\
\mathbf{T}_X^{e = (k, l)}(\tau) &= \exp(\tilde{K}_{e}^\tau X_{e}Z_{k} Z_{l}).
    \end{aligned}
\end{equation}
As the transfer matrices are written in the spin-edge basis given in \cref{eq: spin_edge_basis},  notice that $Z_a Z_b$ transforms the basis the same way as $X_i$ with $i = \langle a , b \rangle $
\begin{equation}
     X_i \ket{s_a s_b}_i \ket{s_a}_a\ket{s_b}_b  = Z_a Z_b \ket{s_a s_b}_i \ket{s_a}_a\ket{s_b}_b,
\end{equation}
which yields the equivalence relation $Z_a Z_b \sim X_i$. The transfer matrix is further written as
\begin{equation}
\begin{aligned}
\mathbf{T}_{ZZ}^{i = \langle a, b \rangle}(\tau) &= \exp(K_{i} X_i),\\
\mathbf{T}_X^{e = (k, l)}(\tau) &= \exp(\tilde{K}_{e}^\tau X_{e}Z_{k} Z_{l}).
\end{aligned}
\end{equation}
$X$-measurement on the boundary vertex qubits is described by projection
\begin{equation}
    \mathcal{P}_X^\pm =\frac{1}{2^{L}} \prod_{e\in \partial \mathcal M } (1 \pm X_e)
\end{equation}
where $e$ labels the boundary vertex sites, and $L$ is the number of boundary vertex sites. Such projection operator introduces an equivalence relation 
\begin{equation}
    X_e Z_k Z_l \sim Z_k Z_l\quad e = (k, l)
\end{equation} 
in the following manner
\begin{equation}
    \mathcal{P}_X^\pm X_e Z_k Z_l = \mp \mathcal{P}_X^\pm Z_k Z_l, \quad \forall e = (k, l),
\end{equation}
since $(1 \pm X_e)X_e Z_k Z_l = \mp (1 \pm X_e) Z_k Z_l$. The transfer matrices are now modified to
\begin{equation}
    \begin{aligned}
        \mathbf{T}_{ZZ}^{i = \langle a, b \rangle}(\tau) &= \exp(K_{i} X_i),\\
\mathbf{T}_X^{e = (k, l)}(\tau) &= \exp(\tilde{K}_{e}^\tau Z_k Z_l ).
    \end{aligned}
\end{equation}
Such matrices can be further written in the bi-linear form of majorana operators $\gamma_i$'s via Jordan-Wigner Transformation $i\gamma_{2i -1} \gamma_{2i} = X_i$, $i\gamma_{2k} \gamma_{2k + 1} = Z_kZ_{k + 1}$ with $\{\gamma_i, \gamma_j\} = 2 \delta_{i, j}$,
\begin{equation}
    \begin{aligned}
        \mathbf{T}_{ZZ}^{i = \langle a, b \rangle}(\tau) &= \exp\bigg(i K_{i} \gamma_{2i -1} \gamma_{2i}\bigg),\\
\mathbf{T}_X^{e = (k, l)}(\tau) &= \exp(i\tilde{K}_{e}^\tau \gamma_{2k} \gamma_{2k + 1}),
    \end{aligned}
\end{equation}
where $e$ labels the edge connecting vertex $k$ and $k + 1$.
As for random  Pauli measurement on the edge, the transfer matrices are, up to some Pauli rotations,
\begin{equation}
    \begin{aligned}
        \mathbf{T}_{ZZ}^{i=\langle a, b \rangle}(\tau) &\propto (1 \pm X_i) \\
        \mathbf{T}_{X}^{e = (k, l)}(\tau) &\propto (1 \pm Z_{k} Z_{l}),
    \end{aligned}
\end{equation}
with $i$, $k$, and $l$ labeling the edge qubits and $k$, $l$ being neighboring edges. Here $\mathbf{T}_{ZZ}^{i=\langle a, b \rangle}(\tau)$ is induced by Pauli-X measurement on the spatial edge $i$ at time slice $(\tau)$, while $\mathbf{T}_{X}^{e = (k, l)}$ is induced by Pauli-Z measurement on the temporal edge connecting vertices on time slice $\tau$ and $\tau + 1$. Bulk $Y$ measurements are effectively
\begin{equation}
    \begin{aligned}
    \mathbf{T}_{ZZ}^{i = \langle a, b \rangle }(\tau) &= \exp(\pm i \frac{\pi}{4} X_i)\\
    \mathbf{T}_{X}^{e = (k, l)}(\tau) &= \exp(\pm i \frac{\pi}{4} Z_{k} Z_{l})
    \end{aligned}.
\end{equation}
It is easy to identify them as the braiding gates of the Majoranas in terms of the free fermion representation of the Ising model. Such dynamics is the same as in \cite{PRXQuantum.2.030313}. As Pauli-X measurement on the vertex qubits of the Lieb lattice model projects the Lieb cluster state $\ket{\Psi}$ to the toric code state $\ket{\Psi}_{TC}$ up to some Paulis, in this limit, we have the same model as a recent work \cite{negari2023measurementinduced}, where the boundary  entanglement structure due to random measurements in the bulk of toric code model is investigated.

\subsection{Tensor Network Construction}
In this section, we present an alternate interpretation of cluster state sampling as a {\blue \((1+1)\mathrm{D}\) } quantum circuit. The resulting circuit consists of one- and two-qubit unitary gates, interspersed with weak single-qubit measurements along the Z axis. Akin to previous sections, the setting is as follows: we consider a cluster state defined on a square lattice $\mathcal{M}$ of dimensions $L_x\times L_y$, containing $N\equiv L_x\times L_y$ qubits. Measurements are performed along an arbitrary direction $\hat{n}(x,y)$, where $(x,y)$ denotes the coordinates of the qubits. The state of interest is the boundary state $\ket{\Psi}_{\partial\mathcal{M}}$, supported on one row $(x,y=L_y)$ of qubits, obtained after performing measurements on all other qubits. We then show that the boundary states for different bulk measurement outcomes can be interpreted as a 1D state, to which a particular family of quantum circuits of depth $L_y$ have been applied.
	
	The cluster state $\ket{\Psi}$ can be obtained by initializing all the qubits on a square lattice in the $\ket{+}$ state, and applying controlled-Z gates to every pair of nearest neighbors $i$ and $j$, denoted $CZ_{ij}$.
	
	\begin{equation}
		\begin{aligned}
			\ket{\Psi} &= \prod\limits_{\expval{ij}} CZ_{ij} \ket{+}^{\otimes N}\\
			&= \frac{1}{\sqrt{2^N}}\prod\limits_{\expval{ij}} CZ_{ij} \sum\limits_{n_1\dots n_N} \ket{n_1, n_2 \dots n_N}\\
			&= \frac{1}{\sqrt{2^N}} \sum\limits_{n_1\dots n_N} \prod\limits_{\expval{ij}} (-1)^{n_i n_j}\ket{n_1, n_2 \dots n_N}
		\end{aligned}
	\end{equation}
     where $n_i = 0, 1$ for $i = 1,\ldots ,N$. We relabel the qubits by their physical coordinates $(x,\tau)$, with the $y-$coordinate serving as an effective time direction, and define the shorthand $\qty{n}_\tau\equiv\qty(n_{1,\tau}, n_{2,\tau},\ldots,n_{L_x,\tau})$ to label the coordinates on each row (or timestep) $\tau$. Next, the pairs of nearest neighbors are enumerated and divided into two types -- those that share a vertical edge (sites at $(x,\tau)$ and $(x,\tau+1)$), and those that share a horizontal edge (sites at $(x,\tau)$ and $(x+1,\tau)$), so that    \begin{equation}
             \prod\limits_{\expval{ij}} (-1)^{n_i n_j} = U^{\qty{n}_{L_y}}_2\prod\limits_{\tau<L_y} U^{\qty{n}_{\tau+1},\qty{n}_\tau}_1 U^{\qty{n}_\tau}_2,
     \end{equation}

where
\begin{equation}
    \begin{aligned}
        U^{\qty{n}_{\tau'},\qty{n}_\tau}_1&\equiv \prod\limits_x (-1)^{n_{x,\tau'},n_{x,\tau}}&&\text{(vertical edges)}\\
             U^{\qty{n}_\tau}_2&\equiv\prod\limits_x (-1)^{n_{x,\tau},n_{x+1,\tau}}&&\text{(horizontal edges)}.
    \end{aligned}
\end{equation}
 $\ket{\Psi}$ can now be expressed as
 \begin{equation}
 \begin{aligned}
     \ket{\Psi} = \sum\limits_{\qty{n}} &U^{\qty{n}_{L_y}}_2 \ket{\qty{n}_{L_y}} \\
     &\prod\limits_{\tau<L_y} \qty(U^{\qty{n}_{\tau+1},\qty{n}_\tau}_1 U^{\qty{n}_\tau}_2\ket{\qty{n}_\tau})
 \end{aligned}
 \end{equation}
Given a specific set of bulk measurement outcomes $\qty{\mu_{x,\tau}, w_{x,\tau}}; \tau<L_y$, the (unnormalized) boundary state can be written as
	
	\begin{equation}
		\begin{aligned}
			\ket{\Psi}_{\partial\mathcal{M}} \propto &\braket{\qty{\mu_{x,\tau}, w_{x,\tau}}}{\Psi}\\
   =&\sum\limits_\qty{n} U^{\qty{n}_{L_y}}_2\ket{\qty{n}_{L_y}}\times\\
			&\prod\limits_{\tau<L_y} \qty(U^{\qty{n}_{\tau+1},\qty{n}_\tau}_1 U^{\qty{n}_\tau}_2 M^{\qty{n}_\tau}),
		\end{aligned}
	\end{equation}
 where
 \begin{equation}
     M^{\qty{n}_\tau} \equiv \prod\limits_x \braket{W_{x,\tau}}{n_{x,\tau}} = \prod\limits_x\frac{W_{x,\tau}^{n_{x,\tau}}}{\sqrt{1 + |W_{x,\tau}|^2}}.
 \end{equation}

	Recall from \cref{eq: local_basis} that $\braket{W}{n} =\frac{W^n}{\sqrt{1 + |W|^2}}$, where $\ket{W} \equiv \frac{1}{\sqrt{1 + |W|^2}} \qty(\ket{0} + W\ket{1})$. To complete the mapping to a quantum circuit, we introduce factors of $\braket{\qty{n}_\tau}=1$ and rewrite the expression for $\ket{\Psi}_{\partial\mathcal{M}}$ as
	\begin{equation}
            \ket{\Psi}_{\partial\mathcal{M}} \propto \widetilde{U}\qty(L_y)\ket{+}^{\otimes L_x} 	
	\end{equation}
	where the nonunitary circuit $\widetilde{U}\qty(L_y)$ of depth $L_y$ is
	\begin{equation}
		\begin{aligned}
			\widetilde{U}\qty(L_y,0) &\equiv\sum\limits_\qty{n}U^{\qty{n}_{L_y}}_2\dyad{\qty{n}_{L_y}}\times \\
			&\prod\limits_{\tau<L_y} \Bigg\lbrace\qty(U^{\qty{n}_{\tau+1},\qty{n}_\tau}_1\dyad{\qty{n}_{\tau+1}}{\qty{n}_\tau})\times\\
			&\qty(U^{\qty{n}_{\tau}}_2\dyad{\qty{n}_\tau}) \qty(M^{\qty{n}_{\tau}}\dyad{\qty{n}_\tau})\bigg\rbrace
		\end{aligned}
        \label{eq:U_def}
	\end{equation}

	The terms in each of the parentheses can readily be interpreted as quantum gates and measurements by following the prescription
	
	\begin{equation}
		\begin{aligned}
			U_1 &\rightarrow \prod\limits_xH_x \text{ (Hadamard gates)}\\
			U_2 &\rightarrow \prod\limits_x CZ_{x,x+1}\\
			M &\rightarrow \text{Weak measurements } P_\pm(\beta) \sim e^{\pm\beta Z}.
		\end{aligned}
	\label{eq:prescription}
	\end{equation}

    The circuit, written as a sequential application of $L_y-1$ layers of quantum gates, is
    \begin{equation}
    \begin{aligned}
        \widetilde{U}(L_y) &= \prod\limits_x CZ_{x,x+1}\prod\limits_\tau \widetilde{U}_\tau\\
        \widetilde{U}_\tau &\equiv \prod\limits_x H_x \prod\limits_x CZ_{x,x+1} \prod\limits_x P_{\mu_{x,\tau}}.
    \end{aligned}
    \end{equation}
    Through $P_{\mu_{x,\tau}}$, the outcomes of the measurements performed on cluster state fittingly influence the measurements that are to be applied in the {\blue \((1+1)\mathrm{D}\) } circuit. An illustration of one layer $\widetilde{U}_\tau$ is shown in \cref{fig:1dCirc}. We elaborate on the prescription \cref{eq:prescription} below.

    \begin{figure}
        \centering
        \includegraphics[width=0.45\textwidth]{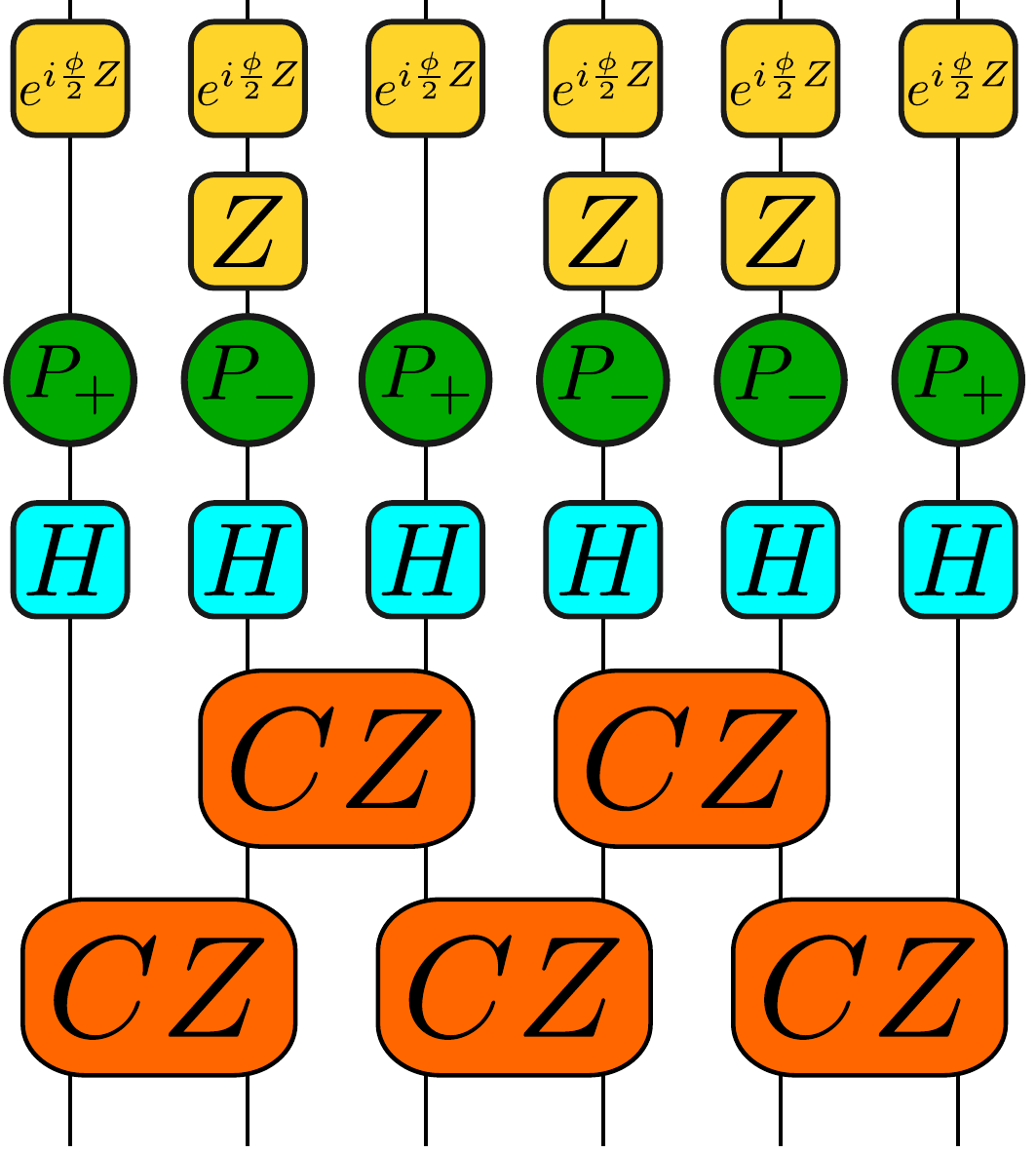}
        \caption{One layer of the circuit defined in \cref{eq:U_def}, for a particular set of measurement outcomes. A circuit with $L_y$ such layers, with $P_\pm$ chosen according to the measurement outcomes for each row of the cluster state, results in $\ket{\Psi}_{\partial\mathcal{M}}$ -- the state obtained after measuring the bulk qubits of a cluster state.}
        \label{fig:1dCirc}
    \end{figure}

	\textit{$H$ gate}: The term $U^{\qty{n}_{\tau+1},\qty{n}_\tau}_1\dyad{\qty{n}_{\tau+1}}{\qty{n}_\tau}$ can be decomposed into a product of single particle operators 
 \begin{equation}
     \prod\limits_x (-1)^{n_{x,\tau+1} n_{x,\tau}} \dyad{n_{x,\tau+1}}{n_{x,\tau}},
 \end{equation} so it suffices to show that each single particle operator with matrix elements of the form $(-1)^{n_{x,\tau+1} n_{x,\tau}}$ corresponds to a $H$ gate.
 A single-qubit operator $U_1$ can be represented in the computational basis as
	\begin{equation}
		U_1 \equiv \sum\limits_{n,n'} \qty(U_1)_n^{n'} \dyad{n'}{n}.
	\end{equation}
	
	By requiring the matrix elements of $U_1$ to be $\qty(U_1)_n^{n'} = (-1)^{nn'}$, 
 
	\begin{equation}
		\qty(U_1)_n^{n'} = \mqty(1&1\\1&-1).
	\end{equation}
  $U_1$ has the same representation as the Hadamard gate $H$ up to a normalization factor $1/\sqrt{2}$, in the computational basis. This establishes the first connection in \cref{eq:prescription}. 
	
	\textit{$CZ$ gate}: Turning to the second term, we now construct a two-qubit unitary gate that is diagonal in the computational basis. It can similarly be represented as
	\begin{equation}
		U_2 \equiv \sum\limits_{n,n'} U_2^{n,n'} \dyad{n,n'}{n,n'}.
	\end{equation}
	
	Again, requiring $U_2^{n,n'} = (-1)^{nn'}$, $U_2$ has the same representation as the $CZ$ gate,
	\begin{equation}
		U_2 \equiv \mqty(\dmat{1,1,1,-1}) = CZ.
	\end{equation}

	\textit{Weak Measurements}: As with the $H$ gate, we consider a possibly non-unitary, diagonal, single-qubit operator $M$ whose representation is given by
	
	\begin{equation}
		\begin{aligned}
			M \equiv \sum\limits_n M^n\dyad{n};\\
			M^n \propto \frac{W^n}{\sqrt{1 + |W|^2}}.
		\end{aligned}
	\label{eq:weak_meas}
	\end{equation}

	We reiterate the parametrization of $W$ according to \cref{eq: measurement parameter,eq: weight_parameter}. This will also enable us to fix the normalization constant for $M$. For a single-qubit measurement performed in the direction $\hat{n} = \cos\theta\hat{z} + \sin\theta\cos\phi\hat{x} + \sin\theta\sin\phi \hat{y}$ with a measurement outcome $+\hat{n}$, the weight $W = \tan\frac{\theta}{2} e^{i\phi}$. The resulting operator $M_+$, constructed according \cref{eq:weak_meas}, is
	
	\begin{equation}
		\begin{aligned}
			M_+ &\propto \frac{1}{\sqrt{1+\tan^2\frac{\theta}{2}}}\qty(\dyad{0} + e^{-i\phi}\tan(\frac{\theta}{2})\dyad{1})\\
			&\propto \exp(i\frac{\phi}{2} Z) \qty(\cos(\frac{\theta}{2})\dyad{0} + \sin(\frac{\theta}{2})\dyad{1})\\
			&\equiv \exp(i\frac{\phi}{2} Z) P_+(\beta).
		\end{aligned}
	\end{equation}
	where, in the last line, we have unambiguously defined the operator $M_+$. $P_+(\beta)$ refers to a weak projector (of strength $\beta$) to the $\ket{0}$ state
	
	\begin{equation}
		P_+(\beta) \equiv \frac{1}{\sqrt{2\qty(1 + \tanh^2(\beta))}}\qty(1 + \tanh(\beta) Z),
	\end{equation}
	
	and the strength $\beta$ is defined via

	\begin{equation}
		\tanh(\beta) = \tan(\frac{\pi}{4} - \frac{\theta}{2}).
		\label{eq:strength}
	\end{equation}

	Similarly, when the outcome is $-\hat{n}$, the operator $M_-$ is defined as
	\begin{equation}
		\begin{aligned}
			M_- &\propto \frac{1}{\sqrt{1+\cot^2\frac{\theta}{2}}}\qty(\dyad{0} - e^{-i\phi}\cot(\frac{\theta}{2})\dyad{1})\\
			&\equiv \exp(i\frac{\phi}{2} Z) Z P_-(\beta),
		\end{aligned}
	\end{equation}

	where $P_-$ analogously refers to a weak projector to the $\ket{1}$ state
	\begin{equation}
		P_-(\beta) \equiv \frac{1}{\sqrt{2\qty(1 + \tanh^2(\beta))}}\qty(1 - \tanh(\beta) Z).
	\end{equation}
	
	At their ``strongest", $P_\pm(\beta\to\infty)$ reduce to the projectors to the $\ket{0}$ and $\ket{1}$ states
	\begin{equation}
		P_\pm(\beta\to\infty) = \frac{1\pm Z}{2},
	\end{equation}
	whereas at their weakest, $P_\pm(0) \propto 1$. Thus, these weak measurements generalize the idea of projective measurements, where weaker measurements (i.e., with smaller $\beta$) obtain less information from the system. $P_\pm$ describe a POVM and obey $P_+^\dagger P_+ + P_-^\dagger P_- = 1$. It follows that $M_\pm$ also constitute a valid set of Kraus operators since they are unitarily related to $P_\pm$; hence, $M_\pm(\beta)$ describe weak measurements followed by a single-qubit rotation about the Z axis.
	
	The boundary state is now equivalent to a 1D state obtained after a hybrid quantum circuit of depth $L_y$, consisting of measurements and unitary operations, is applied to the state $\ket{+}^{\otimes L_x}$. The unitary gates in this circuit are dual unitary, and known to be maximally chaotic for generic $\phi$ \cite{PhysRevLett.123.210601,PhysRevLett.128.060601}. This dynamical interpretation is also recast in terms of tensor network representations in \cref{ap:TContDyn}. While the method pioneered in \cite{PhysRevX.12.021021} provides a generic algorithm to sample from 2D shallow circuits -- such as the circuits used to prepare cluster states -- we use a more direct method that utilizes the exact form of the tensors that constitute the cluster state, to construct the boundary state.
	
	In conventional setups that have studied the MIPT in {\blue \((1+1)\mathrm{D}\) } random circuits, the tuning parameter is the rate $p$ at which projective measurements are performed on the qubits. In our interpretation, weak measurements are performed on every qubit at every time step, so $p=1$. However, an MIPT is observed even in such a setting \cite{PhysRevB.100.064204}. The entanglement scaling of $\ket{\Psi}_{\partial\mathcal{M}}$ is expected to exhibit a transition from volume-law to area-law behavior as the strength $\beta$ is increased. An increase in $\beta$ is achieved by choosing the measurement axis to align closer to the Z axis (i.e. by decreasing $\theta$), according to \cref{eq:strength}. When $\theta=0 \qty(\beta=\infty)$, we expect an area-law scaling for $\ket{\Psi}_{\partial\mathcal{M}}$, whereas when $\theta=\frac{\pi}{2}\qty(\beta=0)$, a volume-law scaling is expected, since the circuit is purely unitary (and chaotic). In subsequent sections, we present numerical evidence that a transition in the entanglement scaling behavior is observed as $\theta$ is varied. Using this mapping, the physics underlying this transition has been shown to be innately connected to the physics of MIPTs.

\section{Boundary States, Sampling  and Ising Partition Functions}\label{sec:connections}

We have shown thus far that the amplitudes of specific outcomes upon performing single qubit measurements on certain 2D states can be expressed as 2D Ising partition functions. Moreover, the measurement directions and outcomes only alter the strength of the couplings of partition functions, not their form. In this section, we explain the connections amongst the evaluation of these partition functions, the entanglement scaling of the boundary state and the overarching problem of resource state sampling.

Given an Ising partition function $\mathcal{Z}$ of the forms discussed above, we can appropriately choose the measurement directions and outcomes so that the overlap $\braket{\mu,\omega}{\Psi_\text{resource}} = \mathcal{Z}$. MPS-based algorithms as in \cref{ap:TContDyn} can successfully evaluate such overlaps, provided the entanglement entropy of the boundary state does not follow a volume-law. Therefore, in the regimes where $\psi_{\partial\mathcal{M}}$ obeys an area-law, the corresponding partition functions can be easily approximated, classically. We note that the converse need not be true, as reconstructing a cluster state from partition functions generally requires the evaluation of an exponential (in the number of qubits) number of partition functions.

On the other hand, many resource states are constructed from shallow circuits; the cluster state, for instance, involves only the application of commuting $CZ$ gates on qubits, its preparation involves a depth-1 circuit. As noted in the Introduction, a tensor-network based proposal to classically sample from shallow 2D circuits was put forth in \cite{PhysRevX.12.021021}, by mapping them to an effective {\blue \((1+1)\mathrm{D}\) } dynamics with random unitary gates and projective measurements -- a mapping that is elaborated in \cref{sec: pauli_numerics}. This method succeeds provided the boundary state remains area-law entangled. We extend the results of \cite{PhysRevX.12.021021} by specifically considering 2-qubit $CZ$ gates, 1-qubit rotations about the Z-axis and weak measurements, all of which can be obtained by preparing a cluster state and then performing 1-qubit measurements in arbitrary directions. Therefore, by identifying measurement directions which result in $\psi_{\partial\mathcal{M}}$ with area-law entanglement, we determine ``easy" regimes where cluster states can be efficiently sampled. The generality of the directions renders Clifford methods inefficient, since the stabilizer rank of $\psi_{\partial\mathcal{M}}$ is in general exponential in $L$\cite{Qassim2021improvedupperbounds}. 

We conclude this section with the following caveats -- our results only assert that these computational tasks are easy in certain settings; efficient sampling could be performed classically even beyond these ``easy" regimes. Moreover, these three tasks are generally disparate, and our work does not establish an equivalence between them. We have, instead, determined that classically sampling from certain cluster states and the classical evaluation of the Ising partition function can be performed using our boundary-state-based algorithm only when the boundary state can itself be simulated accurately; the boundary state must have a non-volume-law scaling of entanglement, or be produced through bulk measurements in the $X$, $Y$ or $Z$ bases alone. \cref{fig: road_map} offers visualization of these connections.

\section{Numerical Results}\label{sec: numerics}

We now turn to the numerical characterization of the boundary state obtained after performing single-qubit measurements on all the qubits in the bulk of a 2D graph state. We consider two types of bulk measurements: (A) measurements in arbitrary directions, and (B) Pauli X/Y/Z measurements. Broadly, in (A), we vary the direction along which the measurements are performed, whereas in (B), we alter the fraction of X/Y/Z measurements. We are specifically interested in determining the presence of an entanglement transition as these parameters are varied. We further seek to establish connections between the entanglement of the boundary state and the ease of approximating certain families of partition functions.

\subsection{Entanglement Scaling at Arbitrary Measurement Directions}

\begin{figure}
    \centering
    \includegraphics[width=0.45\textwidth]{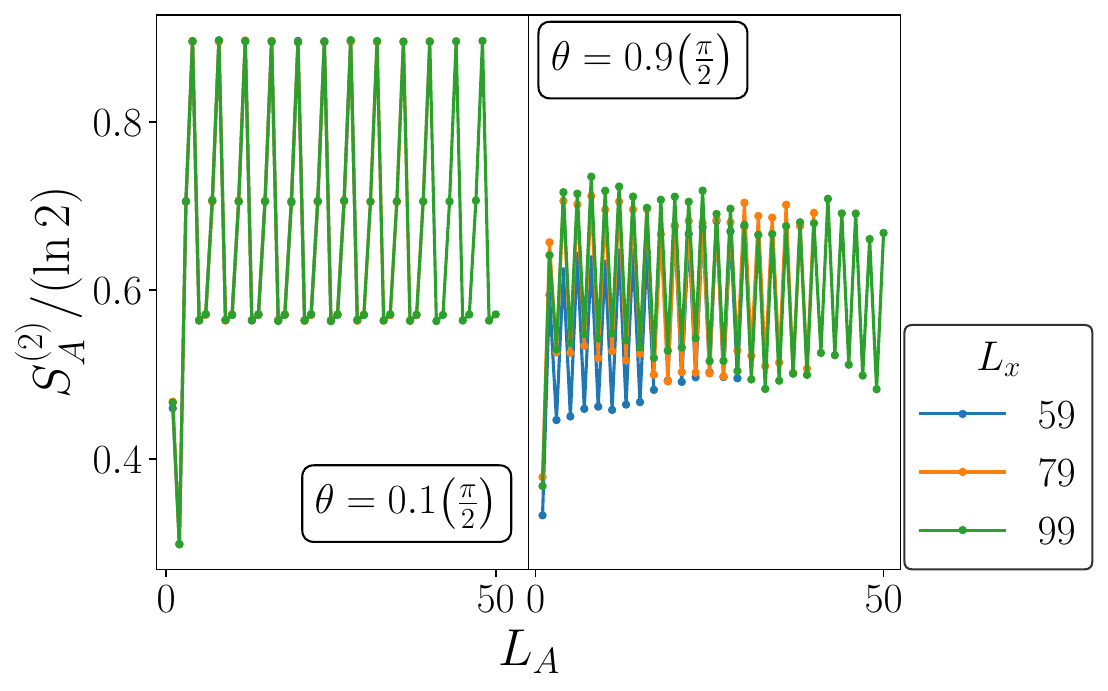}
    \caption{The entanglement entropy in the boundary state of a Lieb lattice with dimensions $L_y=500$ and  given $L_x$. Measurements are performed along $\hat{n} = \cos(\theta) \hat{z} + \sin(\theta) \hat{x}$. In both cases, $S_A$ shows no dependence on subsystem-size $L_A$, nor does it change with $L_x$, indicating an area-law phase.}
    \label{fig:LL_XZ}
\end{figure}

\begin{figure}
    \centering
    \includegraphics[width=0.45\textwidth]{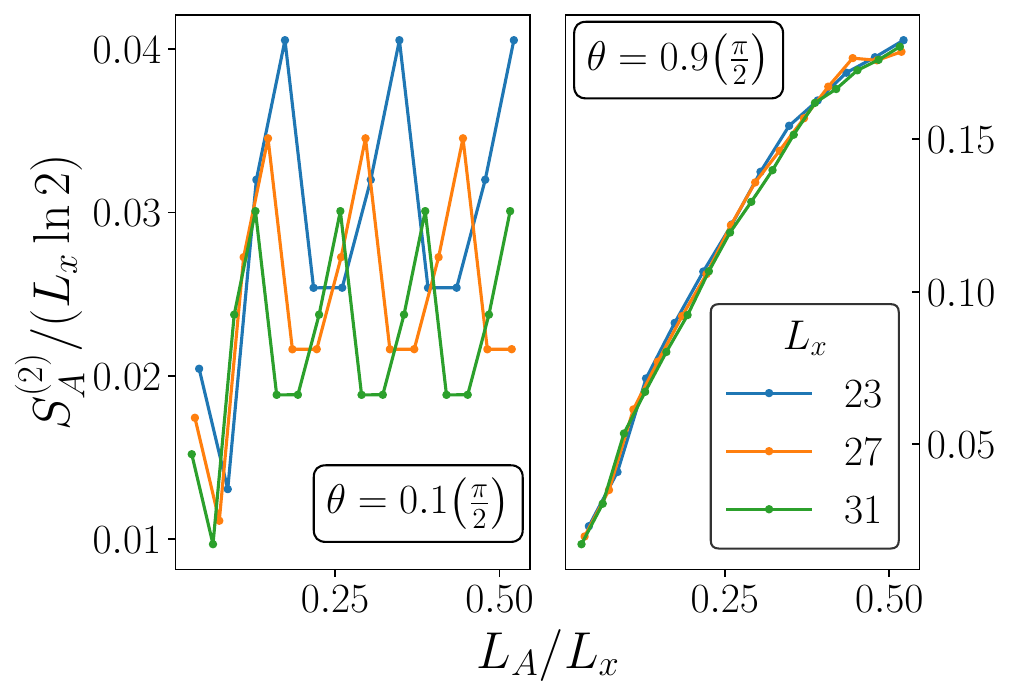}
    \caption{The scaling of entanglement entropy in the boundary state of a Lieb lattice with dimensions $L_y=500$ and  given $L_x$. Measurements are now performed along $\hat{n} = \cos(\theta) \hat{z} + \sin(\theta) \hat{y}$. (Left) $S_A/L_x$ trends downward as $L_x$ increases, indicating that the entanglement scales sub-linearly. When $S_A$ is plotted as a function of $L_A$ (as in \cref{fig:LL_XZ}, not shown), the curves superimpose one-another, indicating an area-law phase. (Right) $S_A$ scales linearly with $L_x$, and the data is amenable to a collapse indicative of a volume-law phase.}
    \label{fig:LL_YZ}
\end{figure}

\begin{figure}[!ht]
    \centering
    \includegraphics[width=0.45\textwidth]{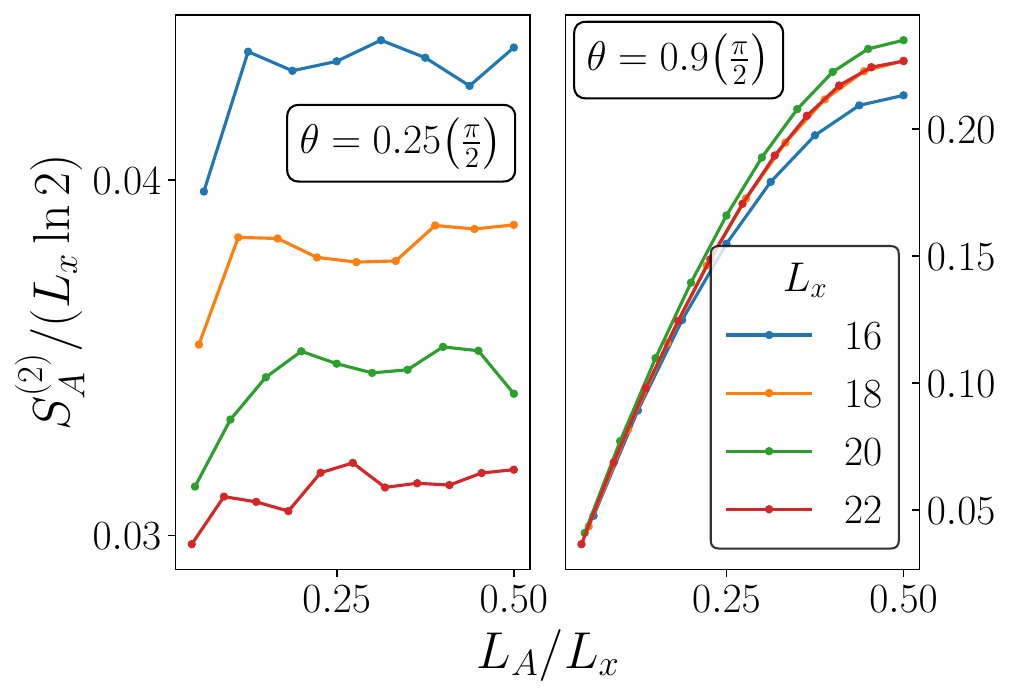}
    \caption{The scaling of entanglement entropy in the boundary state of a square lattice with dimensions $L_y=500$ and  given $L_x$. Measurements are performed along $\hat{n} = \cos(\theta) \hat{z} + \sin(\theta) \hat{x}$. As in \cref{fig:LL_YZ}, (Left) $S_A/L_x$ trends downward as $L_x$ increases, while (Right) $S_A$ scales linearly with $L_x$, indicating a volume-law phase.}
    \label{fig:SqLat}
\end{figure}

Results are presented here for the entanglement of the boundary state after forced measurements are performed along an arbitrary direction $\hat{n}$ on the qubits in the bulk. The entanglement is quantified using the $2^{\rm nd}$ R\'{e}nyi entropy $S^{(2)}_A$, defined as
	\begin{equation}
		S^{(2)}_A = -\ln\tr\rho^2_A,
	\end{equation}
where $\rho_A = \tr_{\overline{A}}\dyad{\psi_{\rm edge}}$ is the reduced density matrix describing subsystem $A$ of the boundary state $\ket{\psi}_{\partial\mathcal{M}}$, and $\overline{A}$ is the complement of $A$.
	
We consider the cluster state defined on (1) the Lieb lattice and (2) the square lattice as initial states. Our simulation emulates the following setup -- on the initial state, a single qubit measurement is performed on the qubit located at position $\vec{r}$, along the (possibly position-dependent) direction $\hat{n}(\vec{r})$. Following a measurement, the qubit points along $\hat{n}$ or $-\hat{n}$. The outcome, however, is ``post-selected", in that the outcome is chosen with a predetermined probability distribution; the specific distributions will be described later in this section.
	
We denote by $L_x$ the horizontal width of the lattice, and by $L_y$, its vertical length. By employing the tensor network scheme described in \cref{ap:TContDyn}, these dimensions have the alternate interpretation of $L_x$ qubits undergoing a nonunitary, stroboscopic evolution for $L_y$ ``time steps".  While tensor network algorithms usually fix the bond dimension for the tensors, we only require that the normalized sum of the truncated singular values remains lower than a threshold ($10^{-9}$ in our case); the bond dimension can grow up to a maximum of $2^{\frac{L_x}{2}}$. This ensures that the error in the numerically simulated boundary states (and thus the Ising partition functions) is bounded \cite{MPSTruncError}. 
The entanglement scaling behavior is determined by the following criteria: for a given $\hat{n}$, we first plot the scaled entanglement entropy $S_A/L_x$ as a function of fractional subsystem size $L_A/L_x$ for various sizes $L_x$. If the curves approximately collapse onto a single curve, the boundary state obeys a volume-law scaling, $S^{(2)}_A \sim |A|$, and is hard to simulate using a tensor network scheme, since the bond dimension has to grow exponentially in $L_x$ to maintain an $\mathcal{O}(1)$ truncation error. If the curves instead trend downward with increasing $L_x$, this signals an area-law scaling, $S^{(2)}_A \sim {\rm const}$, of the entanglement entropy, and the boundary state is easily simulable. These simulations were implemented using the ITensor Julia package \cite{itensor,itensor-r0.3}. 
	
In this section, we first explore the effects of uniformly measuring all the bulk qubits along a direction $\hat{n}(\vec{r})\equiv\hat{n}$, independent of their position. $\hat{n}$ is parameterized by the angles $\theta$ and $\phi$ as $\hat{n} = \cos\theta\hat{z} + \sin\theta\cos\phi\hat{x} + \sin\theta\sin\phi \hat{y}$. The outcomes at each site are chosen independently and randomly to be $\pm\hat{n}$ with equal probability, and $S^{(2)}_A$ for the boundary state is finally averaged over 20 such sets of outcomes, so as to ameliorate the noise, and bring the underlying scaling with $|A|$ to the fore.  It should be noted that the equally probable measurement outcomes mentioned here are artificially assigned and differ from the probabilities obtained from Born’s rule. We believe that our results capture the typical behavior of the boundary states, despite this post-selection. For the cluster state defined on the Lieb lattice, we separately consider the cases where $\hat{n}$ lies in {\blue $x$-$z$} plane ($\phi=0$), and in the {\blue $y$-$z$} plane ($\phi=\frac{\pi}{2}$), whereas in the square lattice, we only consider the case where $\phi=0$. 
	
Under certain conditions, we find that the entanglement of the boundary state undergoes a phase transition as the measurement angle $\theta$ is varied. This is interesting in its own right, since this transition adds to the panoply of measurement-induced phase transitions. Additionally, such a transition also bears interesting consequences for the feasibility of using an effective {\blue \((1+1)\mathrm{D}\) } dynamics to approximate the partition functions of 2D Ising models. As described in \cref{sec: boundary_states}, every boundary state has an associated classical Ising partition function $\mathcal{Z}$. In the area-law phase, $\mathcal{Z}$ can be computed efficiently for large system sizes with controlled approximations, but in the volume-law phase, these approximations to $\mathcal{Z}$ are uncontrolled, and thus, the calculations of $\mathcal{Z}$ requires resources exponentially large in $L_x$. In the {\blue \((1+1)\mathrm{D}\) } dynamics, the approximations correspond to retaining only a fixed number of the largest singular values in each of the tensors that constitute $\ket{\psi_{\rm virt}}$. This can be done with a controlled error only if $\ket{\psi_{\rm virt}}$ is area-law entangled.
	
We further utilize this connection to explore the complexity of approximating partition functions of specific classical models. This proceeds by restoring the position dependence of $\hat{n}(\vec{r})$ and postselecting for the $+\hat{n}(x,y)$ outcome. The measurement protocols that we consider correspond to classical Random Bond Ising Models (RBIMs), as in \cref{eq: lieb_boundary_state,eq:sq_lattice_boundary_state}. We begin by studying RBIMs where the absolute values of the interaction parameters ($|K|$) are fixed to be constant, but their signs ($K/|K|$) take random values $\pm1$ for each set of interacting spins. The magnetic field $h$ is taken to be a constant as well. We proceed to study a few different types of coupling, and thereby attempt to uncover the factors that could play a key role in making the calculation of partition functions hard.  We posit that the complexity of calculating $\mathcal{Z}$ (using a {\blue \((1+1)\mathrm{D}\) }-like algorithm) depends crucially on the form of the interactions in the Hamiltonian of $\mathcal{Z}$.

\subsubsection{Lieb Lattice}

In the Lieb lattice, a volume-law entangled phase is absent in the boundary state of a Lieb lattice when $\hat{n}$ lies in the {\blue $x$-$z$} plane. $\hat{n}$ is parameterized by the angle $\theta$ that it makes with the z-axis, as $\hat{n}\equiv \cos{\qty(\theta)}\hat{z} + \sin{\qty(\theta)}\hat{x}$. In this case, the boundary state is always area-law entangled{\blue, as shown in \cref{fig:LL_XZ}}. The $\mathcal{Z}$ corresponding to this case has positive and negative weights, but we see that this does not lead to a change in the complexity of its calculation. When $\hat{n}$ is instead chosen to lie in the {\blue $y$-$z$} plane, now parameterized as $\hat{n}\equiv \cos{\qty(\theta)}\hat{z} + \sin{\qty(\theta)}\hat{y}$, an area-law to volume-law transition is observed as $\theta$ is tuned through $\theta_c\sim0.75\qty(\frac{\pi}{2})$. $\mathcal{Z}$ now includes complex weights, but it is only above a specific value of $\theta$ that this calculation becomes hard. {\blue This transition is shown by noting that the entanglement entropy of a subsystem scales linearly with its size above $\theta_c$, as in \cref{fig:LL_YZ}.}

These results can be explained by studying the effective \((1+1)\mathrm{D}\) circuit on the Lieb lattice, analogous to that on the square lattice (\cref{fig:1dCirc}). Measurements on the qubits situated on the vertices and vertical links translate to (single-qubit) rotations and weak measurements, as in the square lattice. However, the two-qubit gate acting on a pair of neighboring vertex qubits is determined by the measurements on the horizontal link joining them. These two-qubit gates are still diagonal in the computational basis, but their unitarity depends now on the measurement angles. 

For a general measurement of a qubit on a horizontal link, with an outcome $m=\pm$, the computational basis representation of the gate acting on qubits $i$ and $j$ is
\begin{equation*}
\begin{aligned}
    G^m_{ij}(\theta,\phi) &\equiv \text{diag}\qty(g^m_+, g^m_-, g^m_-, g^m_+);\\
    g^{m=+}_\pm &\equiv \cos\qty(\frac{\theta}{2}) \pm e^{-i\phi}\sin\qty(\frac{\theta}{2}), \\
    g^{m=-}_\pm &\equiv \sin\qty(\frac{\theta}{2}) \mp e^{-i\phi}\cos\qty(\frac{\theta}{2}).
\end{aligned}
\end{equation*}

\textbf{{\blue $x$-$z$} Measurements:} When $\phi=0$,
\begin{equation*}
    \begin{aligned}
        G^m_{ij} &\propto e^{m\alpha Z_i Z_j};\\
        \tanh(\alpha) &= \exp(-2\beta), \\
        \tanh(\beta) &= \tan(\frac{\pi}{4} - \frac{\theta}{2}).
    \end{aligned}
\end{equation*} $G$ is clearly a nonunitary weak measurement on $Z_iZ_j$, with the strength increasing  as $\theta$ increases. The vertical measurements, on the other hand, correspond to weak single-qubit measurements of $X$ (links) and $Z$ (vertices). When the vertex qubits are measured in the X direction, the circuit consists solely of weak measurements that can be mapped to free fermionic measurements, as elucidated in \cref{sec: exact_solvable}. It is further known that such nonunitary free fermion dynamics do not admit a volume-law phase \cite{XiaoFF}. Upon allowing the vertex qubits to be measured along a generic angle the {\blue $x$-$z$} plane, the system remains in an area-law phase, since single-qubit $Z$ measurements are incapable of increasing the entanglement in this case. Thus, the partition function of the Ising model on the Lieb lattice is always easily approximated.

\textbf{{\blue $y$-$z$} Measurements:} When $\phi=\frac{\pi}{2}, G^m_{ij} \propto e^{-im\theta Z_i Z_j}$ -- a unitary entangling gate. Measurements on the vertices result in unitary rotations along $Z$ (a non-free fermionic gate), in addition to weak measurements. A volume-to-area-law transition is expectedly observed here, since the \((1+1)\mathrm{D}\) dynamics realizes a (non-free fermionic) quantum circuit with weak measurements, which, as in the square lattice, undergoes a volume-to-area-law MIPT.

\subsubsection{Square Lattice}

In contrast to the Lieb lattice, we find an area- to volume-law transition in the scaling of the entanglement entropy of the boundary state in the square lattice, even as $\hat{n}$ is varied in the {\blue $x$-$z$} plane. For small $\theta$, the state is area-law entangled, but becomes volume-law entangled as $\theta$ exceeds $\theta_c \sim 0.7\frac{\pi}{2}$, as shown in \cref{fig:SqLat}. Again, here, $\mathcal{Z}$ is entirely of real weights, but the corresponding Ising model now has 4-body interactions \cref{eq:sq_lattice_boundary_state}, as opposed to 2-body interactions in the Lieb lattice. Using the mapping \cref{eq:sq_lattice_boundary_state}, we can obtain the following intuition for the transition. When $\theta = 0$, the (real part of) Ising interaction strength is $0$, while the contribution from the magnetic field to the partition function is 0, unless all the outcomes are $-$. In the opposite limit $\theta=\frac{\pi}{2}$, the Ising interactions are equally likely to be $\pm\infty$. The increasing relevance of the frustrations arising from the 4-body interactions as $\theta$ is varied could be responsible for the poor performance of the \((1+1)\mathrm{D}\) algorithm.

\begin{figure}
    \centering
    \includegraphics[width=0.48\textwidth]{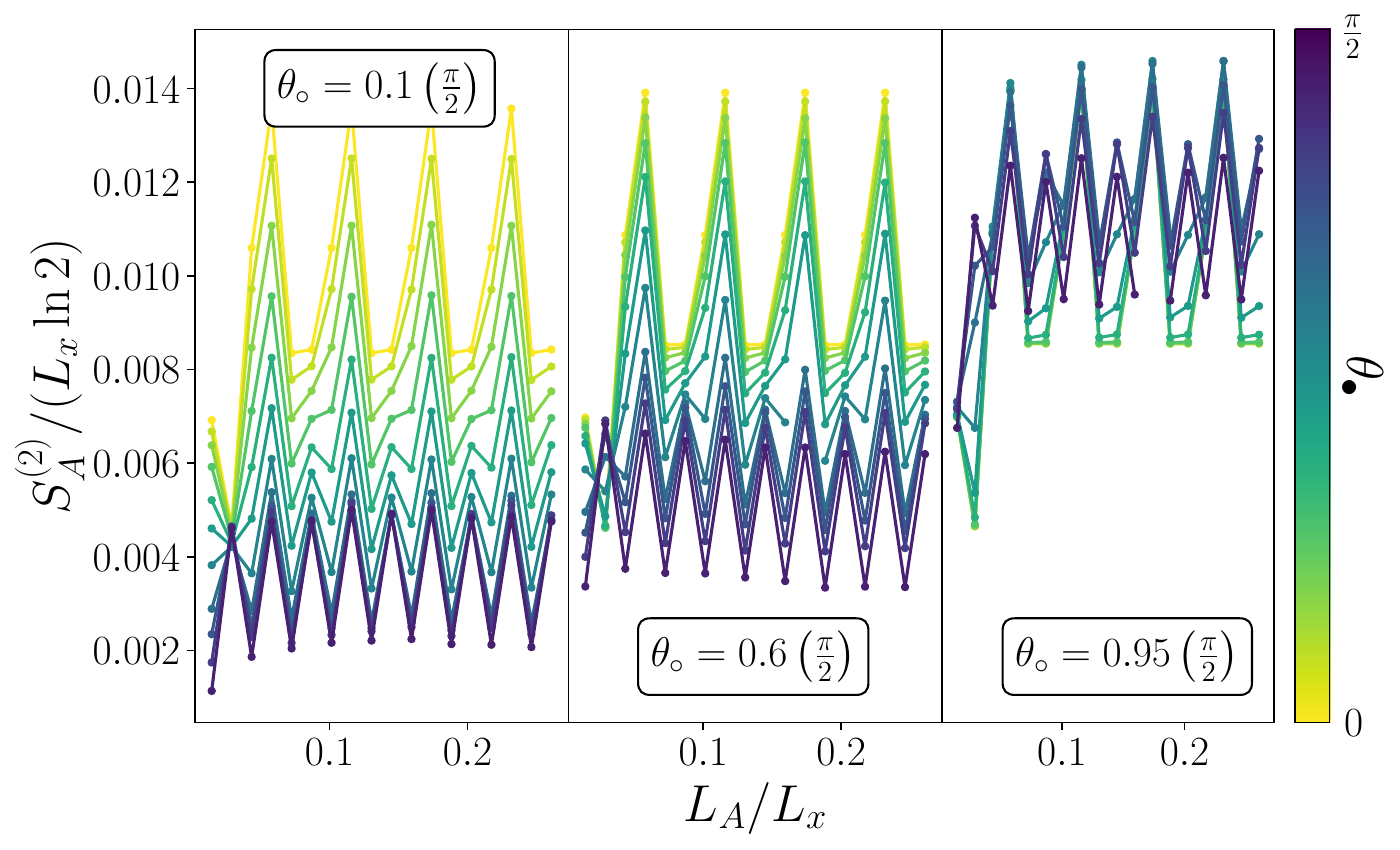}
    \caption{$S^{(2)}_A$ for $\ket{\psi}_{\partial\mathcal{M}}$ obtained after measurements on the Lieb lattice, for different values of $\theta_\bullet$ and $\theta_\circ$. The measurement outcomes are post-selected to be $+\hat{n}_{\circ/\bullet}$, where $\hat{n}$ lies in the {\blue $x$-$z$} plane on both $\circ$ and $\bullet$ sublattices. $\theta_\circ$, which sets the magnetic field in $\mathcal{Z}$, is held constant, while only $|\theta_\bullet|$ is fixed (i.e. $\hat{n}$ is randomly chosen to make an angle $\pm\theta_\bullet$ with the $Z$ axis). The boundary state is always in an area law, indicating that the associated Ising partition function, with 2-spin random interactions on the square lattice, consists only of positive weights and is easy to approximate.}
    \label{fig:rbim_ll}
\end{figure}

\begin{figure}[h!]
    \centering
    \includegraphics[width=0.475\textwidth]{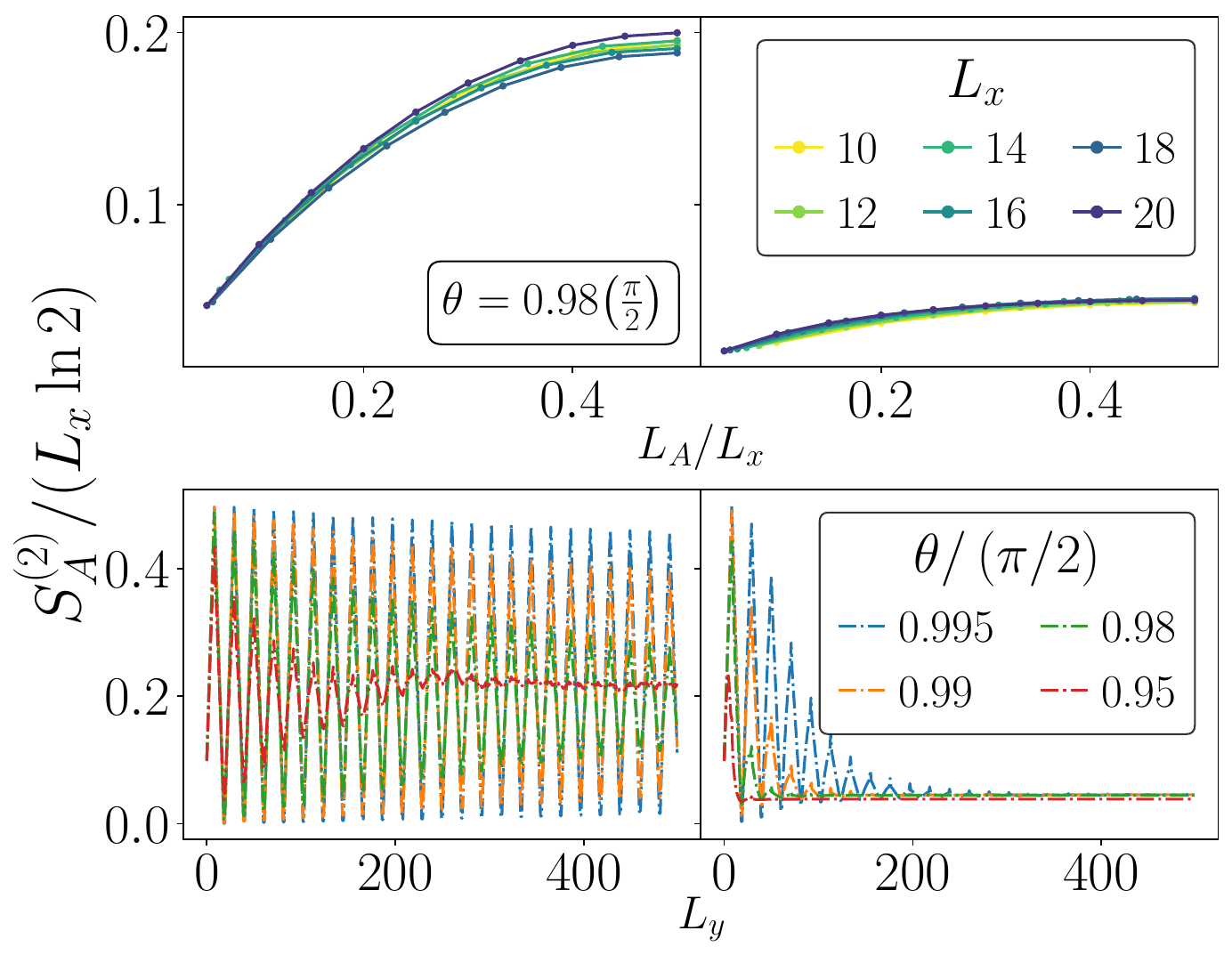}
    \caption{Entanglement entropy density of the boundary state when the measurement outcomes of the bulk are (left) random and (right) fixed to be along $+\hat{n}$. The greater entanglement when there is randomness in the measurement outcomes is accompanied by a higher degree of frustration in the corresponding classical model. (Top) Scaling of $S_A$ as a function of subsystem size for different $L_x$ (Bottom) ``Evolution" of the entanglement of the boundary state with vertical depth $L_y$ (or imaginary time $\tau$). In the absence of randomness, {\blue $S^{(2)}_A$} decays strongly, ultimately saturating in values much smaller than in the case with randomness as $L_y\to\infty$.}
    \label{fig:randomness}
\end{figure}

Randomness in the measurement outcomes is crucial to observing any non-area-law phase on the square lattice {\blue (the Lieb lattice (\cref{fig:rbim_ll}) is discussed below)}. If the measurement outcomes are all chosen to be equal ($+1$, for instance), the boundary state is area-law entangled in the thermodynamic limit $L_y\to\infty$. The finite-size scaling of entanglement is shown in the presence and absence of randomness in \cref{fig:randomness}, in the square lattice, making evident the significance of stochasticity. Frustration is important to making the calculation of $\mathcal{Z}$ hard, and cannot exist if all measurement outcomes and directions -- consequently, the Ising interaction strengths -- are the same. This explains the differences between our results and those found in \cite{guo:2023}, where the boundary state obtained after measurements on the square lattice is never in a volume-law phase, except at $\theta=\frac{\pi}{2}$.

\subsubsection{Connections to Classical Partition Functions}
We have, thus far, discussed the entanglement scaling as the measurement direction is kept uniform across the lattice, while outcomes are random. We have also demonstrated that the resulting boundary state $\ket{\psi}_{\partial\mathcal{M}}$ can be described in terms of a classical partition function $\mathcal{Z}$. In this subsection, we instead use a different type of post-selected measurement outcomes to glean information about the computational complexity of the corresponding $\mathcal{Z}$ that describes the $\ket{\psi}_{\partial\mathcal{M}}$. We focus on the specific cases of $\mathcal{Z}$ that describe RBIMs, where the randomness is only in the sign of the Ising interaction. The implementation is described below.

We begin by considering the case where the measurement outcomes are all $+1$, meaning that the measurement outcomes all point along $\hat{n}(x,y)$, but $\hat{n}$ can now be position-dependent. We also choose $\hat{n}$ to lie in the {\blue $x$-$z$} plane (so $\phi=0$). From \cref{eq: ising_para}, we see now that the $K$ and $h$ parameters are given by

\begin{equation}
    \begin{aligned}
        &\exp(-2 h_k) = \tan\qty(\theta_\circ/2) &\quad k \in \circ\\
        &\exp(-2 K_i) =\frac{ 1 - \tan\qty(\theta_\bullet/2)}{1 + \tan\qty(\theta_\bullet/2)}&\quad i \in \bullet
    \end{aligned}
    \label{eq:forcedPlus}
\end{equation}

Given a value of the magnetic field $h_k$ at site $k$, we can instead set that term to be $-h_k$ by setting $\theta_\circ\to \pi-\theta_\circ$. This, in turn, ensures that the $\mathcal{Z}$ that describes $\ket{\psi}_{\partial\mathcal{M}}$ has the magnetic field at site $k$ flipped.

Similarly, to toggle only the sign of the interaction $K_i$ from $+$ to $-$, only the change $\theta_\bullet$ to $-\theta_\bullet $ needs to be effected. We then consider an evolution where the measurement outcomes are all chosen to be $+1$, but the angle is drawn randomly from $\qty{\theta_\bullet, -\theta_\bullet}$ with equal probability. These angles are independently drawn for each site on the $\bullet$ sublattice. This models interactions of the form 
\begin{equation}
    |K|\sum_{\expval{i,j}} {\rm sign}\qty(K_{ij}) s_i s_j
\end{equation}
in the Lieb lattice, and 

\begin{equation}
    |K|\sum_{i\in\bullet} {\rm sign}\qty(K_{i}) s_a s_b s_c s_d
\end{equation}
in the square lattice. As in \cref{eq: square_ising}, $s_a, s_b, s_c$ and $s_d$ refer to the 4 sites belonging to the $\circ$ sublattice that enclose site $i$. The sign of the coupling ${\rm sign}\qty(K) = \pm 1$ corresponds to $\hat{n} = \qty(\cos\theta) \hat{z} \pm \qty(\sin\theta)\hat{x}$. Measurements on the $\circ$ sublattice are performed along a fixed direction $\hat{n}$ and the outcomes are fixed to be +1, corresponding to a $\mathcal{Z}$ with a fixed magnetic field.

\begin{figure}
    \centering
    \includegraphics[width=0.48\textwidth]{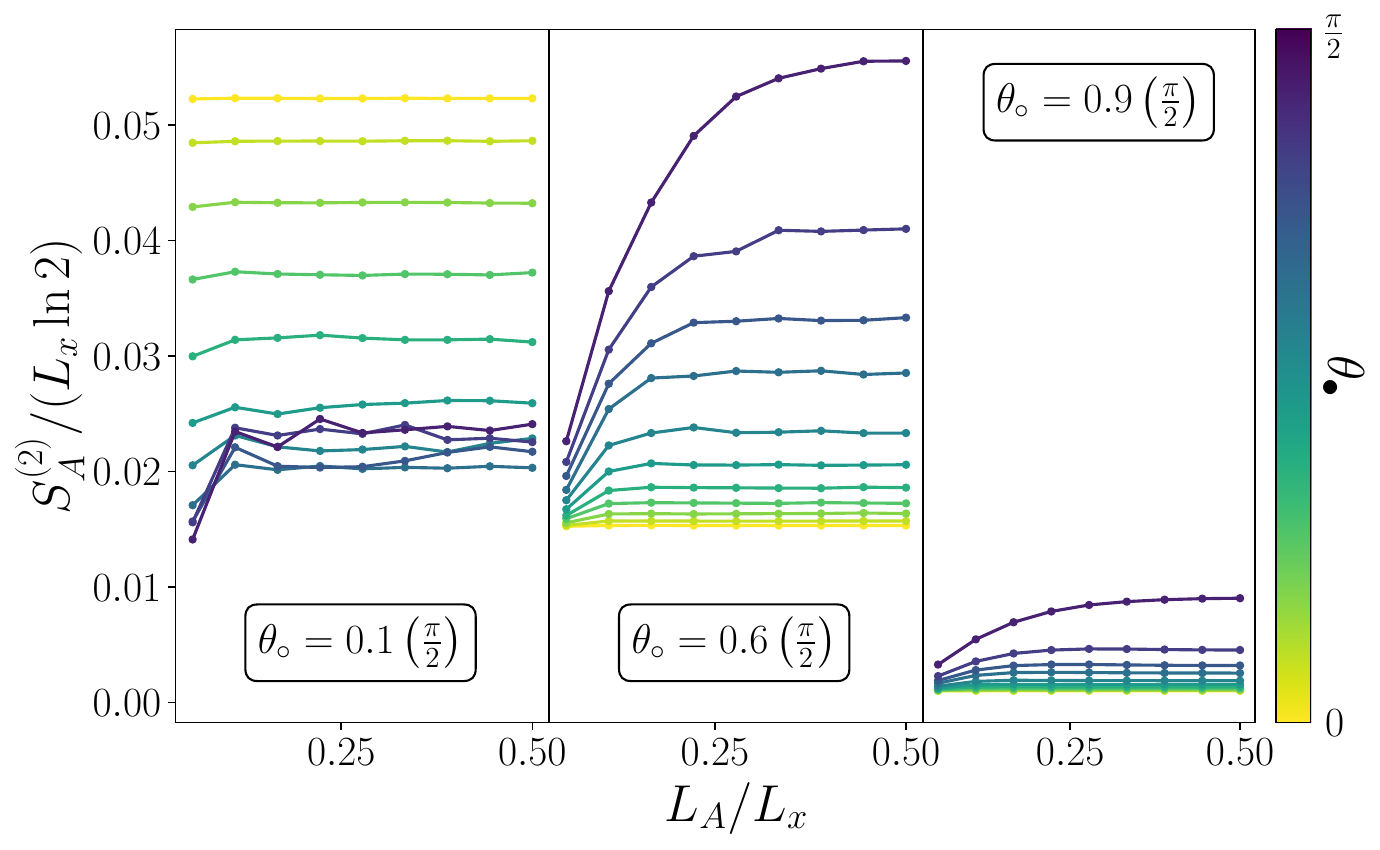}
    \caption{$S^{(2)}_A$ for $\ket{\psi}_{\partial\mathcal{M}}$ obtained after measurements on the square lattice. The set up is identical to that used for \cref{fig:rbim_ll}. The associated Ising partition function, with 4-spin random interactions on the square lattice, but still consisting only of positive weights, is also easy to approximate.}
    \label{fig:rbim_sql}
\end{figure}

In the Lieb lattice, we find that the state is \textit{always} in an area-law, as shown in \cref{fig:rbim_ll}, which suggests that the partition function of an RBIM with 2-body interactions should always be easily approximated, independent of $\theta$ (or $|K|$). Surprisingly, we find this to be the case in the square lattice as well. The plots presented in \cref{fig:rbim_sql} show that $\ket{\psi}_{\partial\mathcal{M}}$ is found to obey an area law, unlike in the case with randomized measurement outcomes. Using the results of our numerical simulations, we find certain regimes where $\mathcal{Z}$ can be efficiently approximated, for the specific interactions that we have considered in this work.

\begin{figure}
    \centering
    \includegraphics[width=0.48\textwidth]{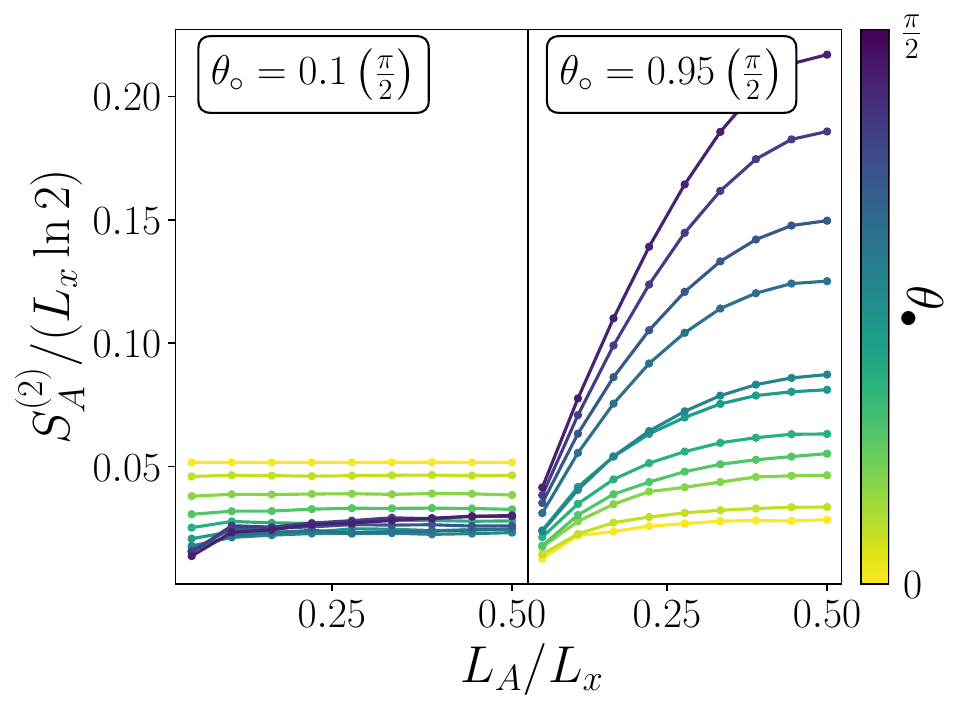}
    \caption{$S^{(2)}_A$ for $\ket{\psi}_{\partial\mathcal{M}}$ obtained after measurements on the square lattice. Unlike in \cref{fig:rbim_sql}, $\theta_\bullet$ is held fixed, but the measurement outcomes are now randomly chosen to be $\pm\hat{n}$, but only on the $\bullet$ sublattice. The boundary state entanglement now shows an area- to volume-law transition provided $\theta_{\circ/\bullet}$ are sufficiently high. The associated $\mathcal{Z}$ is now composed of both positive and negative weight terms. This suggests the necessity of negative weights in observing a transition in the complexity of approximating $\mathcal{Z}$.}
    \label{fig:rbim_sql_1l}
\end{figure}

When $\theta_\bullet$ is either $\pm\theta$, with the $+\hat{n}$ outcome post-selected and $0\leq\theta\leq\frac{\pi}{2}$, the corresponding weights in $\mathcal{Z}$ are always positive. Since the boundary state is in an area law regardless of the number of spins that participate in each interaction (2 in the Lieb lattice, 4 in the square lattice), we posit that positive weights with such interactions lead to a partition function whose approximation is always easy. 

With a fixed $\theta$, $0\leq\theta\leq\frac{\pi}{2}$, negative weights in $\mathcal{Z}$ correspond to a measurement outcome along $-\hat{n}$, since in this case, \cref{eq:forcedPlus} is modified to give the following relations

\begin{equation}
    \begin{aligned}
        &\exp(-2 h_k) = -\cot\qty(\theta_\circ/2) &\quad k \in \circ\\
        &\exp(-2 K_i) =-\frac{ 1 + \tan\qty(\theta_\bullet/2)}{1 - \tan\qty(\theta_\bullet/2)}&\quad i \in \bullet
    \end{aligned}
    \label{eq:forcedMinus}
\end{equation}

In the square lattice, we showed earlier that in the presence of randomness in the measurement outcomes, i.e. both $\pm\hat{n}$ outcomes are possible, a volume-to-area-law transition is observed. From \cref{eq:forcedMinus}, this corresponds to a $\mathcal{Z}$ with negative weights. Intriguingly, such a transition is observed, as shown in \cref{fig:rbim_sql_1l}, even when the measurement outcomes are identical on one sublattice ($\circ$, in this case), provided there is randomness in the measurement outcomes on the other sublattice ($\bullet$). This indicates that a finite density of negative weights can make the approximation of $\mathcal{Z}$ generically hard.

Lastly, measurements performed along $\hat{n}$ lying in the {\blue $x$-$z$} plane, on a graph state on the Lieb lattice, always result in an area law phase for the boundary state. Thus, approximating $\mathcal{Z}$ for a 2D Ising model with 2-body interactions only becomes hard in the presence of complex weights. This is the case even in the absence of randomness in the measurement outcomes, provided $\theta>\theta_c$.

In summary, partition functions describing these specific interactions and with positive weights appear easy to approximate. Negative weights in $\mathcal{Z}$ can lead to a transition in the complexity of its approximation with 4 body interactions, but evidently not with 2 body interactions. The presence of complex weights generically can cause a transition, provided the real parts of the interaction strengths $K$ (or equivalently, the magnitude of the weights) are above a certain value.

We conclude this subsection by noting that the results presented so far pertained specifically to a sampling algorithm obtained by treating a 2D quantum state sampling problem as an effective {\blue \((1+1)\mathrm{D}\) } nonunitary dynamics \cite{PhysRevX.12.021021}. However, these results do not forestall the existence of an algorithm that could nonetheless circumvent these difficulties; we only find that the boundary evolution algorithm becomes inefficient in the volume-law phase, by requiring exponentially large (in $L_x$) memory or time-resources to maintain a bounded truncation error, while in the area-law phase, a constant overhead is sufficient.

\subsection{Entanglement Scaling with Random Pauli Measurements}\label{sec: pauli_numerics}

This section investigates the boundary entanglement structure induced by bulk Pauli measurements on a 2D cluster state. We first review the dynamical method introduced in our previous work \cite{PhysRevB.106.144311}. 

For a cluster state defined on a 2D lattice, the measurement in one row will only affect its neighboring rows. Consequently, in the process of generating the boundary state at the top row, we can exchange the orders of CZ gates and single qubit measurements. Drawing inspiration from this insight, we design the algorithm as follows: We first prepare a three-layer cluster state, perform Pauli measurement in the middle layer, and remove the measured qubits. Next, we use the CZ gates to entangle the remaining two-layer state with an additional third layer of cluster state. Iterating this process and, at the final time, measuring all the non-boundary qubits, we thus obtain the boundary state $\ket{\Psi}_{\mathcal M}$ by avoiding the need for the construction of the 2D cluster state.

 We observe that as we randomly measure the qubits along the $X/Y/Z$ directions, by tuning the measurement probabilities $p_x, p_y, p_z$, we encounter both volume-law-to-area-law and area-law-to-area-law phase transitions. The critical exponents we extract match those reported in previous studies \cite{PhysRevB.100.134306, skinner2019measurement}. The area-law-to-area-law phase transition aligns with earlier findings \cite{PRXQuantum.2.030313}, further confirming the free fermion nature as discussed in Section~\ref{sec: exact_solvable}. Notably, for random bulk measurements in the $Y/Z$ directions, the boundary critical exponents also match those found in prior work as shown in Table.~\ref{tab: exp}, suggesting that such phase transition and the random Clifford measurement-induced phase transition (MIPT) belongs to the same universality class.

\begin{table}[ht]
    \begin{tabular}{|l|*{6}{p{0.8cm}|}}
    \hline
    Model       & $\nu$  & $c$    & $\Delta$ \\ \hline
    Lieb        & $1.3$  & $3.05$ & $2$      \\ \hline
    Square      &  $1.3$ & $1.6$  & $2$      \\ \hline
    Cliff. MIPT & $1.3$ & $1.6$  & $2$      \\ \hline
    \end{tabular}
    \caption{\label{tab: exp} Critical exponents of random Y/Z measurement on the Lieb-lattice/square-lattice boundary and Clifford MIPT \cite{PhysRevB.100.134306, skinner2019measurement}.} 
\end{table}

\subsubsection{Volume-law-area-law Transition}

\begin{figure}[ht]
  \centering
  \includegraphics[width=0.4\textwidth]{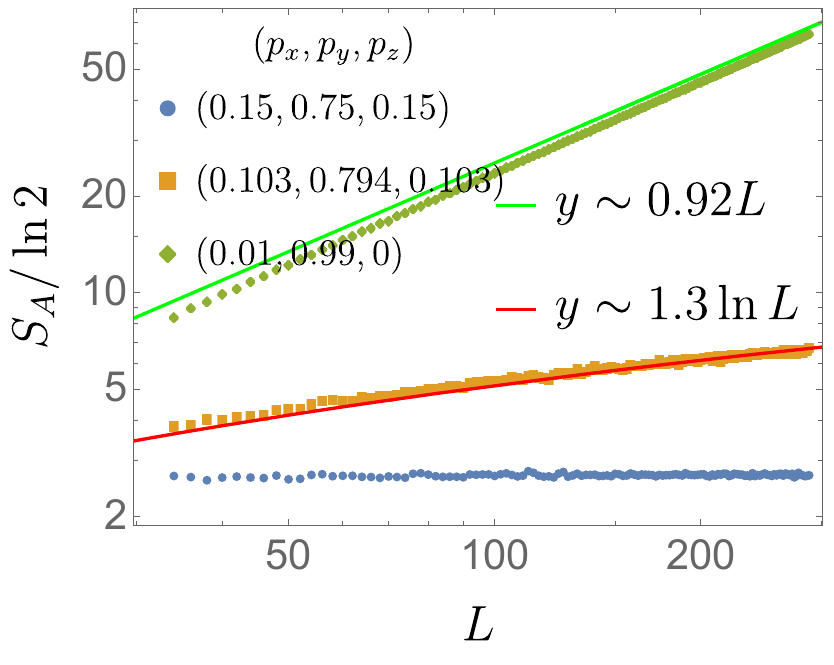}
  \caption{Entanglement entropy $S_A$ of the fixed ratio $L_A/L=1/2$ vs. boundary length $L$ for the boundary state of the Lieb graph state.} \label{fig: vol_area_EE}
\end{figure}

We consider the cluster state defined on the Lieb lattice and perform randomized Pauli $X$, $Y$, and $Z$ measurements on the bulk qubits with probabilities $p_x$, $p_y$, and $p_z$, respectively, restricted by 
\begin{equation}\label{eq: p_constraint}
    p_x + p_y + p_z = 1.
\end{equation}
 For simplicity we only consider the "rough" boundary on the bottom discussed in Sec.~\ref{sec: dynamics}. We first investigate the scaling of the n-R\'enyi entropy 
\begin{equation}
   S_A^n = \frac{1}{1-n}\ln \tr \rho_A^n
\end{equation} on the boundary. Here $\rho_A = \tr_{\overline{A}}\rho$ is the partial trace of the density matrix of the boundary state 
$\rho = \ket{\Psi}_{\partial \mathcal M}\bra{\Psi}_{\partial \mathcal M}$. As a consequence, the post-measurement boundary state is still a stabilizer state, whose entanglement entropy is independent of R\`enyi index $n$. We thus drop the index $n$ in the following parts of this section. Without the $Y$ measurement, the boundary $1$d state is area-law entangled with $S_A \sim \text{const.}$. However, as shown in  Fig.~\ref{fig: vol_area_EE}, when the $Y$ measurement dominates, the entanglement entropy satisfies volume-law scaling, $S_A \sim L$ with $L_A = 1/2L$. By adjusting the measurement rate $p_y$, we induce an entanglement transition from the volume-law phase to the area-law phase. This transition can be mapped to MIPT by the transfer matrix formalism presented in Sec.~\ref{sec: transfer_matrix}, where we demonstrate that $X$ or $Z$ measurements correspond to one or two-qubit projective measurements. In contrast, the $Y$ measurements correspond to entangling unitary gates. The interplay between the entangling $Y$ measurement and the disentangling $X/Z$ measurement drives the entanglement phase transition. In the following part of this section, we take $L$ as the number of qubits on the boundary $\partial \mathcal M$, and $L_A$ as the number of qubits in subsystem $A$.

\begin{figure*}[ht]
\centering
\subfloat[Geometry of subsystem $A,B$, $L_A = L_B = L/ 8$]{\label{fig:mi_setup}\includegraphics[width=0.3\textwidth]{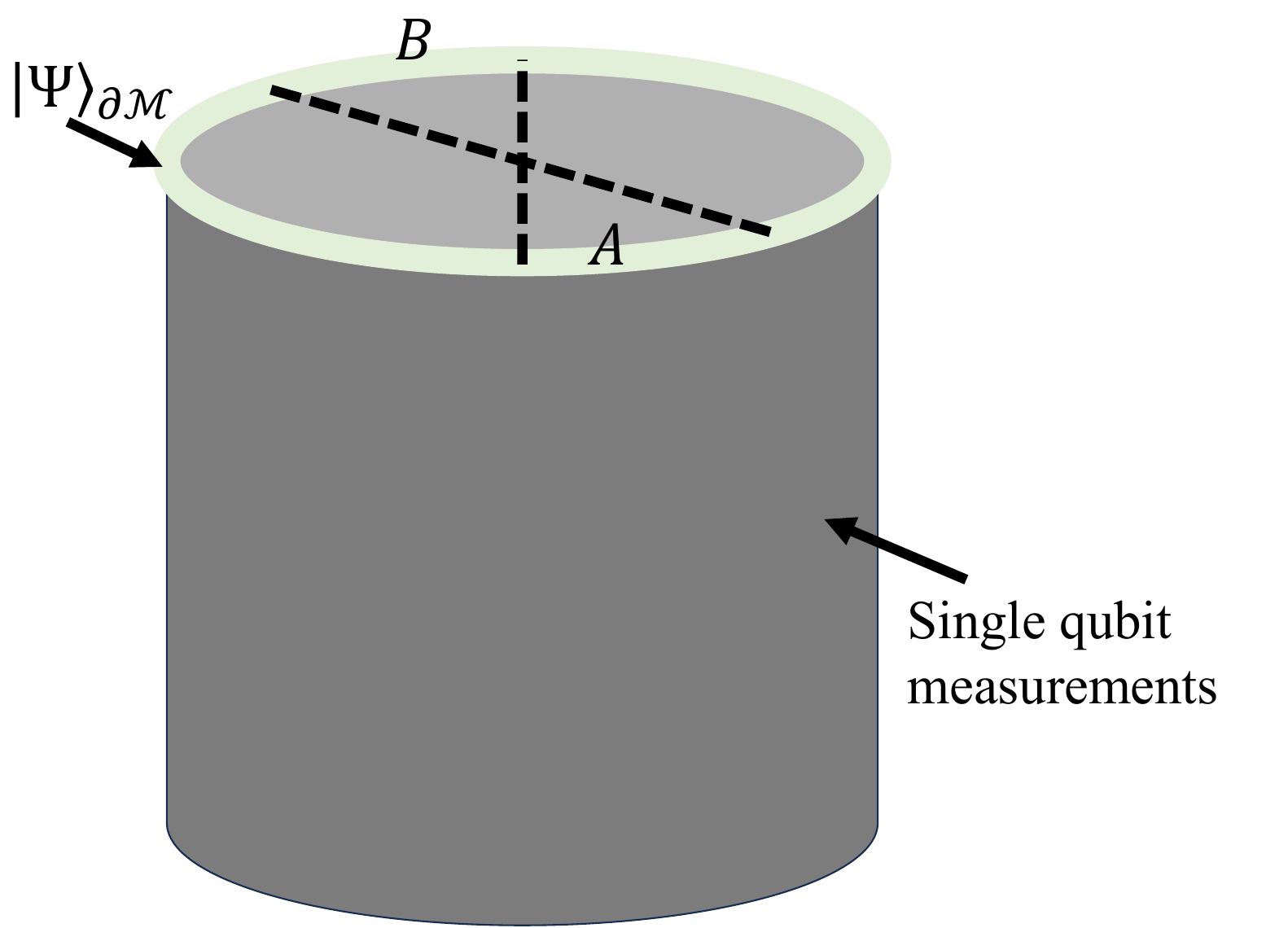}}\qquad
\subfloat[Mutual information data collapse for $p_z = 0$, where $p_y^c = 0.58$, $\nu = 1.78$]{\label{fig: lieb_mi_data_col_xy}\includegraphics[width=0.3\textwidth]{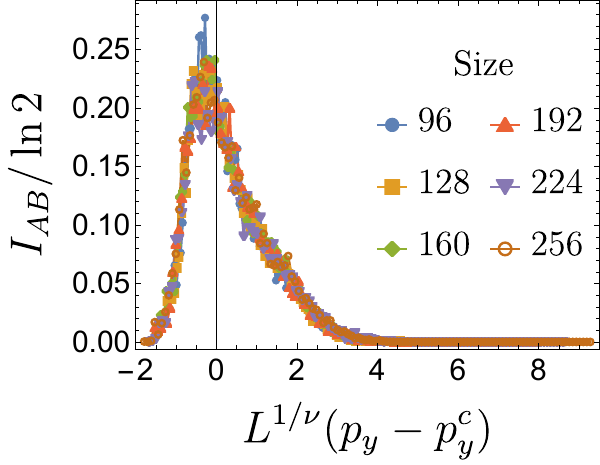}}\qquad
\subfloat[Mutual information data collapse for $p_x = 0$, where $p_y^c = 0.858$, $\nu = 1.3$  ]{\label{fig: lieb_mi_data_col_yz}\includegraphics[width=0.3\textwidth]{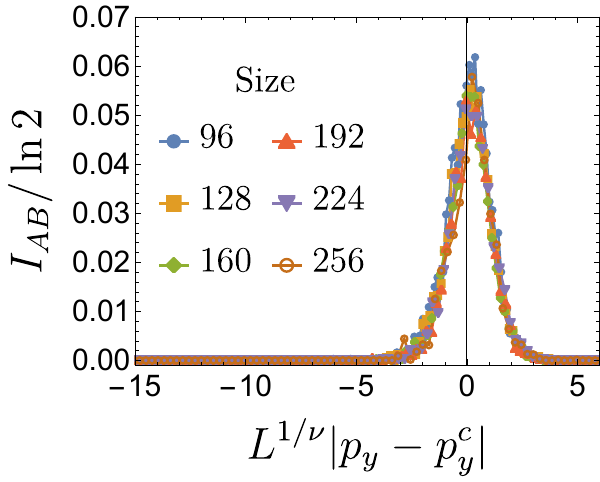}}
\caption{\label{fig: lieb_mi_data_col} Mutual Information data collapse $I_{AB} = f(L^{1/\nu}(p_y - p_y^c))$ for the critical points of the boundary state defined on the Lieb graph state.}
\end{figure*}

\begin{figure}[ht]
  \centering
  \includegraphics[width=0.4\textwidth]{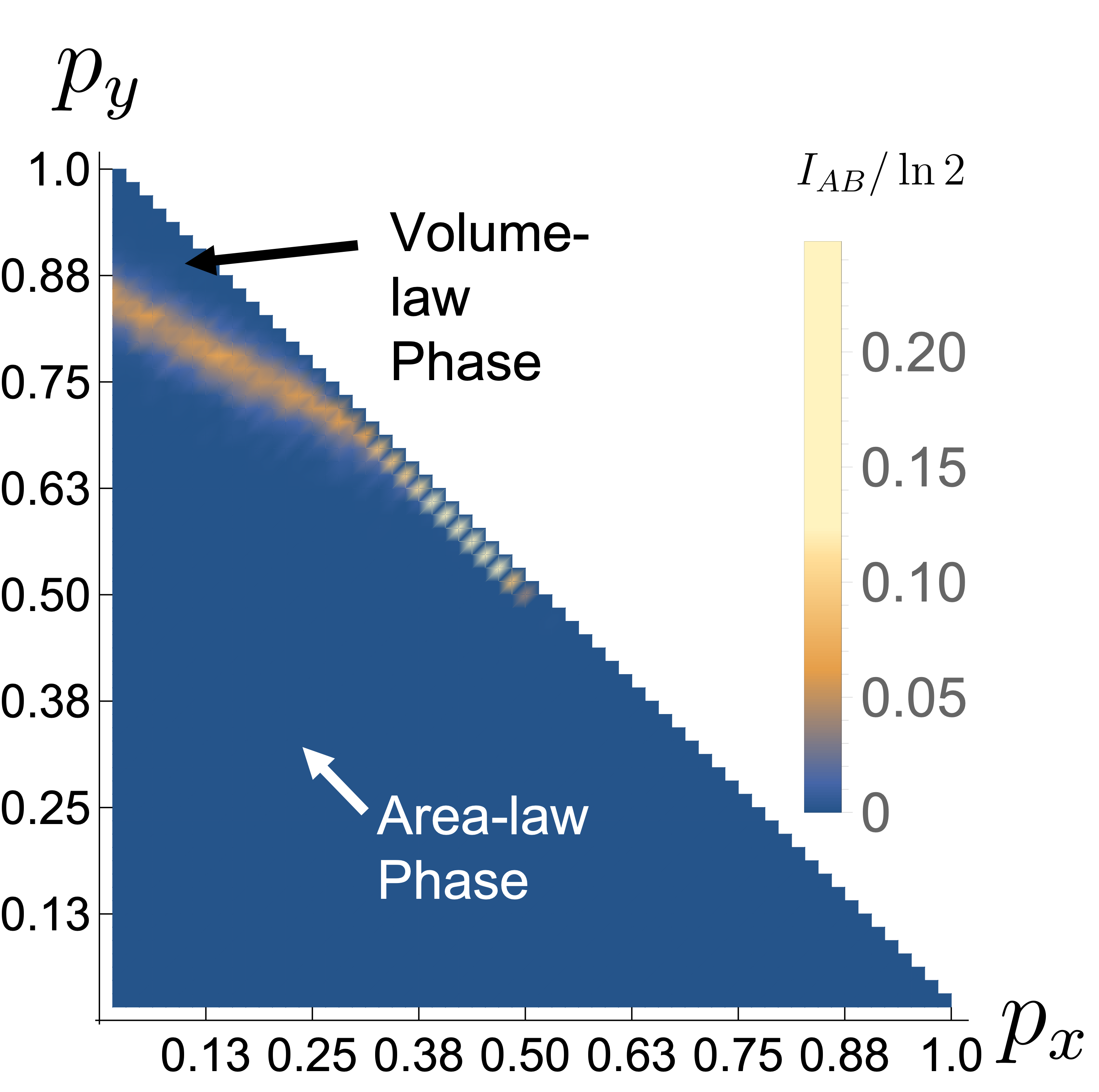}
  \caption{ (Lieb lattice model) Mutual information $I_{AB}$ of antipodal domains $A$ and $B$, with $L_A = L_B = L/8$. The yellow line (mutual information peak) marks the phase boundary.} 
  \label{fig: lieb_mi_phase_diag}
\end{figure}

The position of the volume-law to area-law entanglement phase transition can be determined by identifying the peak of the mutual information~\cite{PhysRevB.100.134306}
\begin{equation}
    I_{AB} = S_A + S_B - S_{AB}
\end{equation}
of two antipodal regions $A$ and $B$ on the boundary $\partial \mathcal M$ with $L_A = L_B = L / 8$ as shown in Fig.~\ref{fig: lieb_mi_data_col}. Employing this method, we construct the phase diagram presented in Fig.~\ref{fig: lieb_mi_phase_diag}.

We further explore the entanglement phase transition of the two specific cases: (1) when $p_z = 0$, allowing only $X$ and $Y$ measurements, and (2) when $p_x = 0$, with only random $Z$ and $Y$ measurements.

\begin{figure*}[ht]
\centering
\subfloat[Geometry of subsystem $A$]{\label{fig:ee_setup}\includegraphics[width=0.3\textwidth]{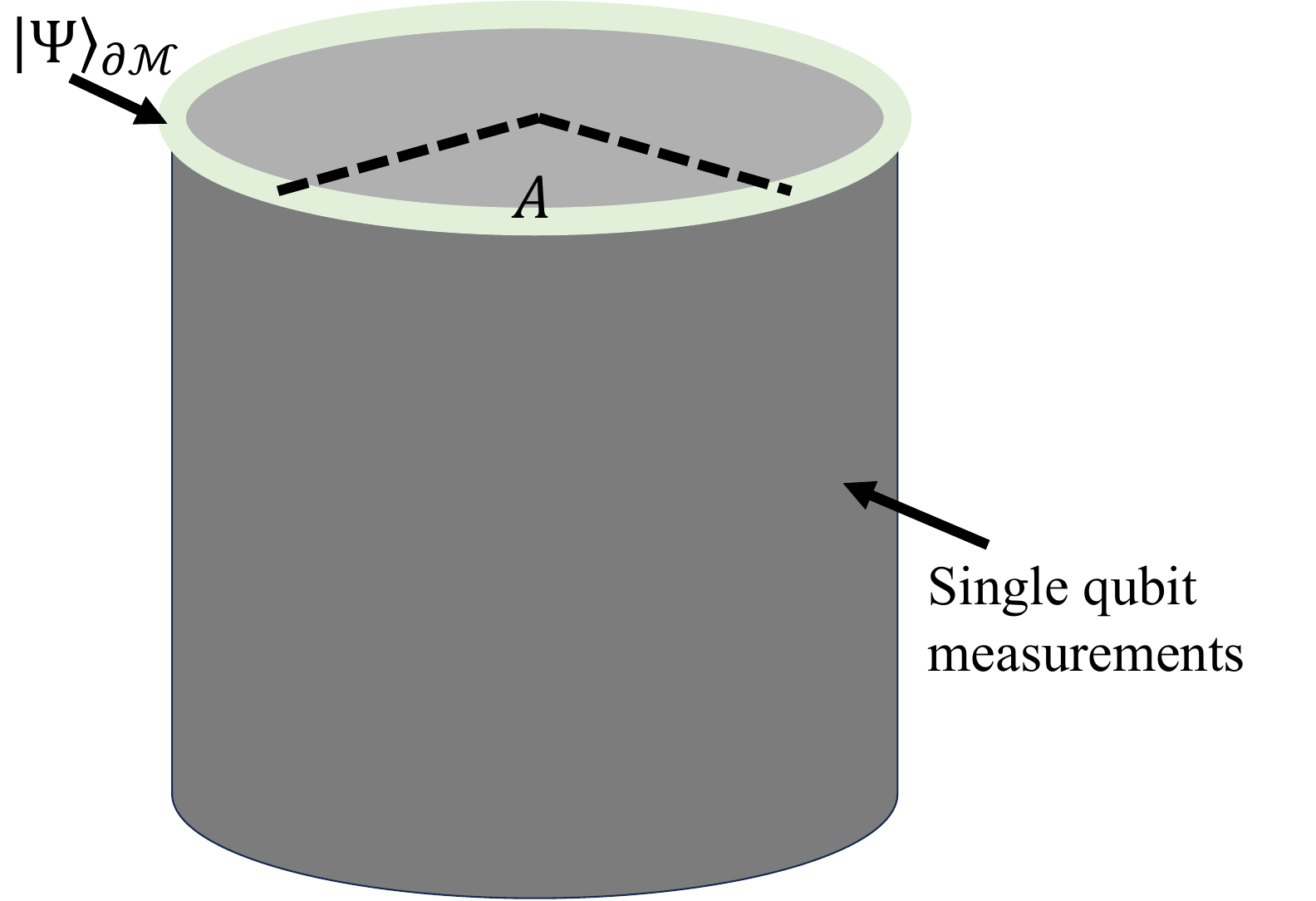}}\qquad
\subfloat[$p_z = 0$, $p_y = p_y^c = 0.58$, $p_x = 0.42$]{\label{fig: ee_critical_scaling_xy}\includegraphics[width=0.3\textwidth]{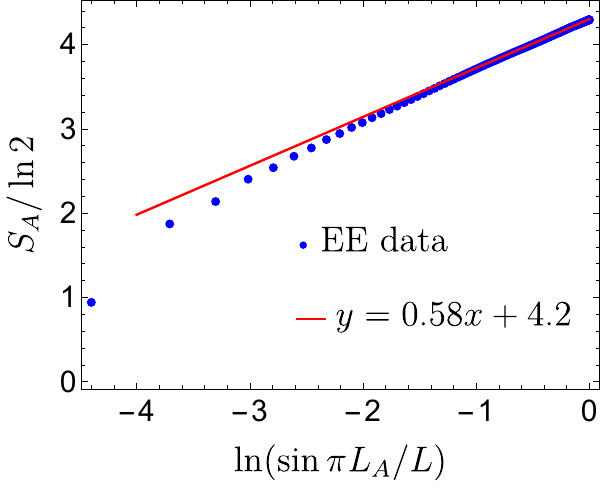}}\qquad
\subfloat[$p_x = 0$, $p_y = p_y^c = 0.858$, $p_z = 0.142$  ]{\label{fig: ee_critical_scaling_yz}\includegraphics[width=0.3\textwidth]{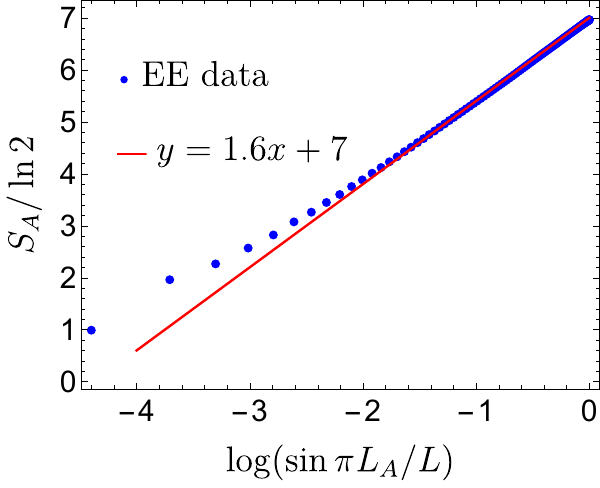}}
\caption{\label{fig: ee_critical_scaling} The entanglement entropy scaling of Lieb lattice model at the critical points, with $L = 256$
}
\end{figure*}

\begin{figure*}[ht]
\centering
\subfloat[Geometry of subsystem randomly chosen disjoint $A$, $B$]{\label{fig:mi_x_setup}\includegraphics[width=0.3\textwidth]{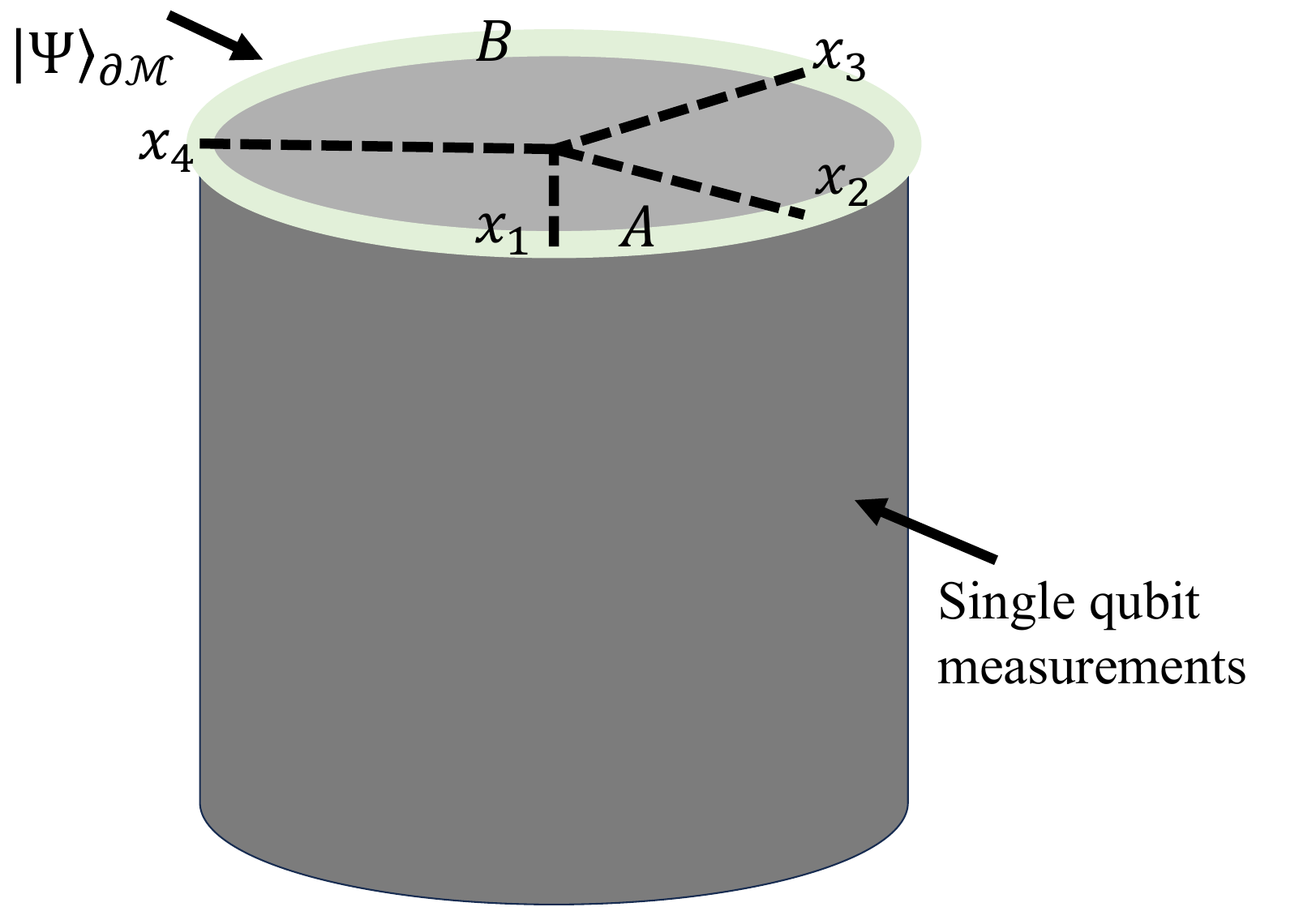}}\qquad
\subfloat[$p_z = 0$, $p_y = p_y^c = 0.58$, $p_x = 0.42$]{\label{fig: lieb_mi_critical_xy}\includegraphics[width=0.3\textwidth]{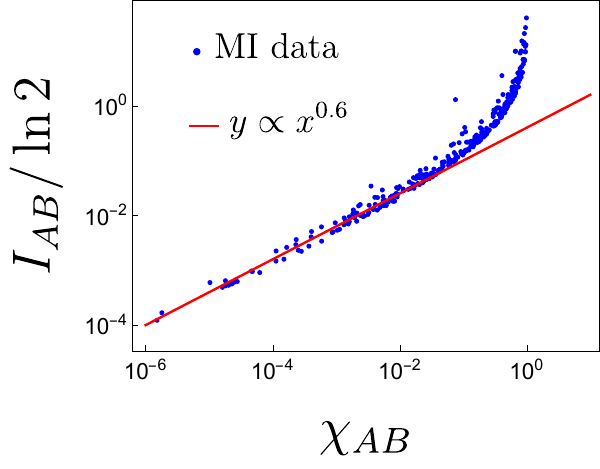}}\qquad
\subfloat[$p_x = 0$, $p_y = p_y^c = 0.858$, $p_z = 0.142$  ]{\label{fig: lieb_mi_critical_yz}\includegraphics[width=0.3\textwidth]{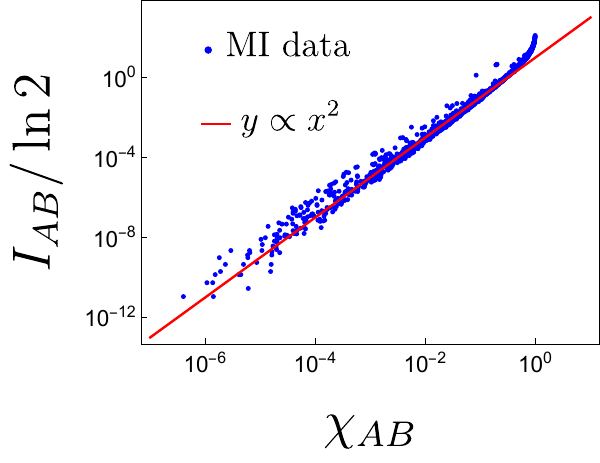}}
\caption{\label{fig: lieb_mi_critical} Critical mutual information (MI) scaling of Lieb lattice model, $I_{AB }\sim \chi_{AB}^{\Delta}$}
\end{figure*}

\begin{figure*}[ht]
\centering
\subfloat[$p_x = p_z = 0.4$, $p_y = 0.2$]
{\label{fig: field_free_ee_py02}
\includegraphics[width=0.3\textwidth]{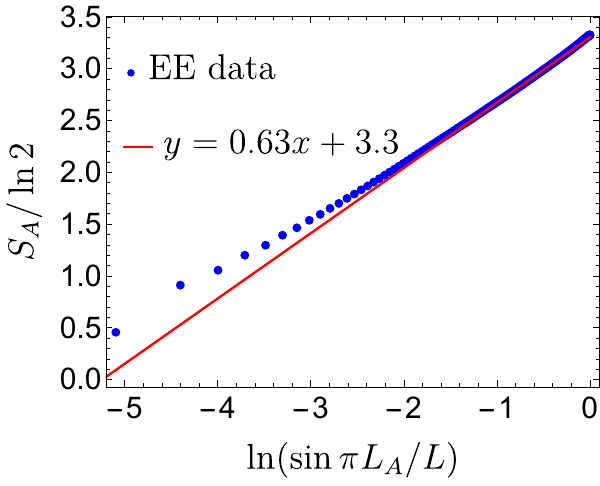}}
\qquad
\subfloat[ $p_x = p_z = 0.2$, $p_y = 0.6$]
{\label{fig: field_free_ee_py06}
\includegraphics[width=0.3\textwidth]{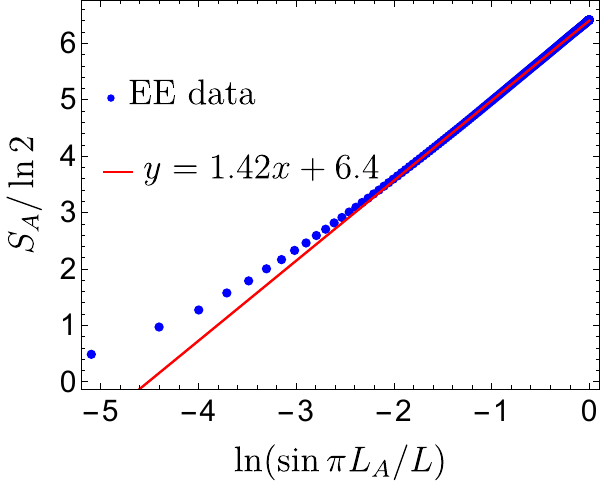}}\qquad
\subfloat[ $p_x = p_z = 0.1$, $p_y = 0.8$]{\label{fig: field_free_ee_py08}\includegraphics[width=0.3\textwidth]{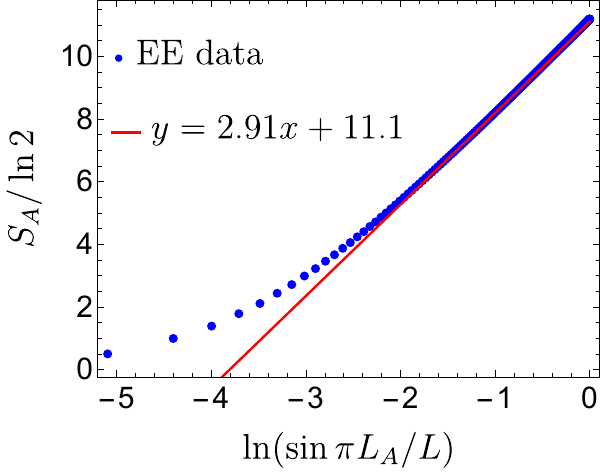}}
\caption{\label{fig: field_free_ee_goldstone} Logarithmic scaling of entanglement entropy in the critical Goldstone phase with system size $L = 512$.}
\end{figure*}

\begin{figure*}[ht]
\centering
\subfloat[Critical entanglement scaling, $p_x = p_z = 0.5$, $p_y = 0$]
{\label{fig: field_free_ee_scale}
\includegraphics[width=0.3\textwidth]{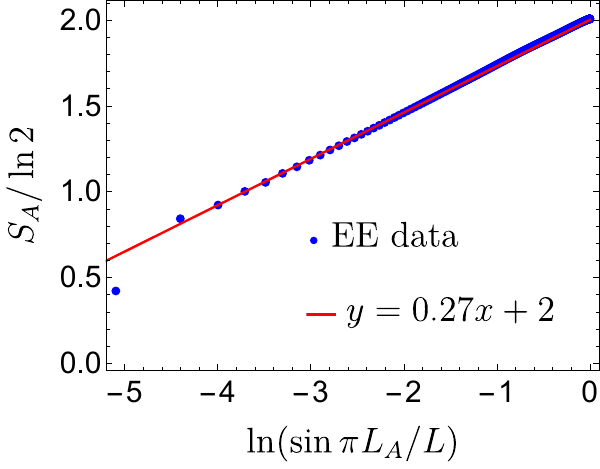}}
\qquad
\subfloat[Critical mutual information scaling, $p_x = p_z = 0.5$, $p_y = 0$]
{\label{fig: field_free_mi_scale_xz}
\includegraphics[width=0.3\textwidth]{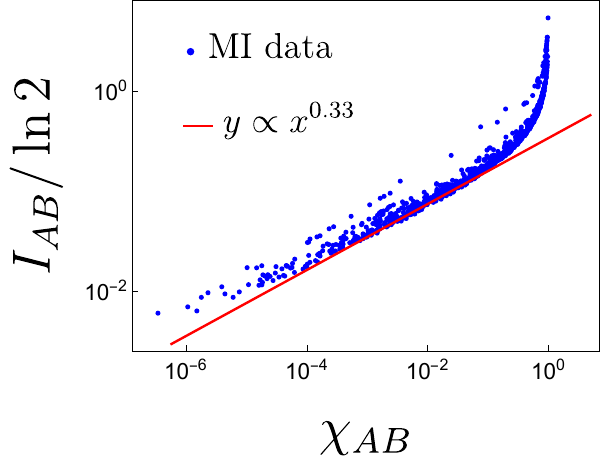}}\qquad
\subfloat[Mutual information data collapse at $p_y =0$, $p_x^c = 0.5$, $\nu = 3/4$]{\label{fig: mic3}\includegraphics[width=0.3\textwidth]{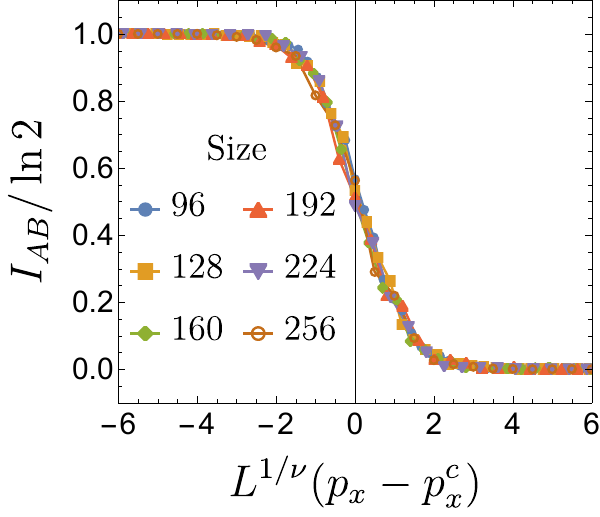}}
\caption{Criticality of the area-law phase to the area-law phase transition at $p_y = 0$, $p_z = p_x = 0.5$.}
\end{figure*}

 For $p_z = 0$, where only $X$ and $Y$ Pauli measurements exist, the criticality is observed at $p_y = 0.58$ and $p_x = 0.42$. The mutual information of two antipodal regions peaks at the critical point and collapses to
\begin{equation}
\begin{aligned}
    I_{AB} \sim f\big(L^{1/\nu }&(p_y - p_y^c)\big)\\
    p_y^c = 0.58,&\quad\nu = 1.78 .
\end{aligned}
\end{equation}
as shown in Fig.~\ref{fig: lieb_mi_data_col_xy}. At criticality, the entanglement entropy scales logarithmically with subsystem size $L_A$
\begin{equation}
    S_A/\ln2 \sim 0.58 \ln \sin (\pi \frac{L_A}{L})
\end{equation}
which is shown in Fig.~\ref{fig: ee_critical_scaling_xy}. At criticality, the mutual information $I_{AB}$ of two disjoint regions $A$ and $B$ is a function of the cross-ratio $\chi_{AB}$. In particular, when $\chi_{AB}$ is close to zero, as shown in Fig.~\ref{fig: lieb_mi_critical_xy}, we have
\begin{equation}
    I_{AB} \sim \chi_{AB}^{0.6}
\end{equation}
with
\begin{equation}
    \chi_{AB} = \frac{x_{12}x_{34}}{x_{13}x_{24}},\quad \mbox{with}~x_{ij} = \frac{L_x}{\pi}\sin\left(\frac{\pi}{L_x}|x_i - x_j |\right),
\end{equation}
where $x_{i = 1,2,3,4}$ are endpoints of subregions $A$ and $B$ as shown in Fig.~\ref{fig:mi_x_setup}. 

Taking $p_x = 0$, the critical point occurs at $p_y = 0.858$ and $p_z = 0.142$. Fig.\ref{fig: ee_critical_scaling_yz}  shows that $S_A/\ln2 \sim 1.6\ln (\sin(\pi L_A / L))$. The mutual information $I_{AB}$ data collapse around the critical point is shown in Fig.~\ref{fig: lieb_mi_data_col_yz}, where we have $\nu = 1.3$. Furthermore, at criticality, Fig.~\ref{fig: lieb_mi_critical_yz} presents the scaling of mutual information between two disjoint intervals $A$ and $B$, characterized by $I_{AB} \sim \chi_{AB}^{2}$ with small $\chi$.

\subsubsection{Absence of the volume-law phase}

As demonstrated in \cref{eq: tc2}, measuring all the vertex qubits along the $X$-direction on the Lieb lattice allows us to generate the toric code state up to some Pauli rotations. We further perform randomized Pauli $X$, $Y$, and $Z$ measurements in the bulk edge qubits of the Lieb lattice cluster state with probabilities $p_x$, $p_y$, and $p_z$, respectively. Using the transfer matrix method detailed in Sec. \ref{sec: exact_solvable}, we can show that this dynamics is equivalent to a free fermion system subject to repeated projective measurements as reported in \cite{PRXQuantum.2.030313}. Similar physics has recently been discussed in Ref.~\cite{negari2023measurementinduced}. The $Y$ measurement generates a unitary free fermion gate, effectively braiding a pair of Majorana fermions, while the $X$ and $Z$ measurements correspond to $X$ and $ZZ$ measurements in the effective {\blue \((1+1)\mathrm{D}\) } dynamics, respectively. This non-unitary free fermion dynamics can be mapped to a loop model, which has been extensively studied in Ref.~\onlinecite{Nahum_2013}. 

When the $Y$ measurement dominates, a critical phase emerges. Its entanglement entropy scales like this
\begin{equation}
\begin{aligned}
     S_A /\ln 2\sim \alpha\ln\sin\left(\frac{\pi L_A}{L}\right) + \\
      \mathcal{O}\bigg(\ln^2\sin\left(\frac{\pi L_A}{L}\right)\bigg),
\end{aligned}
\end{equation}
with $\alpha$ being some non-universal constant varying with measurement rate $p_x, p_y, p_z$. Some entanglement entropy scaling data with $p_x = p_z = (1 - p_y)$ are shown in Fig.~\ref{fig: field_free_ee_goldstone}. 

For nonzero $p_y$, varying $p_x$ or $p_z$ can induce transitions between the area-law and critical phases. In the limit $p_y=0$, the intermediate critical phase disappears. Consequently, varying $p_x$ results in a direct area-law to area-law transition, with the critical point described by the 2D percolation conformal field theory\cite{cardy2001conformal}. This result is confirmed by us numerically. At criticality, as illustrated in Fig.~\ref{fig: field_free_ee_scale}, the entanglement entropy scales as:
\begin{equation}
S_A /\ln 2 \sim \frac{K}{2} \ln\sin\left(\frac{\pi L_A}{L}\right),
\end{equation}
where $K = 0.54\approx \sqrt{3}/\pi$, which agrees with the theoretical prediction \cite{Nahum_2013, cardy2001conformal}. The critical scaling of the mutual information is given by:
\begin{equation}
I_{AB} \sim \chi_{AB}^{\Delta}, \quad \Delta = 0.33 \sim \frac{1}{3},
\end{equation}
as shown in Fig.~\ref{fig: field_free_mi_scale_xz}. Furthermore, we examine the data collapse of the mutual information for two antipodal intervals, $A$ and $B$, with $L_A = L_B = L / 4$. The mutual information collapses to
\begin{equation}
    I_{AB} \sim g(L^{1/\nu}(p_x - p_x^c))
\end{equation}
with $\nu = 3/4$ and $p_x^c = 0.5$, which can be identified as a bond percolation on a square lattice, which agrees with the result shown in Ref.~\onlinecite{PRXQuantum.2.030313}.

\begin{figure}[ht]
  \centering
  \includegraphics[width=0.4\textwidth]{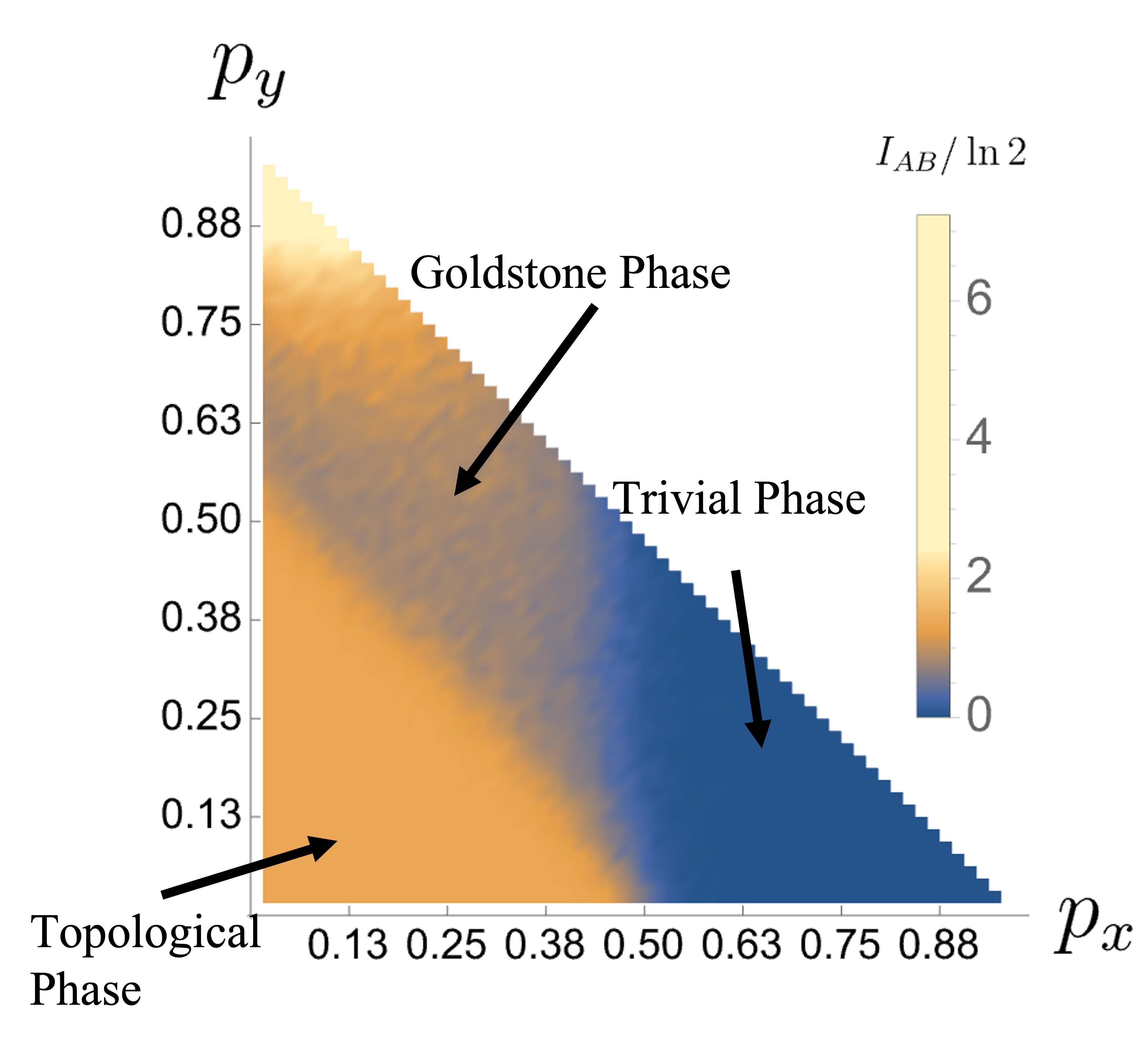}
  \caption{(Lieb lattice with X measurement on vertices) Mutual information $I_{AB}$ of antipodal domains $A$ and $B$, with $L_A = L_B = L/4$. Phases are marked by different colors, as $I_{AB}$ behaves differently in different phases. } \label{fig: field_free_phase_diagram}
\end{figure}

The two area-law phases can be further characterized by their difference in the mutual information $I_{AB}$. Notably, we consider $L_A = L_B = L/4$ and the domain distance $r_{AB} = L / 2$. The phase diagram is presented in Fig.~\ref{fig: field_free_phase_diagram}. In one phase, we have $I_{AB} = 0$. Conversely, in the other phase, the system displays long-range entanglement characterized by $I_{AB} = \ln 2$. This arises from the logical operator  $O_L = \prod_{i\in l_n} X_i$ of the parent toric code state, where $l_n$ denotes the non-contractable loop of the cylinder. When the bulk $Z$ measurement on the edge qubits dominates, this logical operator is pushed onto the boundary and contributes to the $I_{AB} = \ln 2$ mutual information of the boundary state. For the critical phase, one observes non-constant mutual information. No volume-law phase is observed in this model, even with the effective unitaries induced by the bulk $Y$ measurement. The absence of the volume-law phase agrees with statement that the evaluation of the Ising partition function defined on a  planar graph, which in our case is Eq.\eqref{eq: lieb_ising} with $h=0$, only requires polynomial resource and can be efficiently evaluated on the classical computer \cite{FBarahona_1982}.

\subsubsection{Entanglement Transition in Square Lattice Cluster State}

\begin{figure}[ht]
  \centering
  \includegraphics[width=0.4\textwidth]{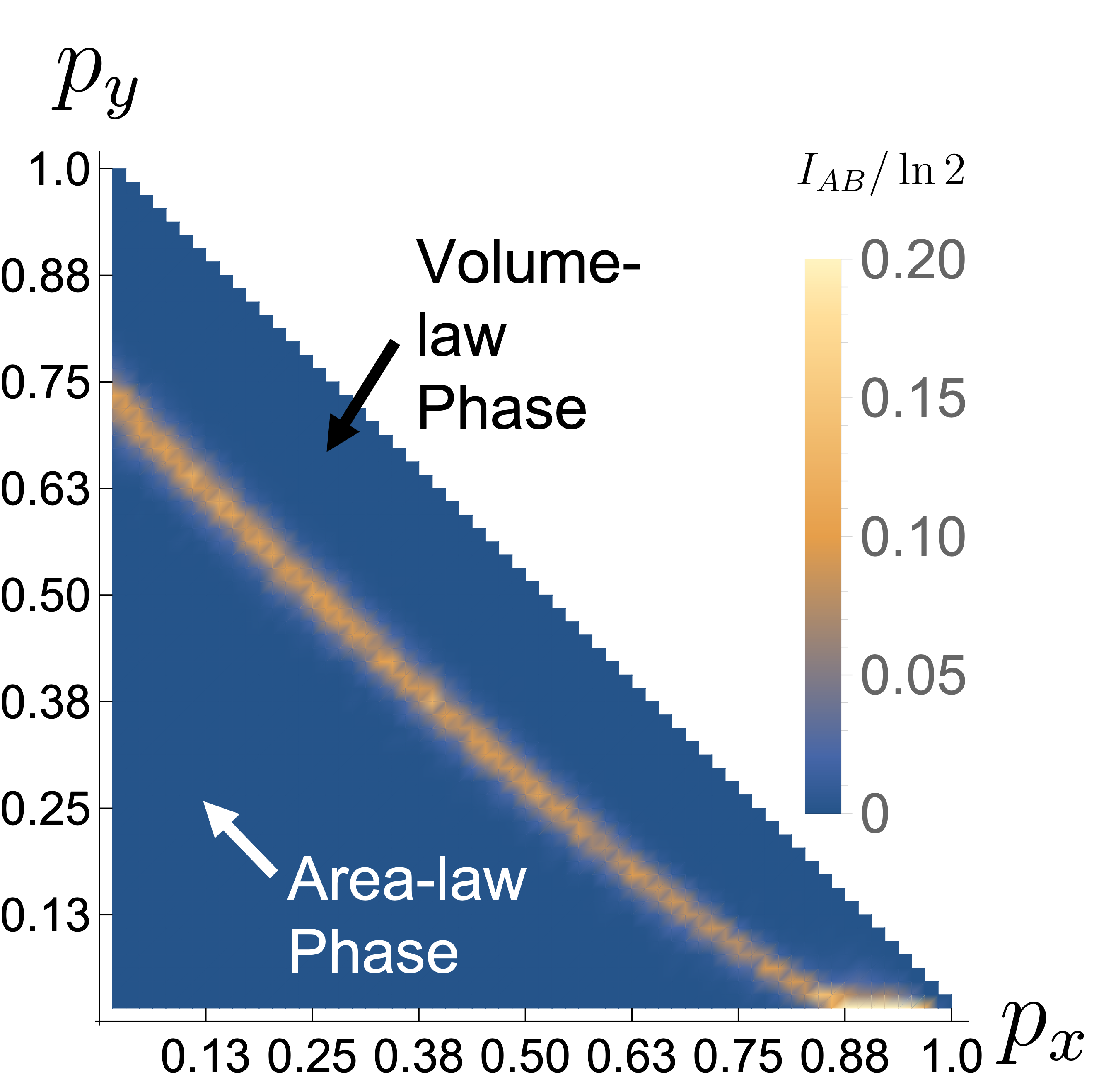}
  \caption{ (Square lattice model) Mutual information $I_{AB}$ of antipodal domains $A$ and $B$, with $L_A = L_B = L/8$. The yellow line (mutual information peak) marks the phase boundary.} \label{fig: sq_mi_phase_diag}.
\end{figure}

We now investigate the boundary state entanglement structure of the square lattice cluster state with random Pauli measurements in the bulk. We take the smooth boundary condition for the Square Lattice cluster state as shown in Fig.~\ref{fig: sq_bounadry_conven}. Consider random $X/Y/Z$ measurement with the measurement rate satisfying $p_x + p_y + p_z = 1$. The phase diagram is shown in Fig.~\ref{fig: sq_mi_phase_diag}. When $X/Y$ measurement dominates, the boundary state is in the volume-law phase. When $Z$ measurement dominates, the boundary state is in the area-law phase.

\begin{figure*}[ht]
\centering
\subfloat[Critical entanglement scaling, $p_y = 0.743$, $p_x = 0$]
{\label{fig: sq_ee_scale}
\includegraphics[width=0.3\textwidth]{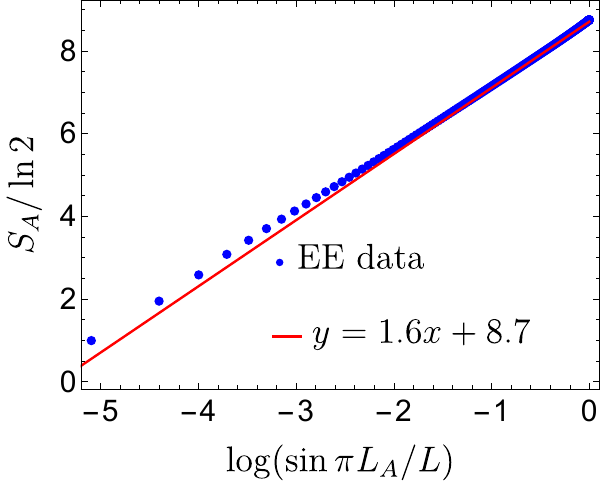}}
\qquad
\subfloat[Critical mutual information scaling, $p_y = 0.743$, $p_x = 0$]
{\label{fig: sq_mi_scale}
\includegraphics[width=0.3\textwidth]{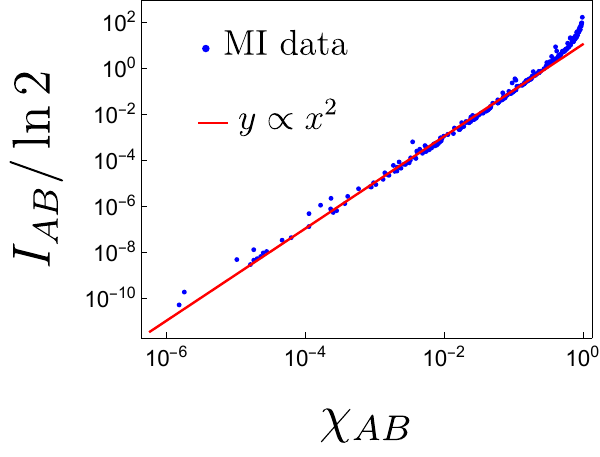}}\qquad
\subfloat[Mutual information data collapse at $p_y^c = 0.743$, $p_x = 0$, $\nu = 1.3$]{\label{fig: sq_mi_data_col}\includegraphics[width=0.3\textwidth]{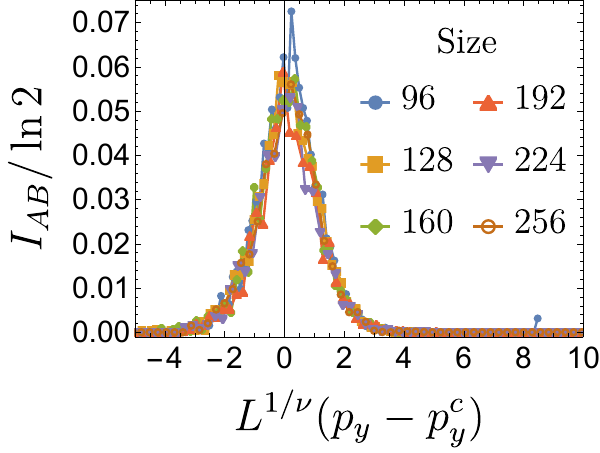}}

\caption{Volume-law to area-law phase transition on the square lattice.}

\end{figure*}

The volume-law to the area-law phase transition along $p_y =0$ is studied in our previous work Ref.~\cite{PhysRevB.106.144311}. Here, we investigate the boundary state entanglement transition along the line with $p_x =0$. The mutual information collapses to 
\begin{equation}
    I_{AB} \sim h(L^{1/\nu}(p_y - p_y^c)), ~\nu = 1.3,
\end{equation}
with data collapse shown in Fig.~\ref{fig: sq_mi_data_col}.
At the critical point $p_y = 0.743$, the entanglement entropy, as shown in Fig.~\ref{fig: sq_ee_scale} is
\begin{equation}
    S_A/\ln2 \sim 1.6 \ln \sin( \frac{\pi L_A}{L})
\end{equation}
and the mutual information is a function of the cross ratio and scales like 
\begin{equation}
    I_{AB} \sim \chi_{AB}^2
\end{equation}
as shown in Fig.~\ref{fig: sq_mi_scale}. 

We argue that the criticality observed along the $p_x = 0$ line of the square lattice model and the Lieb lattice model falls within the same universality class as the volume-law to area-law entanglement transition in the {\blue \((1+1)\mathrm{D}\) } hybrid random Clifford circuit presented in \cite{PhysRevB.100.134306}. Several reasons support this claim:

First, they exhibit the same critical exponents for entanglement entropy scaling, as illustrated in Figs.~\ref{fig: ee_critical_scaling_yz} and ~\ref{fig: sq_ee_scale}
\begin{equation*}
    S_A \sim 1.6 \ln \sin\left(\frac{\pi L_A}{L}\right).
\end{equation*}

\begin{figure}[ht]
  \centering
  \includegraphics[width=0.4\textwidth]{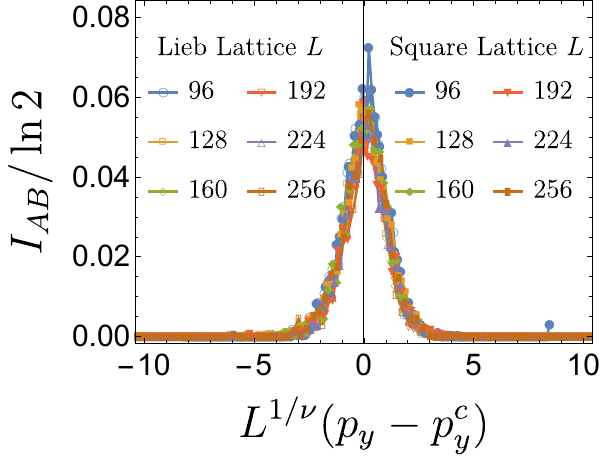}
  \caption{Mutual information data collapse for Lieb/Square lattice model along $p_x = 0$ with $\nu = 1.3$. Empty/Filled icons label the Lieb/Square lattice mutual information data.} \label{fig: mix_mi_col}
\end{figure}

Second, they both display the same critical mutual information scaling,
\begin{equation*}
    I_{AB} \sim \chi_{AB}^2,
\end{equation*}
as shown in Figs.~\ref{fig: lieb_mi_critical_yz} and \ref{fig: sq_mi_scale}. Additionally, their mutual information collapses to the same universal function with the same exponents $\nu = 1.3$, as depicted in Fig.~\ref{fig: mix_mi_col}. These critical exponents agree with the ones extracted in \cite{PhysRevB.100.134306}.

In summary, our numerical observations demonstrate that cluster states defined on the Lieb lattice and square lattice exhibit entanglement phase transition on the 1D boundary induced by the bulk measurement. Such phase transition is closely related to the measurement-induced phase transition observed in the {\blue \((1+1)\mathrm{D}\) } hybrid circuits.\cite{PhysRevB.98.205136, PhysRevB.99.224307, PhysRevB.100.134306,PhysRevLett.125.030505, PhysRevB.101.104301, PhysRevX.10.041020, PhysRevB.101.104302, skinner2019measurement, PhysRevB.103.104306,google2023measurement, PhysRevLett.130.220404, lavasani2021measurement, lavasani2022monitored, PhysRevLett.127.235701, PhysRevResearch.3.023200, PhysRevX.11.011030, PRXQuantum.2.030313}

\section{Conclusions and Discussion}
\label{sec:conclusion}
In this work, we have focused on the entanglement structure of the 1D boundary state and its relevance to the computational complexity of the cluster state sampling problem and the Ising partition function evaluation. We have established a connection between the sampling problem and the 2D Ising partition function, elucidating how this information can be encoded within the 1D boundary state. We have also explored dynamic boundary state generation and its connection to {\blue \((1+1)\mathrm{D}\) } non-unitary evolution. Finally, we numerically studied the boundary entanglement transitions induced by bulk measurements, specifically on the cluster states generated on the Lieb and square lattices. This method allowed us to identify some of the regimes where the Ising partition can be efficiently evaluated.

In \cref{sec: prelim}, we established a crucial connection between sampling the cluster state and evaluating the classical Ising partition function. This connection was unveiled by demonstrating that the inner product between the single qubit product state and the 2D cluster state follows the identical form as the classical Ising partition function with complex variables (as indicated in \cref{eq: cluster_ising}).

In \cref{sec: boundary_states}, we illustrated that the 1D boundary state encodes all the information of the Ising partition function. A 1D wavefunction can be represented as a MPS, whose complexity is determined by its entanglement structure. This observation led us to a significant insight: the task of approximating the Ising partition function can be directly mapped to understanding the entanglement scaling of the 1D boundary state. When the boundary state exhibits area-law entanglement scaling, it indicates that the corresponding Ising partition can be efficiently evaluated using classical computational resources.

In \cref{sec: dynamics}, we provided two dynamical approaches to generate the 1D boundary state. In the first approach, based on the transfer matrix, we explicitly established the correspondence between the single qubit measurement on the Lieb lattice cluster state and the non-unitary gates involved in {\blue \((1+1)\mathrm{D}\) } non-unitary dynamics.  In the second approach, we employed a specific order of contractions of 2D tensor networks to obtain the boundary state, and demonstrated that such an order can be naturally interpreted as a {\blue \((1+1)\mathrm{D}\) } dynamics.

In \cref{sec: numerics}, we studied the boundary entanglement structure using two complementary numerical methods. First, we explicitly implemented the tensor network inspired dynamics to find that the boundary state undergoes an entanglement transition as the measurement direction $\hat{n}$ is varied. This transition is similar to the measurement-induced transition found in {\blue \((1+1)\mathrm{D}\) } non-unitary circuits. We found that the manifolds in which $\hat{n}$ ought to be varied to observe such a transition depended crucially on the geometry of the lattice. We related the presence of this transition, to the nature of the weights -- positive, real or complex -- in corresponding Ising-type partition functions and consequently, draw conclusions about the ease of their approximation. We also showed that randomness in the bulk measurement outcomes was crucial to the existence of volume-law phases in the boundary state. Secondly, we utilized the classical simulability of Pauli gates to study the behavior of the boundary states under random Pauli measurements. We found similar entanglement phase transitions here as well, supporting the generality of such transitions.

As emphasized throughout our work, our research offers a fresh perspective on computational complexity through boundary entanglement. Moreover, this perspective can provide insight into measurement-based quantum computing models (MBQCs), where the computation is done by controlling the measurement pattern on a resource state~\cite{PhysRevLett.86.5188}. A crucial question arises in the realm of MBQCs: Is the resource state universal? Our study provides insights by demonstrating that the Lieb/Square lattice cluster state can prepare a boundary state with stable volume-law entanglement. This observation aligns with the assertion that Lieb/Square lattice cluster states can serve as universal resource states, as previously indicated~\cite{PhysRevA.76.022304, PhysRevLett.122.090501, PhysRevLett.97.150504}. In contrast, non-universal states, such as the toric code state, fail to generate boundary states with volume-law entanglement. This highlights how boundary entanglement analysis can be a valuable tool for determining whether the resource states are universal. 

For further directions, we aim to extend this concept to other 2D resource states and explore whether similar approaches can yield interesting 1D boundary states. Additionally, we intend to investigate whether the computational power of thermal states can be characterized by the boundary entanglement structures.

\begin{acknowledgments}
  We gratefully acknowledge computing resources from Research Services at Boston College and the assistance provided by Wei Qiu. H.L. thanks Tianci Zhou, Shengqi Sang, Timothy H. Hsieh, Tsung-Cheng (Peter) Lu, and Leonardo Lessa for their discussions and comments. H.L. thanks Yizhou Ma for pointing out typos in the draft. We thank the anonymous referee for constructive feedback. This research is supported in part by the Google Research Scholar Program and is supported in part by the National Science Foundation under Grant No. DMR-2219735.
\end{acknowledgments}

\nocite{*}

\appendix
\section{Bipartite Cluster State and Ising model}\label{ap: graph_ground}
This appendix presents an alternative approach to establish the connection between sampling the bipartite cluster state and the Ising partition function, drawing on techniques introduced in \cite{de2009completeness}.

The cluster state $\ket{\Psi}$ serves as a ground state of the following Hamiltonian:
\begin{equation}\label{eq: bipartite_hamiltonian}
    H = -\frac{\Delta}{2} \sum_i K_i \quad \text{where} \quad K_i = X_i \prod_{n\in N_i} Z_n.
\end{equation}
Here, $\Delta$ represents the energy gap, and $N_i$ denotes the neighboring sites of $i$ \cite{PhysRevA.71.062313}. For cluster states defined on a bipartite graph, the Hamiltonian can be expressed as:
\begin{equation}\label{eq:bipartite_hamiltonian}
    H = -\frac{\Delta}{2} \left(\sum_{i\in \circ} K_i + \sum_{j \in \bullet} K_j\right).
\end{equation}
The ground state can be directly expressed using classical Ising spins $s$ defined on $\circ$ as follows:
\begin{equation}
    \ket{\Psi} = \frac{1}{2^{|\circ|}}\sum_{s} \bigotimes_{i\in\circ,j\in \bullet}\ket{s_i}_{i}\otimes\ket{Q_j}_j,
\end{equation}
where $Z_i\ket{s_i} = s_i \ket{s_i}$, $X_j \ket{Q_j} = Q_j \ket{Q_j}$, and $Q_j = \prod_{i \in N_j} s_i$. It can be verified that for any $l\in \bullet$:
\begin{equation}
\begin{aligned}
     K_l \ket{\Psi} = \sum_s Q_l \times \prod_{n\in N_l} s_n\bigotimes_{i\in\circ,j\in \bullet}\ket{s_i}_{i}\otimes\ket{Q_j}_j\\
     =\sum_s \bigotimes_{i\in\circ,j\in \bullet}\ket{s_i}_{i}\otimes\ket{Q_j}_j,
\end{aligned}
\end{equation}
as $Q_l \times \prod_{n\in N_l} s_n = 1$ by definition. For any $m\in \circ$:
\begin{equation}
    K_m \ket{s_m}_{m} \bigotimes_{n\in N_m}\ket{Q_n}_n = \ket{\overline{s_m}}_{m} \bigotimes_{n\in N_m}\ket{\overline{Q_n}}_n,
\end{equation}
which effectively flips the Ising spin $s_m \to \overline{s_m}$. We therefore have $K_m \ket{\Psi} = \ket{\Psi}$ for any $m\in\circ$. It is now evident that $\ket{\Psi}$ is the ground state of \cref{eq:  bipartite_hamiltonian}, hence the cluster state.

For simplicity, we parameterize the measurement wave function as:

\begin{equation}\label{eq: ap_cp_para}
\begin{aligned}
    \ket{\phi}_i =& \sum_{s_i} \exp(h_i s_i) \ket{s_i}_i & \text{for } i\in \circ\\
    \ket{\phi}_j =& \sum_{Q_j} \exp(K_j Q_j) \ket{Q_j}_j & \text{for } j\in \bullet
\end{aligned}
\end{equation}
with $h_i , K_j \in \mathbb{C}$, and again $Z_i\ket{s_i} = s_i \ket{s_i}$, $X_j \ket{Q_j} = Q_j \ket{Q_j}$. The overlap function is then:

\begin{equation}
    \bigotimes_{i\in\circ,j\in \bullet}\bra{\phi}_i \otimes \bra{\phi}_j \cdot \ket{\Psi} = \mathcal{Z}
\end{equation}
where:
\begin{equation}\label{eq: SQ}
    \mathcal{Z} = \sum_{s} \exp\left(\sum_{j\in\bullet} K_j Q_j + \sum_{i\in \circ} h_i s_i \right)
\end{equation}
is the Ising partition function, taking $Q_j = \prod_{n \in N_j} s_n$. Thus, we recover \cref{eq: cluster_ising}. 

Rewriting \cref{eq: ap_cp_para} in the $Z$ basis, we relate the Ising parameters $\{K, h\}$ with the weight parameters introduced in \cref{eq: weight_parameter}:
\begin{equation}
\begin{aligned}
    \exp(-2 h_i) = & W_i & \text{for } i\in \circ\\
    \exp(-2 K_j )= &\frac{1 - W_j}{1 + W_j } & \text{for } j\in \bullet.
\end{aligned}
\end{equation}
This relationship corresponds to \cref{eq: ising_para}.

\section{Boundary Condition and Initial State}
\label{ap: graph_boundary}

In this section, we discuss the relation between the shape of the Lieb lattice boundary and the initial state $\ket{\Psi}_0$ of the dynamics given by the Ising transfer matrix shown in Sec.~\ref{sec: dynamics}. As previously shown in App.~\ref{ap: graph_ground}, the cluster state defined on the Lieb lattice is given by
\begin{equation}
    \ket{\Psi} = \frac{1}{2^{|\circ|/2}}\sum_{s} \bigotimes_{i\in\circ,j\in \bullet}\ket{s_i}_{i}\otimes\ket{\prod_{n\in N_j} s_n}_j,
\end{equation}
where $\prod_{n\in N_j} s_n$ is the product over spins on the neighbouring sites of $j$. Taking into account the boundary of the Lieb lattice, we modify the wave function to
\begin{equation}
    \ket{\Psi} = \frac{1}{2^{|\circ|/2}}\sum_{s} \bigotimes_{i\in\circ,j\in \bullet}\ket{s_i}_{i}\otimes\ket{\prod_{n\in N_j} s_n}_j \otimes \ket{\psi[s_k]},
\end{equation}
where $\ket{\psi[s_k]}$ labels the boundary condition and $s_k$ labels the boundary spin configuration.

For the ``smooth" boundary condition, the boundary stabilizer is $g_m = X_m Z_k Z_l$ where $m = \langle k, l\rangle$ is the edge connecting vertices $k,l$ on the boundary and $g_k = X_k Z_l Z_m Z_p$
where $k$ is the vertex shared by edges $l,m,p$. $\ket{\psi_0[s_k]}$ is naturally
$$
\ket{\psi_0[s_k]} = \bigotimes_{m\in \bullet, l\in \circ}\ket{s_k s_l}_m \ket{s_k}_l
$$
which is physically the ``free" boundary condition of the Ising model. The boundary term of the Ising partition function $\mathcal{Z}_{\text{b.t}}$ is given by overlap $\braket{\phi}{\psi_0[s_k]}$
\begin{equation}
    \mathcal{Z}_{\text{b.t}} = \sum_{s} \exp( \sum_j K_{j} s_l s_m + \sum_k h_k s_k),
\end{equation}
which is the "free" boundary condition of the Ising model. The initial state of the Ising transfer matrix dynamics is thus
$$
 \ket{\Psi_0} \equiv  \frac{1}{2^{|\{s_{\partial \mathcal M}\}|/2}} \sum_{s_{\partial \mathcal M}}\ket{s_{\partial\mathcal M}}.
$$

As for the ``rough" boundary condition, the boundary stabilizer is $g_m = X_m Z_k$ where $m$ is the boundary edge connected to site $k$ and $g_k = X_k Z_l Z_m Z_p Z_q$ where $k$ is the vertex shared by edges $l,m,p,q$, and $m$ is the boundary edge. We may write the boundary term as
\begin{equation}
    \ket{\psi_0[s_k]} = \bigotimes_{m}\ket{s_k}_m.
\end{equation}
where $X_m \ket{s_k}_m = s_k \ket{s_k}_m$. The boundary term is
\begin{equation}
    \mathcal{Z}_{\text{b.t}} = \sum_{s} \exp( \sum_k h_{k}^{\text{ext}} s_k),
\end{equation}
since the boundary qubit is labeled by a single Ising spin. The single-site measurements on the boundary edge result in an external Ising magnetic field in the following form
\begin{equation}
    \exp(-2 h_{k}^{\text{ext}}) = \frac{1- W_m}{ 1 + W_m}.
\end{equation}
where $W_m$ is the measurement weight $m$ on the boundary. In another perspective, we may regard such external field as bulk Ising spins couple to a fixed boundary spin $s_k = + 1$ for all $k$ on the boundary, and the boundary term is written as
\begin{equation}
\begin{aligned}
    \mathcal{Z}_{\text{b.t}} = \sum_{s} \exp( \sum_k h_{k}^{\text{ext}} s_k s_{p}); \\ 
    s_p = 1, ~\forall p ~\text{on boundary}.
\end{aligned}
\end{equation}
The external magnetic field is now interpreted as Ising spins couple to a fixed boundary $s_p = 1$ for all $p$ on the boundary
\begin{equation}
    \ket{\psi_0[s_p]} = \bigotimes_{p}\ket{s_p = + 1}_p
\end{equation}
where $Z_p \ket{s_k = + 1}_p = \ket{s_p = + 1}_p$, and the fixed boundary condition results in the initial state
\begin{equation}
    \ket{\Psi_0} = \ket{s_{\partial_M} = + 1}.
\end{equation}

\section{Overlap function as Ising gauge field}\label{ap: graph_dual}

In this section, we demonstrate that the overlap function on the Lieb lattice can also be expressed using the $\mathbb{Z}_2$ gauge field partition function on a square lattice
$$
     \mathcal{Z} = \sum_{\tau} \exp\left(\sum_{p = \langle a, b\rangle'} \tilde{h_p} \tau_p + \sum_{i} \tilde{K_i} \prod_{q\in \square_i} \tau_q \right),
$$
and $X$ measurements with outcome $m_i$ on vertex $i$ correspond to applying a flux constraint on the gauge field
$$
\prod_{q\in \square_i} \tau_q = m_i.
$$
Directly taking $\tau = \sigma_a \sigma_b$, we recover the Ising partition function
$$
 \mathcal{Z} = \sum_{\sigma} \exp\left(\sum_{p = \langle a, b\rangle'} \tilde{h_p} \sigma_a \sigma_b \right).
$$

\begin{figure}[ht]
    \centering
    \includegraphics[width=0.6\linewidth]{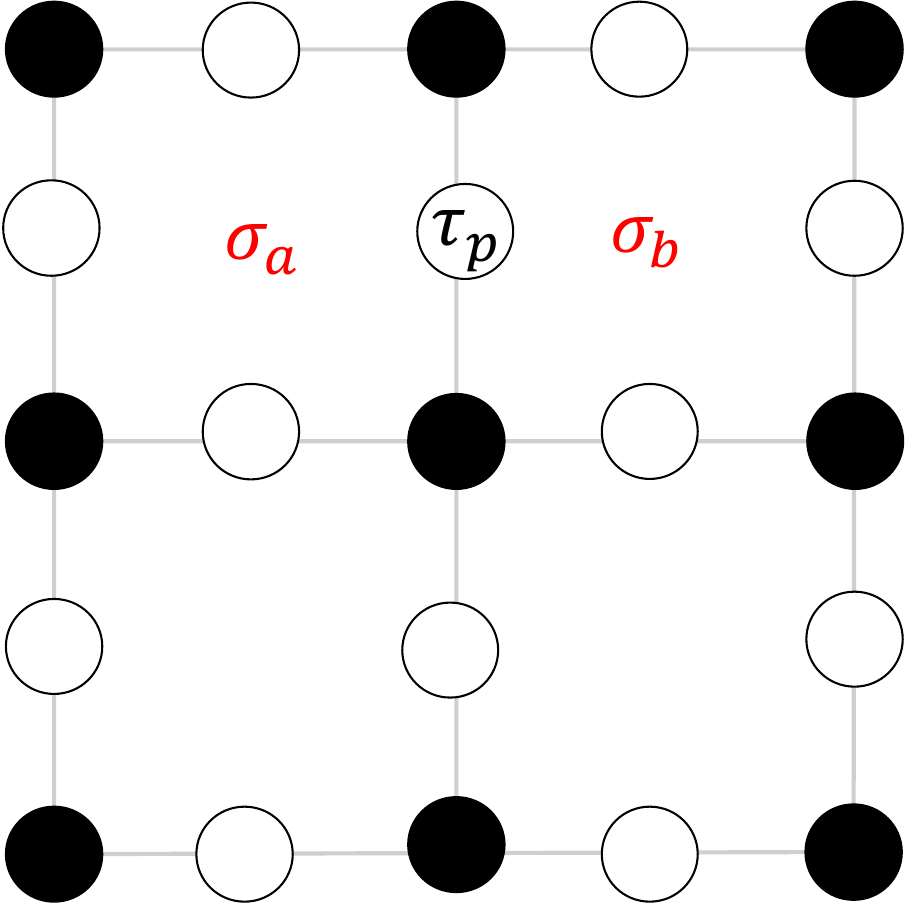}
    \caption{Alternative bipartition of the Lieb lattice. Physical $\mathbb{Z}_2$ degrees of freedom $\tau_p$ are defined on the edge $p$, and the corresponding $\mathbb{Z}_2$ transformations $\sigma_{a},\sigma_{b}$ are assigned to neighboring plaquettes $a$ and $b$.}
    \label{fig: ap_dual}
\end{figure}

As illustrated in Fig.~\ref{fig: ap_dual}, the edges of the Lieb lattice are colored $\bullet$, and the vertices $\circ$. We assign each vertex $i$ a classical spin $s_i$, and each edge $p = \langle j, k\rangle$ a domain-wall variable $Q_{p = \langle j, k\rangle} = s_j s_k$ with $j,k$ being vertices connected by edge $p$. Plugging everything into \cref{eq: SQ}, we recover \cref{eq: lieb_ising}:
\begin{equation*}
    \mathcal{Z} = \sum_{s} \exp\left(\sum_{p = \langle j, k\rangle} K_p s_j s_k + \sum_{i} h_i s_i \right),
\end{equation*}
which corresponds to a 2D classical Ising model with a magnetic field.
Now, we reverse the color of the edges and vertices. The edges are colored $\circ$, and the vertices $\bullet$. We assign each edge $p = \langle a, b\rangle' \in \circ$ a $\mathbb{Z}_2$ variable $\tau_p$, and each site $i$ a $Q_i = \prod_{q\in \square_i} \tau_q$, where $\square_i$ denotes edges enclosing site $i$. To clarify the physical meaning, we parameterize the measurement wave function in \cref{eq: ap_cp_para} as:
\begin{equation}
\begin{aligned}
    \ket{\phi}_p =& \sum_{\tau_p} \exp(\tilde{h_p} \tau_p) \ket{\tau_p}_p & \text{for } p\in \circ\\
    \ket{\phi}_i =& \sum_{Q_i} \exp(\tilde{K_i} Q_i) \ket{Q_i}_i & \text{for } i\in \bullet
\end{aligned}
\end{equation}
where $Z_p\ket{\tau_p}_p = \tau_p \ket{\tau_p}_p$ and $X_i \ket{Q_i}_i = Q_i\ket{Q_i}$. The overlap function becomes:
\begin{equation}\label{eq: ising_gauge}
    \mathcal{Z} = \sum_{\tau} \exp\left(\sum_{p = \langle a, b\rangle'} \tilde{h_p} \tau_p + \sum_{i} \tilde{K_i} \prod_{q\in \square_i} \tau_q \right),
\end{equation}
which gives the partition function for $\mathbb{Z}_2$ gauge theory, as $\mathcal{Z}$ remains invariant under the transformation of $\tau_p$:
\begin{equation}
    \tau_p\to \tau_p \sigma_a \sigma_b \quad \text{where} \quad \sigma\in \mathbb{Z}_2,
\end{equation}
and $a,b$ label the plaquettes sharing edge $p$. The relationship between $\{\tilde{K}, \tilde{h}\}$ and the weight parameters $\{W_i\}$ is given by:
\begin{equation}
\begin{aligned}
     \exp(-2\tilde{K}_i) &= \frac{1 - W_i}{ 1 + W_i} & \text{for } i\in \circ\\
     \exp(-2\tilde{h_p}) &= W_p & \text{for } p\in \bullet.
\end{aligned}
\end{equation}
When measuring the vertex sites along $X$, we have $\tilde{K}_i \to + \infty\times m_i $, where $m_i$ is the measurement outcome on vertex $i$. Consequently, we obtain:
\begin{equation}
    \mathcal{Z} = \sum_{\tau} \exp\left(\sum_{p = \langle a, b\rangle'} \tilde{h_p} \tau_p \right)\prod_{i}(1 + m_i \prod_{q\in \square_i} \tau_q).
\end{equation}
The second term $\prod_{i}(1 + m_i \prod_{q\in \square_i} \tau_q)$ can be understood as a constraint on $\tau$:
\begin{equation}\label{eq: flux}
    \prod_{q\in \square_i} \tau_q = m_i.
\end{equation}

When all measurement outcomes are set to $+1$, we have $\prod_{q\in \square_i} \tau_q = 1$ for all vertices $i$, implying that $\tau = \sigma_a \sigma_b$. Thus, the overlap function becomes:
\begin{equation}
    \mathcal{Z} = \sum_{\sigma} \exp\left(\sum_{p = \langle a, b\rangle'} \tilde{h_p} \sigma_a \sigma_b \right),
\end{equation}
with $\exp(-2\tilde{h_p}) = W_p$. This corresponds to a Kramers-Wannier dual of \cref{eq: lieb_ising} at $h = 0$.

\section{Dynamics from Tensor Contractions}\label{ap:TContDyn}

\begin{figure*}
    \centering
    \includegraphics[width=\textwidth]{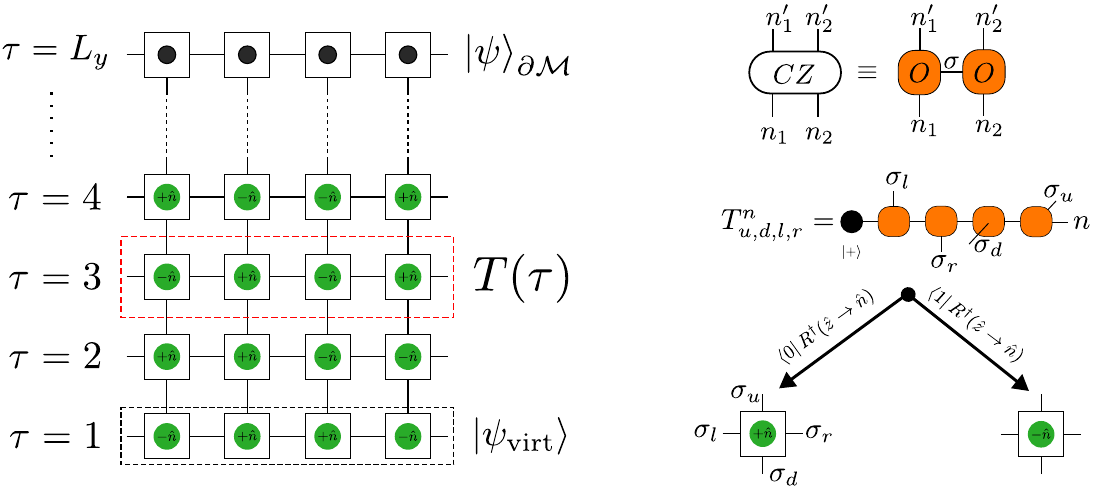}
    \caption{The process by which $\ket{\psi}_{\partial\mathcal{M}}$ is obtained on the square lattice. (Left) Each square represents a tensor of the form given in \cref{eq:T_def}. The green circles denote outcomes (parallel or anti-parallel to the direction $\hat{n}$) for measurements performed on the bulk. The black circles represent the physical degrees of freedom of $\ket{\psi}_{\partial\mathcal{M}}$. (Right) The decomposition of $CZ$ into MPOs as given in \cref{eq:CZTens}, and the consequent iterative construction of $T^n_{u,l,d,r}$ starting from the $\ket{+}$ state.}
    \label{fig:network}
\end{figure*}

Certain 2D states, such as the cluster state, can be represented as a network of tensors \cite{ORUS2014117,RevModPhys.93.045003,PollmannSP2018}, with each index of a tensor represented as a ``leg", with the understanding that legs that are connected correspond to indices that are summed over. Each tensor further has a physical index corresponding to the physical degrees of freedom, qubits, in our case. We distinguish these physical legs from the others by terming the latter ``virtual" legs. The process of measuring a qubit in the Pauli $Z$ basis is equivalent to fixing a specific value $0$ or $1$ of its physical index, which effects a projection $\dyad{0}$ or $\dyad{1}$ on that qubit. Measurements in other bases can be implemented by applying an appropriate single-qubit rotation prior to constraining the value of the physical index. At the end of our protocol, the tensor only has physical legs at the top edge. The boundary state that remains after measuring the bulk qubits is obtained by contracting over all the remaining virtual indices in the bulk of the network. However, owing to the associativity of tensor multiplication and the commutativity of projectors on different sites, we are free to choose the order in which the tensor network can be contracted to obtain the boundary state. A specific pattern of contractions, which can be interpreted as the evolution of a 1D state through a random nonunitary circuit, is described in the following paragraph and in \cref{fig:network}. This procedure can be used to obtain the boundary state following any set of measurement outcomes $\qty(\mu,\omega)$.

Concretely, we begin by considering the lowest row of the cluster state, defined on a square lattice with $L_x$ columns and $L_y$ rows. This row can be treated as an MPS $\ket{\psi_{\rm virt} (\tau=1)}$ where the $L_x$ virtual legs that connect this row to the next serve as physical legs. The tensors in the next row have virtual legs connected to tensors in rows both below and above them. They can be thought of as the Matrix Product Operator (MPO) decomposition of a nonunitary operator. We term this operator $T$, reminiscent of the transfer matrix from the previous section. The state $\ket{\psi_{\rm virt}}$ can now be ``evolved" nonunitarily by contracting the legs it shares with $T$, and this process is iteratively continued over subsequent rows until the boundary state, i.e., the $L_y^{\rm th}$ row, is reached. The vertical direction can be reinterpreted as a discrete effective time direction. After $\tau\leq L_y-1$ such rows have been contracted -- equivalently, $\tau$ ``time steps" -- the state is
\begin{equation}
    \ket{\psi_{\rm virt} (\tau)} = \frac{\prod\limits_{\tau'<\tau}T(\tau') \ket{\psi_{\rm virt}}}{\norm{\prod\limits_{\tau'<\tau}T(\tau') \ket{\psi_{\rm virt}}}}.
\end{equation}

The boundary state $\ket{\psi}_{\partial\mathcal{M}}$ is obtained by contracting $\ket{\psi_{\rm virt}}$ with the $L_y^{\rm th}$ row, resulting in an MPS with physical indices, as shown in \cref{fig:network}.

\subsubsection{Obtaining Tensors for the Cluster State}

The cluster state, as defined in \cref{eq:clusDefn}, is obtained by applying $CZ$ gates across all pairs on neighboring qubits, all of which are initialized in the $\ket{+}$ state. We first derive the tensors for two neighboring qubits, labeled $1$ and $2$. Initially, $\ket{\psi}_1 = \ket{+}_1$, $\ket{\psi}_2 = \ket{+}_2$. Their joint wavefunction can be represented as 
\begin{equation}
    \ket{\psi} = \sum\limits_{n_1, n_2 =0,1} A^{n_1} A^{n_2} \ket{n_1,n_2}
\end{equation}
where $\ket{n_1,n_2}$ is a state in the computational basis of the two qubits. The matrices $A^{n_j} = \frac{1}{\sqrt{2}}$ for $n_j=0,1$.

We proceed by obtaining a representation of the $CZ$ gate in terms of tensors as 
\begin{equation}
    CZ = \sum\limits_{\substack{n_1,n_1'\\n_2,n_2',\sigma}} O_\sigma^{n_1',n_1} O_\sigma^{n_2',n_2} \dyad{n_1',n_2'}{n_1,n_2},
\end{equation} where $\sum\limits_{n_j',n_j} O_\sigma^{n_j',n_j}\dyad{n_j'}{n_j}$ is an operator acting on the $j^{\rm th}$ qubit for each $\sigma$. After the application of the $CZ$ gate, the state 
\begin{equation}
    \ket{\psi}\to CZ\ket{\psi} = \sum\limits_{n_1', n_2', \sigma} B^{n_1'}_\sigma B^{n_2'}_\sigma \ket{n_1',n_2'}
\end{equation}
where $B^{n_j'}_\sigma = \sum\limits_{n_j} O_\sigma^{n_j',n_j}A^{n_j}$. The action of the $CZ$ gate is given by 
\begin{equation}
    CZ\ket{n_1,n_2} = (-1)^{n_1.n_2} \ket{n_1,n_2}.
\end{equation}
$CZ$ can equivalently be written as
\begin{equation}
    \begin{aligned}
        CZ &= P_1^0 + P_1^1 Z_2\\
        &= 2\qty(\frac{1+Z_1}{2})\qty(\frac{1+Z_2}{2}) - Z_1 Z_2\\
        &= \mqty(\sqrt{2}P_1^0& iZ_1) \mqty(\sqrt{2}P_2^0\\
        iZ_2),
    \end{aligned} 
\end{equation}
where $P^{0/1}_j\equiv \frac{1\pm Z_j}{2}$ is the projector onto $\ket{0}_j$or$\ket{1}_j$ and $Z_j$ is the Pauli $Z$ operator, both on the $j^{\rm th}$ qubit. From the last line, $ O_\sigma^{n_j',n_j}$ can be read off as 
\begin{equation}
     O_\sigma^{n_j',n_j} = \begin{cases}
     \sqrt{2}\qty(P^0)^{n_j',n_j}, &\sigma = 0\\
     iZ^{n_j',n_j}, &\sigma=1
     \end{cases},
     \label{eq:CZTens}
\end{equation}
with $M^{n_j',n_j} \equiv \mel{n_j'}{M}{n_j}$ for every 1-qubit operator $M$. Returning to the cluster state, the tensor corresponding to a vertex $v$, $T^n_{\sigma_1, \sigma_2,\dots,\sigma_k}$, whose neighbours are $v_1,v_2\dots v_k$, and $\sigma_j$ labels the leg or index shared between $v$ and $v_j$, is given by
\begin{equation}
    T^n_{\sigma_1, \sigma_2,\dots,\sigma_k} = \frac{1}{\sqrt{2}}\sum\limits_{\qty{n_j}=0,1} O_{\sigma_1}^{n,n_1}O_{\sigma_2}^{n_1,n_2}\dots O_{\sigma_k}^{n_{k-1},n_k}.
    \label{eq:TDef}
\end{equation}
The unnormalized state of $T$ following a projective measurement is obtained by
\begin{equation}
    \qty(P^{0/1})^{n,n'} \qty(R^\dagger(\hat{z}\to\hat{n}))^{n',n''} T^{n''}_{\sigma_1,\sigma_2,\dots,\sigma_k},
    \label{eq:postmeas_T}
\end{equation}
where $R\qty(\hat{z}\to\hat{n})$  is the unitary operator effecting a single-qubit rotation from the $\hat{z}$ to the $\hat{n}$ direction, for measurements performed along $\hat{n}$.

As an illustration, we provide a method by which the tensors for the cluster state on the infinite square lattice may be obtained. Let $T^{n_{x,y}}_{u_{x,y},d_{x,y},l_{x,y},r_{x,y}} (x,y)$ denote the tensor corresponding to the qubit located at the vertex $(x,y)$ of the square lattice. $n$ refers to the physical index, while $u,d,l$, and $r$ refer to virtual indices, respectively, the legs that point up, down, left, and right. The location dependence of these indices will be suppressed in their notation unless there is potential for ambiguity. From \cref{eq:TDef}, with $O$ as in \cref{eq:CZTens},

\begin{equation}
    T^n_{u,d,l,r} = \frac{1}{\sqrt{2}}\sum\limits_{\qty{n_j}=0,1} O_u^{n,n_1}O_d^{n_1,n_2}O_l^{n_2,n_3} O_r^{n_3,n_4}.
\end{equation}

This calculation of $T$ only needs to be performed \textit{once}. Following this, the tensors for all other vertices are straightforwardly obtained by relabelling the indices to match the virtual indices of those vertices. By the geometry of the square lattice, the following indices are to be identified: 
\begin{equation}
    \begin{aligned}
        r_{x,y} &\leftrightarrow l_{x+1,y}\\
        u_{x,y} &\leftrightarrow d_{x,y+1}\\
    \end{aligned}.
    \label{eq:Identify_inds}
\end{equation}

Finally, the cluster state can be written as 
\begin{equation}
    \begin{aligned}
    \ket{\psi_{\rm G}} &= \sum\limits_{\qty{n_{x,y}}}\sum\limits_{\qty{\substack{u_{x,y},d_{x,y}\\l_{x,y},r_{x,y}}}}\\
    &\qty(\prod_{x,y} T^{n_{x,y}}_{u,d,l,r} (x,y) \delta_{r_{x,y},l_{x+1,y}} \delta_{u_{x,y},d_{x,y+1}}) \ket{\qty{n_{x,y}}},
    \end{aligned}
\end{equation}
with the Kronecker-$\delta$s enforcing the identification \cref{eq:Identify_inds} in the summation. The last step in the protocol is the measurement of the bulk qubits, leaving behind the boundary state $\ket{\psi}_{\partial\mathcal{M}}$. We choose a uniform direction $\hat{n}$ along which all the bulk measurements are performed. With $\hat{n}$ fixed, the boundary state can uniquely be described by a collection of $L_x \times (L_y - 1)$ bits $\qty{s_{x,y}}$, where $s_{x,y} = 0/1$ refers to a projective measurement resulting in a state parallel $\qty(\ket{+\hat{n}})$ or anti-parallel $\qty(\ket{-\hat{n}})$ to $\hat{n}$ at site $(x,y)$. This step is implemented in two sub-steps. 
First, we rotate each qubit using the unitary 
\begin{equation}
    \qty(R^\dagger(\hat{z}\to\hat{n}))^{n_{x,y},n'_{x,y}}\dyad{n_{x,y}}{n'_{x,y}},
\end{equation}

where $R(\hat{z}\to\hat{n})$ is the unitary operator that effects a rotation from the $\hat{z}$ axis to $\hat{n}$, as in \cref{eq:postmeas_T}. Next, we project that bulk qubit onto the state $\bra{s_{x,y}}$ by multiplying the tensor $T^n_{u,d,l,r}(x,y)$ by an appropriate vector $\qty(P^{0,1})^n$, corresponding to a projection onto $\bra{0/1}$. The unnormalized boundary state $\ket{\psi}_{\partial\mathcal{M}}$ is finally given by
\begin{equation}
    \begin{aligned}
        \ket{\psi}_{\partial\mathcal{M}} \propto \Biggl(\prod_{\qty(x,y)\in {\rm bulk}} \qty(P^{s_{x,y}})^{n_{x,y}}\qty(R^\dagger)^{n_{x,y},n'_{x,y}} \times \\ T^{n'_{x,y}}_{u,d,l,r} (x,y) \delta_{r_{x,y},l_{x+1,y}} \delta_{u_{x,y},d_{x,y+1}}\Biggr)\\
        \qty(\prod_{x,y = L_y} T^{n_{x,y}}_{d,l,r} (x,y) \delta_{r_{x,y},l_{x+1,y}} \ket{n_{x,y}}),
    \end{aligned}
    \label{eq:edge_tens}
\end{equation}
with the understanding that repeated indices are summed over. The absence of $u$ indices in the last line stem from the fact that the topmost row of the square lattice has no neighbors -- and hence no virtual indices -- above it.

\subsubsection{Implementing the {\blue \((1+1)\mathrm{D}\) } Circuit}

The implementation of the {\blue \((1+1)\mathrm{D}\) } circuit that reproduces the boundary state of the square lattice once its bulk degrees of freedom have been measured can now be described in detail. We begin by noting that the expression for $\ket{\psi}_{\partial\mathcal{M}}$ in \cref{eq:edge_tens} consists of several tensor multiplications and will focus first on the terms in the first parentheses. Two types of multiplications are involved in this process -- those involving contractions of the physical indices $n$ and those involving the virtual indices $\qty{u,d,l,r}$.
Crucially, the order in which these contractions are performed is irrelevant to the final solution, since matrix multiplication is associative. No contraction of the physical indices involves more than one site at a given time. By choosing the order of contractions, an interpretation of this process as a quantum circuit follows naturally. This can be demonstrated by considering the terms in the parentheses corresponding to the lowest row and subsequent rows separately. We proceed by expressing $\delta_{u_{x,y}, d_{x,y+1}}$ as the inner-product of basis states as $\delta_{u_{x,y}, d_{x,y+1}}\equiv \braket{d_{x,y+1}}{u_{x,y}}$, and defining, for $1<y<L_y$,
\begin{equation}
    T_{u,d,l,r} (x,y) \equiv \qty(P^{s_{x,y}})^n\qty(R^\dagger)^{n,n'} T^{n'}_{u,d,l,r} (x,y),
    \label{eq:T_def}
\end{equation}
and for $y=1$, since these tensors -- being those on the last row of the lattice -- do not have any virtual indices ``below" them, as shown in \cref{fig:network},
\begin{equation}
    B_{u,l,r} (x) \equiv \qty(P^{s_{x,1}})^n\qty(R^\dagger)^{n,n'} T^{n'}_{u,l,r} (x,1).
    \label{eq:B_def}
\end{equation}

The term in the first parentheses can be rewritten as 
\begin{equation}
    \begin{aligned}
        \prod_{x, 1<y\leq L_y-1} 
        T_{u,d,l,r}(x,y)\delta_{r_{x,y},l_{x+1,y}} \braket{d_{x,y+1}}{u_{x,y}}\times\\
        \prod_{x,y=1} B_{u,l,r}(x) \delta_{r_{x,1},l_{x+1,1}}\braket{d_{x,2}}{u_{x,1}}.
    \end{aligned}
\end{equation}

The order of the bras and kets can be rearranged to give

\begin{equation}
    \begin{aligned}
        &\bra{d_{1,L_y}d_{2,L_y}\dots d_{L_x,L_y}}&\\
        &\qty(\prod_{x, 1<y\leq L_y-1} 
        T_{u,d,l,r}(x,y)\delta_{r_{x,y},l_{x+1,y}}\dyad{u_{x,y}}{d_{x,y}})\\
        &\hspace{5.5em}\qty(\prod_{x,y=1} B_{u,l,r}(x)\delta_{r_{x,1},l_{x+1,1}}\ket{u_{x,y=1}}).
    \end{aligned}
\end{equation}

For each $1<y\leq L_y-1$, we can define an effective time-evolution operator $T(\tau)$ as
\begin{equation}
    T(\tau) = \sum\limits_{\qty{\substack{u_{x,\tau},d_{x,\tau}\\l_{x,\tau},r_{x,\tau}}}} \prod_{x} T_{u,d,l,r} (x,\tau) \delta_{r_{x,\tau},l_{x+1,\tau}}\dyad{u_{x,\tau}}{d_{x,\tau-1}},
    \label{eq:MPO_TMat}
\end{equation}
and thus, express the terms in the first parentheses of \cref{eq:edge_tens} as an inner product $\braket{d_{1,L_y}d_{2,L_y}\dots d_{L_x,L_y}}{\psi_{\rm virt}(L_y)}$, where
\begin{equation}
    \begin{aligned}
        &\ket{\psi_{\rm virt}(L_y)}\equiv T(L_y-1)T(L_y-2)\dots T(2) \ket{\psi_{\rm virt}(\tau=1)}\\
        &\ket{\psi_{\rm virt}(\tau=1)} \equiv \qty(\prod_{x,y=1} B_{u,l,r}(x)\delta_{r_{x,1},l_{x+1,1}}\ket{u_{x,y=1}}).
    \end{aligned}
    \label{eq:MPS_psivirt}
\end{equation}
The vertical coordinate $y$ has been relabelled as an effective temporal direction $\tau$. The time evolution of $\ket{\psi_{\rm virt}}$ can equivalently be seen as an evolution of the $\qty{B_{u,l,r}(x)}_\tau$, according to the rule 

\begin{equation}
    B_{u,\bar{l},\bar{r}}(x,\tau) = T_{u,l,d,r}(x,\tau) B_{u',l',r'}(x,\tau-1)\delta_{d,u'},
\end{equation}
where $\bar{l}$ and $\bar{r}$ denote the combined indices $(l,l')$ and $(r,r')$ respectively.

To obtain the boundary state $\ket{\psi}_{\partial\mathcal{M}}$ at the final step of this dynamical evolution, the operator $T(\tau=L_y)$, which maps the free virtual indices $\qty{u_{x,y=L_y-1}}$ to the physical indices $\qty{n_{x,y=L_y}}$, is given by the last line of \cref{eq:edge_tens} as

\begin{equation}
    T(\tau=L_y) = \prod_{x,y = L_y} T^{n_{x,y}}_{d,l,r} (x,y) \delta_{r_{x,y},l_{x+1,y}} \dyad{n_{x,L_y}}{d_{x,L_y}}.
\end{equation}
The tensors for $T(\tau=L_y)$ are those that constitute the top-most row of a cluster state on a square lattice. Their $d$ indices must be contracted with the $u$ indices of the MPS describing $\ket{\psi_{\rm virt}(\tau=L_y-1)}$ to obtain $\ket{\psi}_{\partial\mathcal{M}}$. \cref{eq:edge_tens} can finally be expressed as
\begin{equation}
    \ket{\psi}_{\partial\mathcal{M}} \propto \prod\limits_{2\leq\tau\leq L_y} T(\tau) \ket{\psi_{\rm virt}(\tau = 1)}.
\end{equation}

We conclude this section by providing a brief algorithm that describes this implementation. Note that \cref{eq:MPO_TMat} describes the MPO representation of $T$, while \cref{eq:MPS_psivirt} describes the MPS representation of the state $\ket{\psi_{\rm virt}}$. This identification allows us to use the extensively studied techniques for the time evolution of 1D quantum states.

\begin{enumerate}
    \item Given a measurement direction $\hat{n}$, the measurement outcomes $\qty{s_{x,y}}$, and the dimensions of the square lattice $L_x$ and $L_y$. Assume $L_y\geq L_x$.
    \item Obtain the unitary gate $R$ which effects a rotation from $\hat{z}$ to $\hat{n}$
    \item Calculate and store $B_{u,l,r}$ for both measurement outcomes $s=0,1$, using \cref{eq:B_def}.
    \begin{itemize}
        \item According to $\qty{s_{x,y=1}}$, create an MPS for $\ket{\psi_{\rm virt}}$, where the tensors at site $x$ are $B^{(s_{x,y=1})}_{u,l,r}$.
        \item Designate $u$ as a physical index, and relabel the virtual indices $l,r$ uniquely, ensuring, however, that $r_{x} = l_{x+1}$ (this enforces $\delta_{r_{x,y},l_{x+1,y}}$).
    \end{itemize}
     
    \item Calculate and store $T_{u,d,l,r}$ for either measurement outcome, using \cref{eq:T_def}. These tensors $T_{u,d,l,r}(x,y>1)$ constitute the MPO representations of $T$. 
    \begin{itemize}
        \item Use standard techniques \cite{PollmannSP2018,itensor} to ``time-evolve" (and truncate, if necessary) the MPS $\ket{\psi_{\rm virt}}$.
        \item Normalize the state $\ket{\psi_{\rm virt}}$.
        \item Repeat this process for $\tau=L_y-1$ time-steps.
    \end{itemize}
    \item Evolve the state with $T(\tau=L_y)$ to  obtain $\ket{\psi}_{\partial\mathcal{M}}$.
\end{enumerate}

\bibliography{apssamp}

\end{document}